%% file: main.tex
\documentclass[a4paper,11pt]{article}
\pdfoutput=1 

\usepackage{jinstpub} 
\usepackage[british]{babel}
\usepackage[separate-uncertainty = true,multi-part-units=single]{siunitx}                
\usepackage{placeins} 
\usepackage{graphicx}
\usepackage{subfigure}
\usepackage{amsmath}
\usepackage[]{units}
\usepackage[colorinlistoftodos]{todonotes}
\usepackage{datatool, filecontents} 
\usepackage{hhline}

\collaboration{MicroBooNE Collaboration} 

\title{A Method to Determine the Electric Field of Liquid Argon Time Projection Chambers Using a UV Laser System and its Application in MicroBooNE}

\input{microboone-author-list-july2019-JINST}

\abstract{
Liquid argon time projection chambers (LArTPCs) are now a standard detector technology for making accelerator neutrino measurements, due to their high material density, precise tracking, and calorimetric capabilities. 
An electric field (E-field) is required in such detectors to drift ionized electrons to the anode to be collected.
The E-field of a TPC is often approximated to be uniform between the anode and the cathode planes. 
However, significant distortions can appear from effects such as mechanical deformations, electrode failures, or the accumulation of space charge generated by cosmic rays.
The latter is particularly relevant for detectors placed near the Earth's surface and with large drift distances and long drift time. 
To determine the E-field in situ, an ultraviolet (UV) laser system is installed in the MicroBooNE experiment at Fermi National Accelerator Laboratory. 
The purpose of this system is to provide precise measurements of the E-field, and to make it possible to correct for 3D spatial distortions due to E-field non-uniformities. 
Here we describe the methodology developed for deriving spatial distortions, the drift velocity and the E-field from UV-laser measurements.

}


\begin{document}
\maketitle
\flushbottom

\section{Introduction}
\label{sec:LaserSys}
Liquid argon time projection chambers (LArTPCs) are suitable detectors for neutrino experiments at many scales, through very massive detectors. 
Several massive LArTPCs have already been built and the technology will also be used for the Deep Underground Neutrino Experiment (DUNE)~\cite{DUNE}. 
MicroBooNE~\cite{MicroBooneDesign} is the first large LArTPC built at the Fermi National Accelerator Laboratory (FNAL) as part of the short baseline neutrino (SBN) program~\cite{antonello2015proposal}.  
As the MicroBooNE detector is placed near the surface, space charge effects induced by comparatively slow drifting ions produced by cosmic ray muon interactions can alter the local electric field (E-field). 
Since the ion drift speeds are comparable to fluid speeds generated by convection in the cryostat, space charge effects are difficult to predict.
These E-field variations can adversely affect event reconstruction, as ionized electrons experience different drift velocities along their path to the anode and initial ionized electrons have different recombination rates.
This further reduces track and energy reconstruction efficiencies of the detector and introduces additional systematic uncertainties. 
To account for these spatial distortions and measure the E-field together with the drift velocity in MicroBooNE, a novel ultraviolet (UV) laser calibration system has been installed.
Unlike cosmic muons, another calibration source, laser beams do not experience delta ray emission or multiple Coulomb scattering in LAr.
Laser beams can also be repetitively pulsed in controllable directions.
Additionally, the UV laser system can be used to investigate detector failures, such as unresponsive or mis-configured wires in the read-out planes.


It has been shown that a UV laser can generate tracks in a LArTPC through multiphoton ionization ~\cite{laserhardware, larlaser, Prototype_UVlaser_LAr, twophoton}. 
Reproducible laser tracks over \SI{5}{\metre} long are observed in ARGONTUBE~\cite{Argontube}.
The attenuation of the laser beam pulse energy is negligible at this scale.
The feasibility of using a UV laser for free electron lifetime measurements in LArTPCs has also been shown~\cite{Argontube}.
Multiphoton ionization strongly depends on the beam intensity.
Argon atoms in the liquid phase can be lifted to an excited state through a laser-induced virtual state by absorption of two UV photons (\SI{266}{\nano\metre}).
An additional UV photon provides the correct energy to ionize the excited argon atom in liquid phase.
A more detailed discussion of this process can be found in ref.~\cite{twophoton}.


In section~\ref{sec:laser_setup}, we describe the hardware set-up of the MicroBooNE laser systems, its operation, and the result of laser beams scanning over the TPC volume and the calibration of true laser track positions, which are the actual paths of laser beams.
The track reconstruction and selection are both optimized for the laser, which is illustrated in section~\ref{sec:Reco_select}.
A simulation of laser tracks is described in section~\ref{sec:TrackSim}.
In section~\ref{sec:Dmap}, we describe a complete methodology for computing track point spatial displacement maps and their uncertainties. 
A bias study of the methodology is demonstrated using laser track simulations in the same section.
Techniques for extracting the E-field and drift velocity over the TPC volume from the spatial displacement maps are explained in section~\ref{sec: Emap}.
The measurement results of MicroBooNE spatial displacement, drift velocity and E-field maps are presented in section~\ref{sec:measurement_results}.
A study of the temporal stability of the E-field is shown in section~\ref{sec:efieldtime_studies}, with the distortion measured continuously for several hours.

\section{Laser System Description and TPC Volume Scans}
\label{sec:laser_setup}

The MicroBooNE laser system consists of two identical UV laser sub-systems.
One is located upstream of the TPC with respect to the Booster neutrino beam, and the other is located downstream of the TPC.
Each sub-system uses a commercial Nd:YAG laser module Surelite I-10~\cite{laser}.
The Surelite I-10 initially generates infrared (IR) light (\SI{1064}{\nano\metre}), which is shifted to green (\SI{532}{\nano\metre}) first, and then UV (\SI{266}{\nano\metre}) through second and fourth harmonic generators.
The output laser light has most of its intensity in the UV, with significant residual green and IR.
The maximum laser pulse repetition rate is \SI{10}{\hertz}.
The pulse duration is 4--\SI{6}{\nano\second}.
The UV laser light output of the Surelite I-10 in each pulse has an energy of \SI{60}{\milli\joule}.

\begin{figure}[htbp]
	\centering
	\includegraphics[width=0.5\textwidth,]{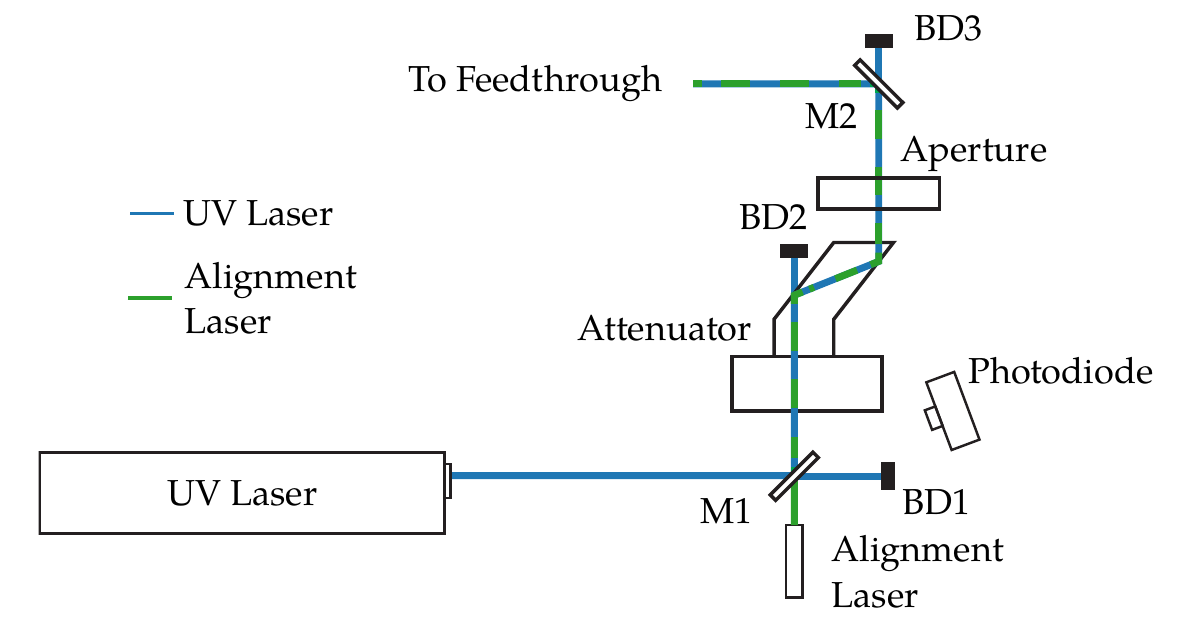}
	\includegraphics[width=0.4\textwidth,]{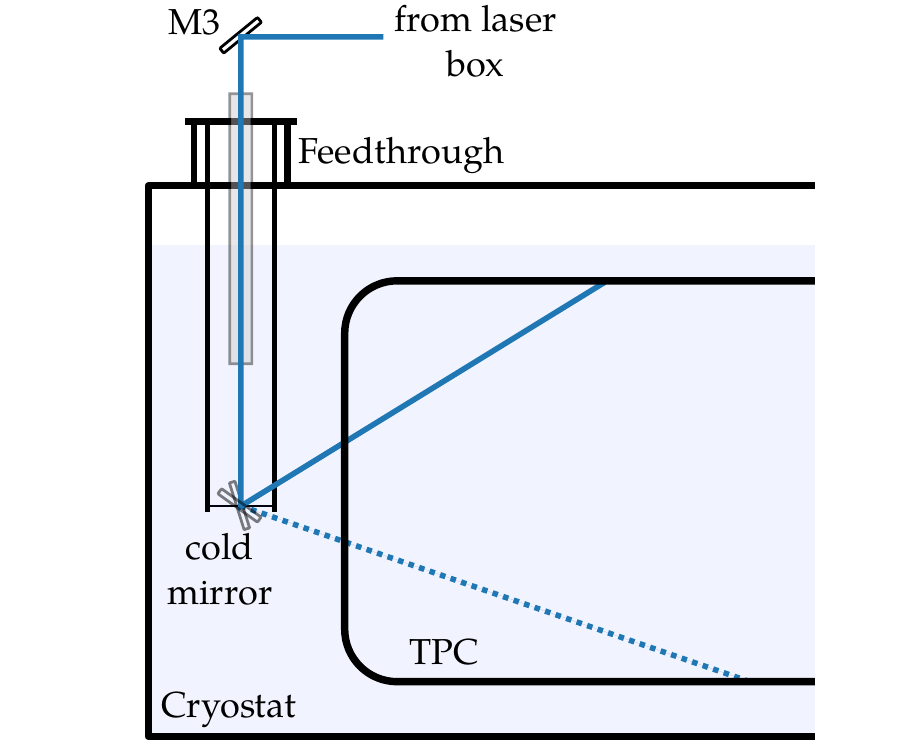}
	\caption{
		The laser system of MicroBooNE;
		the upstream and downstream laser sub-systems have the same layout.
		\textit{Left}: The optical path of the laser beam before entering the cryostat. 
		The \SI{266}{\nano\metre} laser beam is aligned by two mirrors (M1 and M2), and directed towards the feedthrough. 
		An attenuator for beam energy control and an aperture for beam size control are placed in between the two mirrors M1 and M2.
		A photodiode provides a trigger for the readout.
		\textit{Right}: Schematic of the laser feedthrough in the cryostat. 
		The UV laser beam reflects at the dichroic mirror M3 and then enters an evacuated quartz tube which serves 
                as light guide to avoid defocusing of the laser beam at the LAr surface. 
		The cold mirror can rotate horizontally together with the feedthrough assembly.
		A movable rod extending to a cogwheel allows the cold mirror to rotate vertically. 
		The supporting structure of the cold mirror is mounted to the feedthrough flange. }
	\label{fig:Laser_Sys_Schematic}
\end{figure}

To select only UV light, we use a wavelength separator which is composed of two dichroic mirrors, an attenuator, and an aperture. 
Figure~\ref{fig:Laser_Sys_Schematic} shows a schematic of the laser systems.
The first mirror (M1) has high transmittance for IR light and high reflectance for green and UV light.
A beam dump (BD1) at the transmitting side of M1 stops the IR laser light.
The second mirror (M2) has high reflectance for UV light while most of the green light is transmitted, terminating at another beam dump (BD3).
M2 is mounted on a ZABER T-OMG~\cite{gimbal}.
The gimbal has actuators which provide remote control. 
Both dichroic mirrors are supplied by Continuum~\cite{lasermirror}.
In between M1 and M2, there is an Altechna Enhanced Watt Pilot attenuator~\cite{attenuator1,attenuator2}, which is operating in transmission mode.
The attenuation is motorized and can be controlled remotely, with a range of 0.5\% to 95\% transmittance at \SI{266}{\nano\metre}. 
It is used to lower and stabilize the laser beam pulse energy. 
If the energy of the laser pulse is too high or too low, there will be too many or too few ionized electrons to leave a well defined track.
A beam dump (BD2) is at the backside of the attenuator to terminate the passing laser.
In front of M2, a remote controlled aperture is used to limit the diameter of the transverse laser beam to \SI{1}{\milli\metre}.

A green laser is placed at the backside of M1 in order to reach the same optical path as the UV laser beam.
It can give a visual guide for the UV laser beam during alignment and maintenance.
A photodiode is used to trigger the data acquisition (DAQ) for laser events when its signal crosses a threshold.
Thus, each laser event contains one laser track, unless the beam pulse is blocked by the field-cage rings, which are used to shape E-field of the TPC along the drift distance.

All the optical components mentioned above are contained within a light-tight aluminum "laser box".
The output of the laser box is the UV-only beam which is guided to the last warm mirror (M3).
It uses the same gimbal as M2, enabling remote control of the mirror angle.
The last mirror (cold mirror) is in liquid argon and reflects the UV-laser beam into the TPC volume.
It is supported by a \SI{2.5}{\metre}-long feed-through.
The cold mirror is controlled by two motors~\cite{motors}, and can rotate both vertically (polar) and horizontally (azimuthal). 
Two independent encoders measure the azimuthal and polar angles with high precision.
A \SI{2}{\metre}-long evacuated quartz light guide allows the UV-laser beam to enter the LAr without disturbance at the liquid surface.
The cold mirror is within the cryostat but outside of the TPC. 
It is mounted close to the field-cage rings, which block some beam paths into the TPC.
A detailed description of the laser system can be found in refs.~\cite{MicroBooneDesign} and~\cite{laserhardware}.

\subsection{Laser Scan}
\label{subsec:LaserScan}

We use the MicroBooNE coordinate system where the $X$ coordinate is the drift direction, with the anode (readout plane) at $X =\SI{0}{\centi\metre}$, and the cathode at $X$ = \SI{254.4}{\centi\metre};
The $Y$ coordinate is vertical with a range from \SIrange{-161.25}{161.25}{\centi\metre}, from the bottom to the top of both the cathode and the anode planes; 
The $Z$ coordinate is along the beam direction, where $Z =\SI{0}{\centi\metre}$ is at the upstream end of the TPC, closest to the neutrino source, and $Z = \SI{1036.8}{\centi\metre}$ is at the downstream end. 
These TPC boundaries are the physical limits of the true spatial coordinates.
This is right-hand coordinate system, and the anode plane is to the right when looking downstream in $Z$.

\begin{figure}[htbp]
	\centering
	\includegraphics[width=\textwidth]{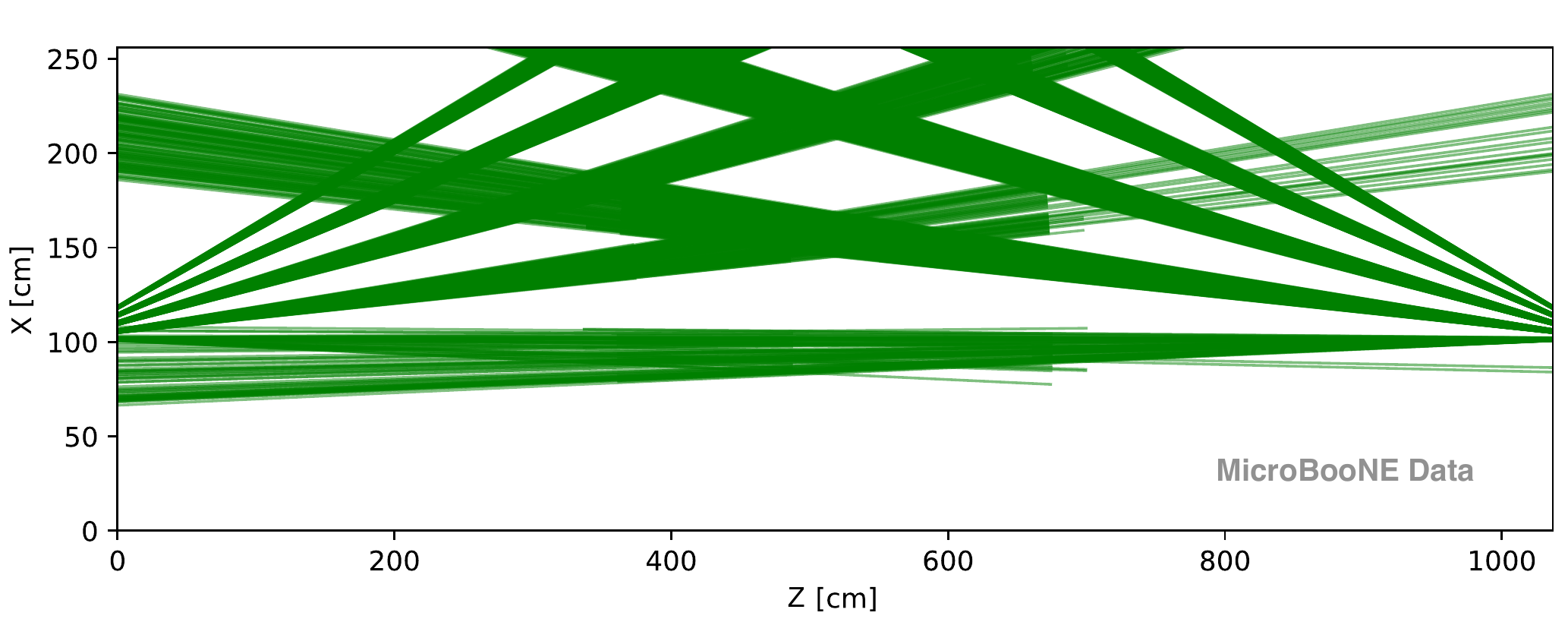}
	\caption{Top view ($X$-$Z$ projection) of reconstructed laser tracks (green lines) for both laser sub-systems in the TPC active volume.
	The laser beams are sent into the TPC from the upstream end (left) and the downstream end (right).
	 The laser tracks shown in the plot pass the selection criteria described in section \ref{subsec:Selection}.
	Some tracks look like they end because the corresponding laser light exits the TPC at the top and bottom surfaces. 
	Gaps in the coverage arise due to the presence of field-cage rings in front of the cold mirror.}
	\label{fig:coverage}
\end{figure}

A full laser scan consists of a range of measurements from both laser sub-systems.
The top view ($X$-$Z$ projection) of a full laser scan pattern with reconstructed laser tracks can be seen in figure~\ref{fig:coverage}. 
Field-cage rings located between the cold mirror and TPC volume obstruct part of the laser beam paths, which results in the gaps in the laser scan.
The TPC corners close to the cathode at the upstream and the downstream ends are not accessible because of the field-cage rings.
Laser beams are not aimed in the direction of the anode since photomultiplier tubes (PMTs) and corresponding wavelength shifting plates are mounted behind the anode.
To maintain the efficiency of the wavelength shifting plates installed in front of the PMTs, exposure to intense UV light should be avoided.
In a laboratory measurement at the University of Bern, the effect of the beam of the same Surelite UV-laser as employed in MicroBooNE was tested on a wavelength plate coated with tetraphenyl butadiene (TPB).
Its effect was found to be negligible.
To obtain a noticeable degradation, the full TPB area (5 cm in diameter) had to be directly illuminated at full laser pulse power.
Even in this unlikely scenario, only a 20\% decrease in light conversion was observed after more than 14,000 UV-laser pulses.

In MicroBooNE, the laser scan coverage from both sides is almost symmetric. 
Slight differences arise from small variations in the attenuator settings between the two sub-systems.
In the case of the upstream laser, the laser beams are aimed from \SIrange{30}{140}{\degree} with respect to the $Y$ axis. 
The mechanics of the feedthrough and characteristics of the dichroic mirror limit the range of accessible vertical angle.
Horizontally, the laser beams sweep from \SIrange{45}{93}{\degree} with respect to the $X$ axis.
Smaller angles with respect to $X$ are obscured by the field-cage rings.
Larger angles are limited by the field-cage rings and the need to avoid the anode.

To achieve a dense scan pattern and to minimize the E-field distortions from the ions induced by the laser beam itself,
 the pulse repetition rates are limited to \SI{4}{\hertz}.
The laser pulses continually while the cold mirrors move.
At the beginning of a scan, a laser sub-system is set to an extreme of both horizontal and vertical ranges.
Then a horizontal sweep is applied with a speed set to 2000 micro-steps per second ($\sim$\SI{0.16}{\degree \per \second}). 
When the horizontal sweep is complete, the cold mirror tilts vertically without the laser pulsing and then starts another horizontal sweep in the opposite direction.
The attenuator setting is updated based on the laser beam incident angle with respect to the cold mirror, to obtain an appropriate laser beam pulse energy.
The cold mirror is a dichroic mirror which has different reflectance at different angles of incidence.

Every time the laser pulses, the photo diode triggers an event.
Meanwhile, the position of the cold mirror is read out by two encoders and stored on a local server. 
Data fragments from the laser DAQ and the TPC DAQ are merged offline based on the coincidence of their time stamps. 
Both of their DAQ time clocks are provided by a common Network Time Protocol (NTP) server.

\subsection{True Laser Track Position}
\label{subsec:TrueLaserPosition}

True laser tracks represent the paths of laser beams in the TPC that are equivalent to the intrinsically straight laser tracks without any E-field distortions.
In the TPC, laser tracks are similar to tracks from charged particles.
Ionized electrons drift in the same E-field, so reconstruction of ionized electrons from the same position always leads to a fixed read-out position.
Thus, laser tracks can be used to calibrate distortions in the TPC related to the E-field.

True laser track positions are provided by the reflection point on the cold mirror and the angles at which the laser beams cross the TPC.
The reflection point and crossing angle provide laser entry and exit points of the TPC.

\begin{figure}[htbp]
	\centering
	\includegraphics[width=0.8\textwidth]{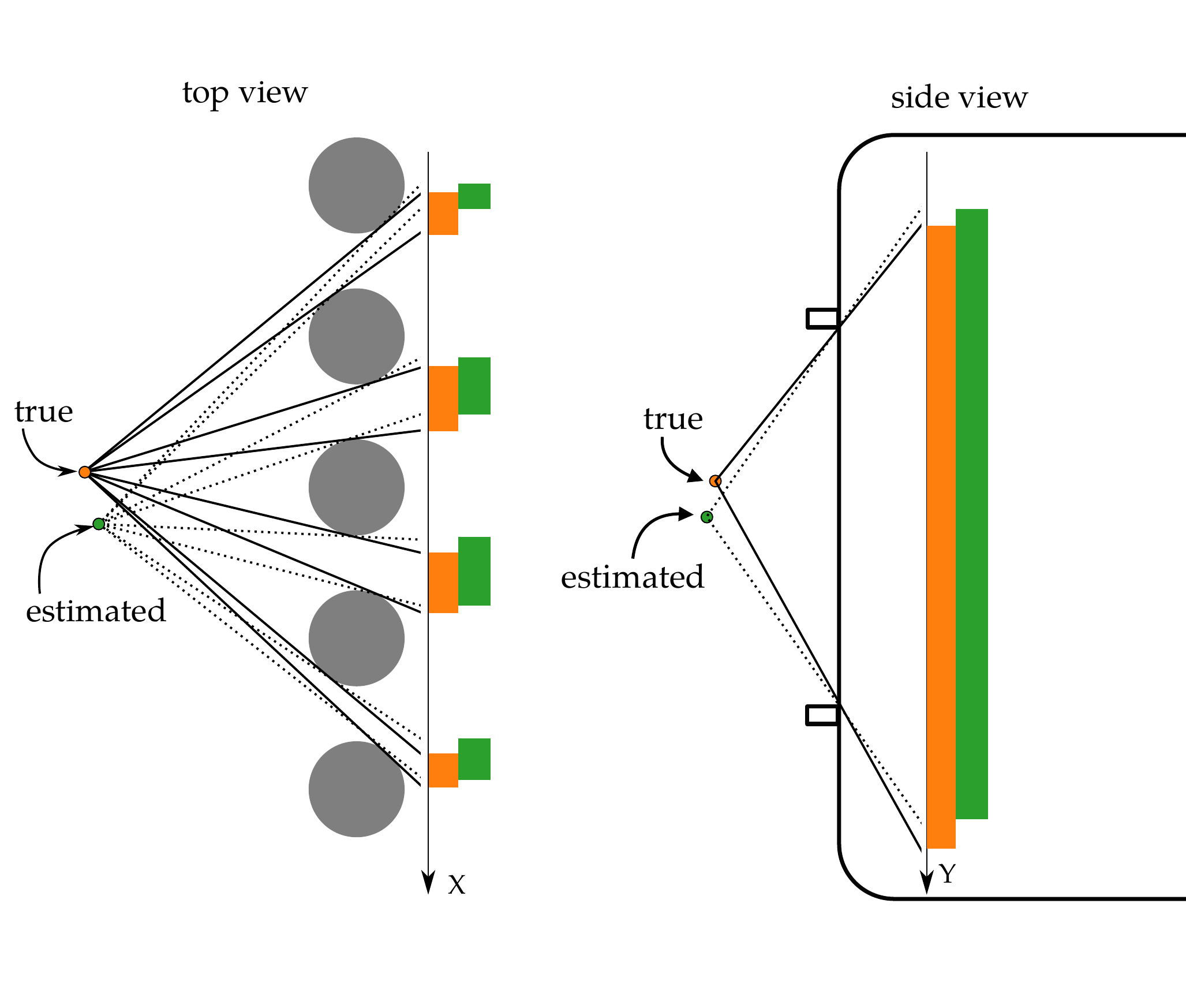}
	\caption{
	Schematic of the calibration procedure used to determine the position of the laser reflection point on the cold mirror.
	\textit{Left:} Top view near the cold mirror showing the determination of the laser reflection point on the cold mirror horizontally in the $X$-$Z$ plane.
	The gray circles are the cross-section of the field-cage rings.
	A horizontal scan produces a pattern in the TPC represented by the the orange boxes at the TPC edge.
	Any different reflection point in the $X$-$Z$ plane would leave a different pattern in the horizontal scan, for example, the green boxes.
	By fitting the $X$ and $Z$ coordinates of the reflection point, the position of the reflection point on the cold mirror is determined.
	\textit{Right:} Side view next to the cold mirror showing the determination of the laser reflection point on the cold mirror in $Y$.
	The two small boxes on the left hand side with solid lines are the G10 supporting bars.
	A vertical laser scan leaves a pattern in the TPC represented by the orange (green) bar for the orange (green) beam reflection origin.
     A fit to the pattern reveals the true reflection point coordinate in $Y$.
	The determination of the laser reflection point follows the order of horizontal coordinates ($X$ and $Z$) first, and then the vertical coordinate ($Y$).
	}
	\label{fig:MirrorPosition}
\end{figure}

The precise position of the reflection point is determined by measuring the ionization pattern caused by the shadow that the field-cage rings cast on the laser scan.
This procedure is illustrated in figure~\ref{fig:MirrorPosition}.
From the engineering drawings, the location of the cold mirror is known to a precision of $\mathcal{O}$(1~\SI{}{\cm}), which provides a starting point for the fit of the exact position.
In MicroBooNE, there are 63 field-cage rings supported by horizontal G10 beams~\cite{MicroBooneDesign}.
A slow horizontal scan is applied, producing a dense scan pattern. 
From this pattern, the sharp edges of the field-cage rings and their separations are apparent.
A 2-D ($X$-$Z$) fit then determines the $X$ and $Z$ positions of the laser reflection point on the cold mirror.
The vertical position ($Y$) is determined using the same technique with a 1-D fit of the horizontal G10 beams' shadow.
The calibrated positions of the laser reflection points are listed in table~\ref{tab:MirrorPos}.

\begin{table}
\caption{
  Coordinates of the laser reflection point on the cold mirror of the upstream and downstream laser sub-systems, with uncertainties.
}
\begin{center}
    \begin{tabular}{  c | c | c | c }
    \hline
    Sub-system & $X$ [\SI{}{\centi\metre}] &  $Y$ [\SI{}{\centi\metre}] &  $Z$ [\SI{}{\centi\metre}] \\ 
    \hline
    Upstream & $ 103.8 \pm 0.1$ & $8.6 \pm 0.1$ & $-35.6 \pm 0.1$ \\
	Downstream & $ 102.5 \pm 0.1$ & $8.2 \pm 0.1$ & $1080.2 \pm 0.1$ \\
    \hline
    \end{tabular}
\end{center}
\label{tab:MirrorPos}
\end{table}


The angles of the laser beams are deduced from the cold mirror positions.
As described previously, the cold mirror angles (azimuth and polar) are measured by two independent encoders to a high precision.
Position accuracy of true laser tracks is $\mathcal{O}$(1~\SI{}{\milli\meter}) at the full distance of \SI{10}{\meter} in the TPC.

\section{Reconstruction and Selection of Laser Tracks}
\label{sec:Reco_select}

\subsection{Hit and Track Reconstruction}
\label{subsec:Reconstruction}

The laser reconstruction procedure first defines a region of interest (ROI), in order to reduce computing time. 
Hits are then reconstructed within the ROI. These hits are then used for track reconstruction. 

The MicroBooNE charge readout consists of two induction-wire planes and one collection-wire plane.
The collection wires are oriented vertical, and wires of the two induction planes are angled $\SI{\pm 60}{\degree}$ with respect to the collection wires.
For all three planes, the wire pitch is \SI{3}{\milli\meter}.
The read-out wires are indexed by wire IDs.
Raw signals are waveforms on each wire with respect to the drift time; details of signal processing are described in ref.~\cite{adams2018ionization1, adams2018ionization2}.
The hit reconstruction identifies hits as the peak amplitudes and times of waveforms associated to wire IDs.
The track reconstruction then employs the hits as an input, grouping them to form 3D space points.
Reconstructed track objects are a sequence of 3D points with spatial coordinates in $X$, $Y$ and $Z$.

Hit and track reconstruction for the laser employs LArSoft~\cite{snider2017larsoft}, the standard software toolkit for simulation and reconstruction used in MicroBooNE. 
 
An ROI is defined around the true laser track wire IDs and drift time.
Given a spatial position in the TPC, LArSoft provides the expected wire ID for each plane.
Thus, a range of wire IDs along the true laser track can be identified.
The ROI is extended by $\pm100$ wires (\SI{30}{\centi\metre}) around each wire to account for E-field distortions.
Similarly, the ROI along the drift axis (time) is also determined by the $X$ coordinates of the true laser track, and extended by $\pm$\SI{20}{cm}.
For simplicity, the extensions to the ROI for both wires and drift time remain the same across all events.

A hit contains information about the waveform peak amplitude, width, time, and the associated uncertainties.
Typical signals in the collection plane are a single positive peak, whereas signals in induction planes are bipolar peaks.
After a baseline correction, a hit is recognized if the waveform exceeds threshold.
In the collection plane, the time of the highest peak amplitude is defined as the hit time.
In the induction planes, the average of the two bipolar peak times is taken as the hit time.
For a laser track, the charge deposition is large and the waveform is broadened to a less well-defined peak.
In this case, the local center of the peak is taken as the hit time.
If the ratio of height to width is too low, the waveform is not reconstructed as a hit.
This hit reconstruction algorithm was developed specifically for laser track reconstruction.
The conventional hit reconstruction involves waveform deconvolution which has difficulties with the large charge deposition of laser tracks.
Minimum ionizing particles have narrower and shorter waveforms than laser waveform signals which typically are three times larger.

The laser tracks are reconstructed by Pandora~\cite{acciarri2018pandora} tools embedded in LArSoft using the collection of hits as an input. 
The tracks are groups of 3D space points which have separate primary ionization origins.


\subsection{Laser Track selection}
\label{subsec:Selection}

\begin{figure}[htbp]
	\centering
	\includegraphics[width=\textwidth]{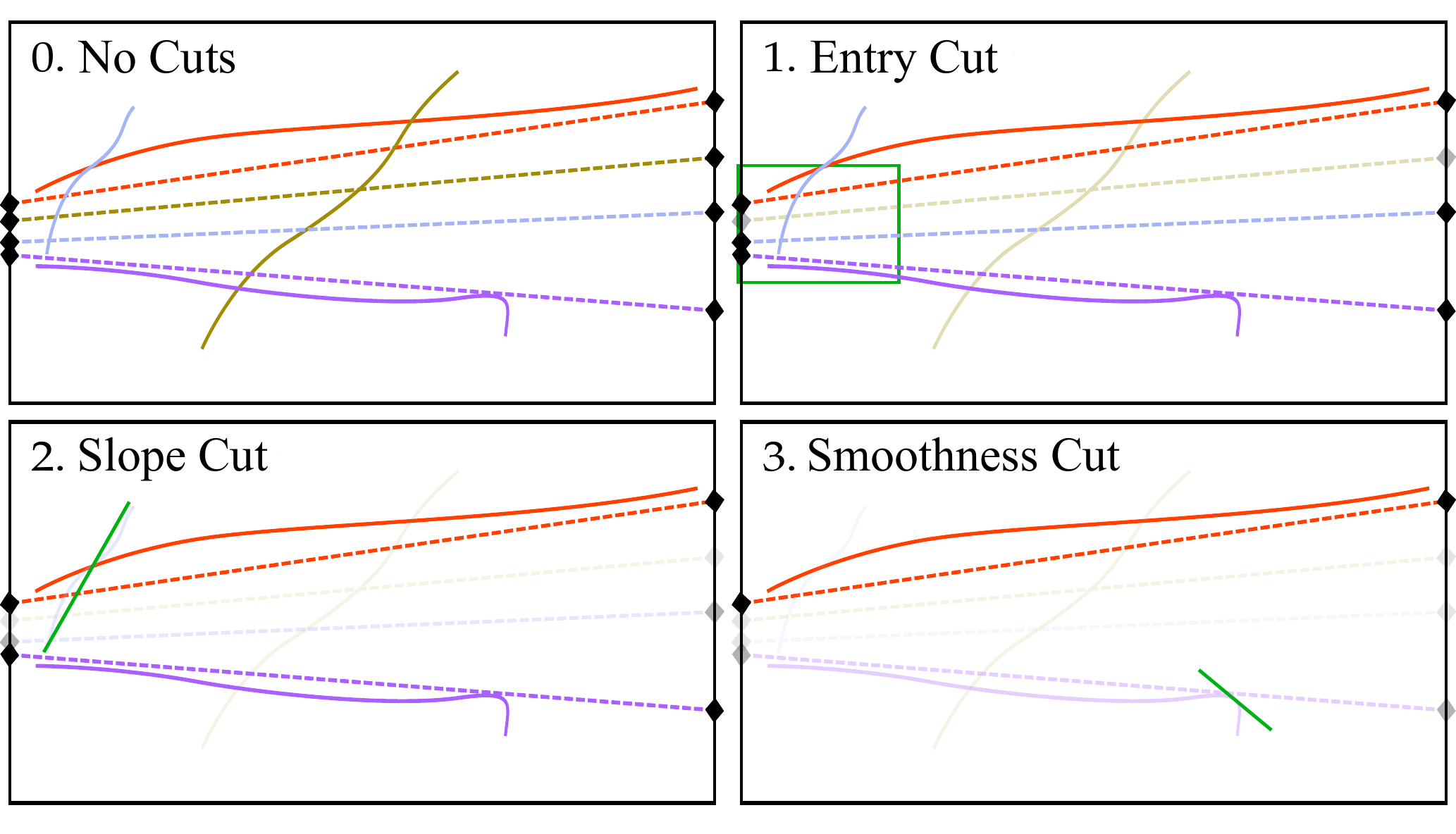}
	\caption[Track selection illustration]{
	Illustration of the track selection steps.
	This example uses laser events originating from the upstream laser sub-system.
	In these illustrations, the $Z$ axis is along the horizontal direction, and the vertical axis can be $X$ or $Y$.
	The black points on the left hand side of each TPC projection are the laser entry points in the events, and the ones on the right side are the laser exit points.
	The dashed lines connecting these points are the true laser beam positions.
	The solid lines with the matching colors represent the possible reconstructed tracks in the corresponding events, which are different from the dashed lines due to E-field distortions.
		(0) Reconstructed tracks with no selection applied.
		(1) Tracks pass this step if one of the track ends is within a \SI[product-units=repeat]{20x20x20}{\centi\metre} region from the laser entry.
		(2) Tracks pass this step if the angles of the first and last hundred space points are within \SI{40}{\degree} of the true laser track angle. 
		(4) Tracks pass this cut if no hard kinks or wiggles, due to mis-reconstruction, are present.
		}
	\label{fig:track_selection}
\end{figure}

Hits induced by cosmic rays can enter ROI and contaminate track reconstruction.
To select a pure laser track, crossing cosmic muons and poorly- or mis-reconstructed laser tracks should be removed.
Three selection criteria are applied to the reconstructed tracks as illustrated in figure~\ref{fig:track_selection}. 

The first two selection cuts are used to eliminate cosmic tracks.
In the first, tracks only pass if one of the reconstructed track-ends is close to the true laser start.
The permitted differences in $X$, $Y$, and $Z$ are \SI{20}{\centi\meter}.
No constraint is applied to the true laser end point, as unresponsive wires or cosmic muons may cause difficulties reconstructing a full laser track.

The second selection cut compares the angle of the track to the true laser track.
Tracks have 3D space points defined by the wire pitch of \SI{3}{\milli\meter}.
For example, a \SI{5}{\meter}-long track has about 1600 space points.
To account for curvature along the track, the first and last 100 space points are used to fit two straight lines.
The angles of the lines from each end of the track are then compared to the true laser track angle.
The track is selected if both angles are less than \SI{40}{\degree} from the true track angle.

Only laser tracks survive the above two selection cuts. 
The final selection step provides a quality control for the reconstructed laser tracks.
In MicroBooNE the high voltage system is fully functional, hence the possible sources of E-field distortion are space charge and detector deformation.
Therefore, no drastic E-field distortion is expected, which means any hard kinks or wiggles in reconstructed laser tracks are not due to E-field related artefacts (figure~\ref{fig:track_selection}).
The large charge deposition and broad waveforms of the laser introduces uncertainties in the hit reconstruction which can eventually lead to large deformities in the reconstructed track.
We sort 3D space points in the laser tracks by their ascending $Z$ coordinate.
Using the true angle and the separation in $Z$ between neighboring 3D space points, expected separations in $X$ and $Y$ can be computed.
A track fails to pass the selection if $X$ or $Y$ separations are more than three times larger than expected for ten or more space points.
For simplicity, mis-reconstructed tracks are removed from the selection, but one could potentially recover the well reconstructed portion of the tracks in future iterations of this work.

We only use laser track spatial information for calibration, so no requirement on reconstructed laser energy has been applied.
From a laser scan taken in the summer of 2016 over a few hours, there are 911 upstream and 1204 downstream tracks after all selection cuts.
These laser tracks are the input for the following calibration.
    
\section{Laser Track Simulations}
\label{sec:TrackSim}

Two simulations are used: a toy simulation evaluates the performance of the methodology with ideal tracks and a perfectly known spatial distortion. A more complete laser simulation verifies the calculation of the spatial-displacement map and the E-field map, and provides an estimate of the systematic uncertainties of the calibration maps.

In both simulations, a ``muon gun'' is used to replicate true laser tracks in the \textit{Geant4}~\cite{agostinelli2003geant4} stage of LArSoft (\textit{LArG4}), with processes such as delta electron emission and multiple Coulomb scattering disabled.
True tracks are the simulated laser tracks, and the distorted tracks are the reconstructed laser tracks.
No additional cosmic ray muons are simulated.

Initially, an E-field distortion is simulated with a charge distribution of positive ions accumulating along $X$ towards the cathode.
The corresponding spatial displacement is simulated parametrically.
The simulated E-field and spatial displacement are embedded in LArSoft, with spatial distortions applied to true tracks via LArSoft.

\subsection{Toy Simulation}
\label{subsec:ToySim}

To avoid bias from the uncertainties of the laser system and track reconstruction, we use a toy simulation to test the methodology.
We only use track positions to calculate the spatial displacement, the corresponding local drift velocity and the E-field.

Spatial distortions are applied directly to the true tracks, with the applied offsets deduced from the known E-field.
The raw waveform generation and their reconstruction are not simulated.
However, simulated tracks are still processed through the track selection described in section~\ref{subsec:Selection}.

\subsection{Complete Laser Simulation}
\label{subsec:LaserSim}

To verify the complete chain of the E-field calculation using the laser system, a complete laser simulation is introduced to mimic the reconstructed laser data.

This provides more comprehensive track information than the toy simulation by including the reconstruction and selection procedure, which allows us to estimate the systematic uncertainties.
To avoid additional systematic uncertainties, the simulated tracks are given the same track angles as the laser data scan, described in section~\ref{subsec:LaserScan}.

In practice, laser beams have a diameter of about $\SI{3}{\milli\metre}$.
The longitudinal diffusion of drifting electrons in the complete laser simulation is increased to be comparable with the larger beam diameter.
To compensate for the larger diffusion of ionized electrons and to represent the greater ionization of laser beams, a scaling factor is applied to the ionization yield.
The scaling factor is chosen to allow the simulated laser signal (raw waveform from wires) to match typical laser track data, with a peak voltage of $U_\mathrm{max} \approx \SI{80}{\milli\volt}$ and time spread of $t_{s} \approx \SI{10}{\micro\second}$.

Afterwards, the simulated laser signals are processed through the reconstruction and the track selection criteria, as described in section~\ref{sec:Reco_select}.

\section{Spatial Displacement Map}
\label{sec:Dmap}
In a distorted E-field, the positions of reconstructed charges in the TPC differs from their true positions. 
Spatial displacement maps are used to show the difference between the reconstructed and true positions. 

There are two types of spatial displacement maps used: a distortion map and a correction map.
The distortion map is based on the true spatial coordinates, in which the TPC boundary follows the known mechanical structure. 
It shows the expected reconstructed space points given the true space points.
It can be used in the simulation of TPC events including E-field distortion.
We use it to verify the method of extracting the spatial displacement by comparing the calculated spatial displacement to the simulation truth. 
The correction map is based on the reconstructed spatial coordinates, in which the TPC boundary may be irregular. 
It shows the expected true space points given by the reconstructed space points.
It can be used to calibrate the position of reconstructed TPC objects in a distorted E-field and it is used to derive the local E-field.

For an irregular E-field, it is not trivial to convert from the distortion map to the correction map.
However, both maps contain related information and have many similarities.

For simplicity, the displacement maps are arranged as a regular grid.
Therefore, any reconstructed position within the TPC can be translated from or to the corresponding true position by interpolation between points on the displacement grid.
We set 26 bins in $X$, 26 bins in $Y$ and 101 bins in $Z$, with the first and last bins centered at the TPC boundaries.
For example, along the $X$ axis, the first bin center is at $X =\SI{0}{\centi\metre}$, and the last bin center is at $X = \SI{254.8}{\centi\metre}$.
The bin size is about \SI[product-units=repeat]{10x10x10}{\centi\metre}.

\subsection{Spatial Displacement Vectors}
\label{subsec:correctionVec}

Spatial displacement vectors represent the difference between the reconstructed and true positions. 
The reconstructed positions are deduced from the wire waveforms, with the assumption that electrons drift in a nominal, and uniform, E-field.
In practice, the E-field in MicroBooNE is distorted. 
The true positions are where the charged particles ionize LAr in the TPC.  
The distortion vectors are measured by comparing points on the true laser tracks to their corresponding reconstructed points.
Alternatively, the correction vectors are measured by comparing points on the reconstructed tracks to their corresponding true points.
At this stage, the distortion vectors and the correction vectors have opposite direction with the same length.

One unambiguous way of calculating the spatial displacement vectors would be to use crossing tracks.
If the E-field does not change, a true position should always be reconstructed in the same position.
Because E-field lines cannot cross each other, the true positions and the reconstructed positions have a one-to-one correspondence.
The intersection point of two true tracks, with a small tolerance allowed, must correspond to the intersection point of their associated reconstructed tracks.
It is therefore straightforward to extract the spatial displacement for those intersection points.
However, given the limited TPC coverage and total numbers of laser tracks, it is rare to have crossing tracks, and this method cannot be used.

Without crossing tracks, the difficulty in determining the displacement vectors comes from the lack of a point-to-point correspondence.
Even though the true laser track positions are known exactly, it is not straightforward to establish where an associated reconstructed laser track should be.
This subsection introduces an alternative method using closest-point projection and track iteration to determine the spatial displacement vectors. 

\subsubsection{Closest-point Projection}
\label{subsubsec:proj}

\begin{figure}[htbp]
\centering 
\includegraphics[width=.7\textwidth]{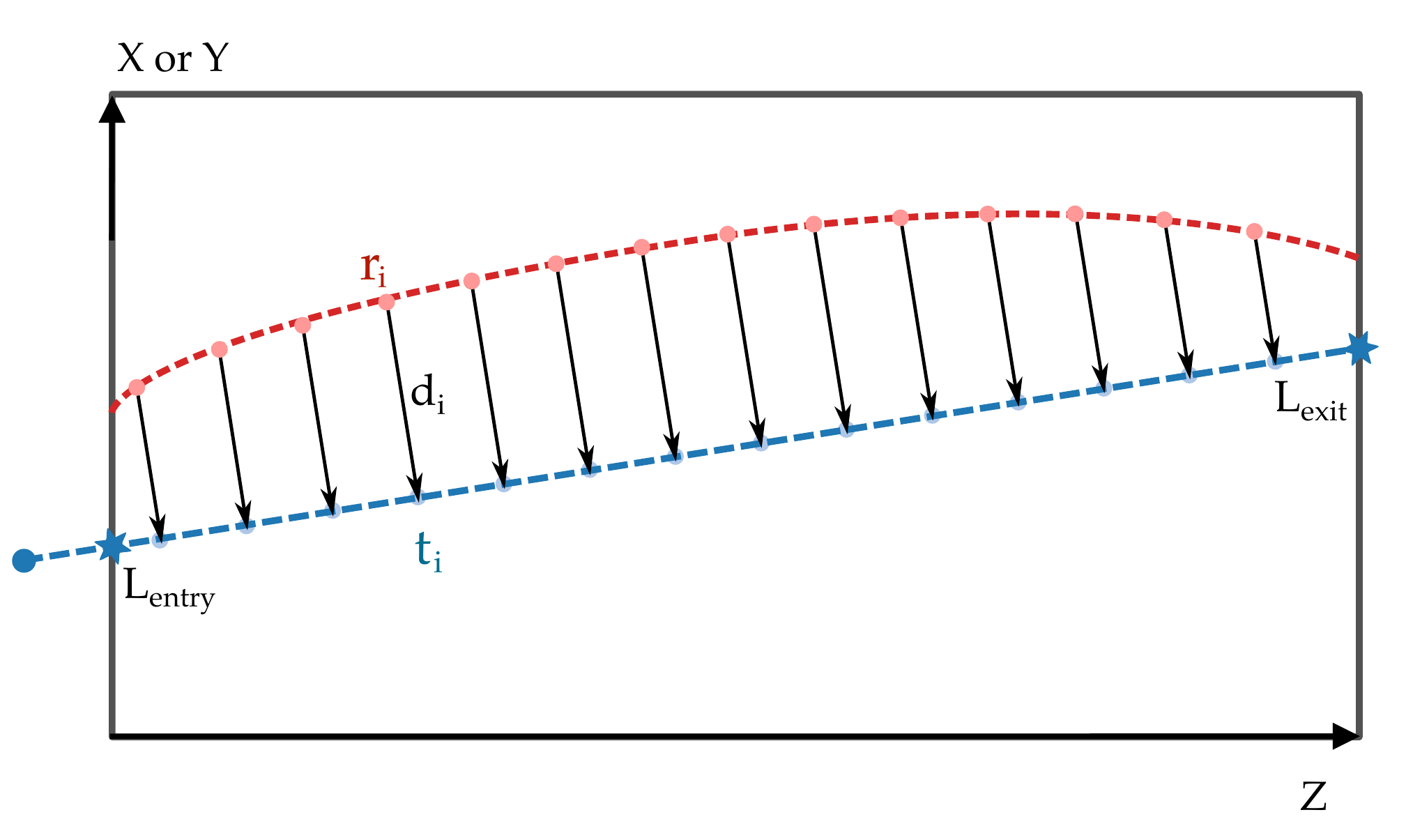}
\qquad
\caption{\label{fig:projection}
Diagram illustrating the closest-point projection used as the first step in calculating the spatial displacement vectors.
The black box is the TPC active volume in true spatial coordinates.
$Z$ is along the horizontal direction, and the vertical direction can be either $X$ or $Y$.
The example shown here has a laser beam originating from the upstream sub-system, 
with the true laser track shown as the blue dashed line.
The blue dot is the reflection point on the cold mirror which is outside of the field-cage rings.
The collection of red dots is the reconstructed track.
Each reconstructed point (red dot) \textbf{$r_i$} is perpendicularly projected to the true laser track in 3D.
Its corresponding point on the true track is \textbf{$t_i$}.
The displacement vector \textbf{$d_i$} starting from \textbf{$r_i$} to \textbf{$t_i$} is the correction vector, as shown in the diagram.
The distortion vector has the same length but is in the opposite direction in to the correction vector.
}
\end{figure}

The closest-point projection is a simple way to give a good initial estimation on the spatial displacement vectors.
We project reconstructed track points perpendicularly to the true track in 3D; an example is shown in figure~\ref{fig:projection}.
The vectors from the reconstructed track points (red) to their closest point on the true track (blue) are the correction vectors.
The vectors with opposite directions starting from the true track (blue) to the reconstructed track points (red) are the distortion vectors.

Calculating spatial displacement vectors by closest-point projection alone introduces a dependency on the laser beam angles.
It forces the displacement vectors to be perpendicular to the corresponding true laser tracks.
We use a track iteration method to reduce the bias from the initial laser beam angles.

\subsubsection{Track Iteration}
\label{subsubsec:iter}


A complete description of the angular acceptance of laser tracks used in the analysis can be found in section~\ref{subsec:LaserScan}.
To maximize the spread in track angles for the correction, we split the laser tracks into two samples.
Laser tracks initiating from the upstream sub-system are assembled in track sample \textbf{A}, and laser tracks initiating from the downstream sub-system are assembled in track sample \textbf{B}.
Essentially, we use tracks from one sub-system to calculate the displacement vectors of the other in several iterative steps.

%

We illustrate the procedure using the example of an \textit{n}-step iteration with track samples \textbf{A} and \textbf{B}.
At the beginning of the first step, we calculate the correction vectors of track samples \textbf{A} and \textbf{B} using closest-point projection. 
Then $\frac{1}{\textit{n}}$ of the calculated correction vectors of track sample \textbf{A} are assigned to the corresponding reconstructed track points.
We interpolate the partial correction vectors of track sample \textbf{B} with the mesh of all the reconstructed track points in sample \textbf{A} and their partial correction vectors.
Details of meshing and interpolation methods are explained in section~\ref{subsec:Interploation}. 
The inverse procedure is applied here.
The partial correction vectors of track sample \textbf{A} are interpolated from track sample \textbf{B}.
By the end of a step, all the track points of both track samples are moved to the intermediate positions along the partial correction vector directions.
At each step of the iteration the above procedures are repeated.
The correction vectors calculated at the beginning of a step are taken from the intermediate track positions.
For normalization, the partial correction vectors of the track sample \textbf{A} (\textbf{B}) in a particular step $i$ are interpolated from $\frac{1}{\textit{n}-(\textit{i}-1)}$ of the calculated correction vectors.
Eventually, in the final step, all the track points from both samples are projected to their closest points on the corresponding true tracks.
This guarantees that the displacement vectors link the reconstructed track points to the corresponding points on the true tracks.
After the \textit{n}-step track iteration, an association of the reconstructed track points to the points on the true tracks is established.
The correction vectors point from the reconstructed track points to the points on the true tracks, and the distortion vectors from the points on the true tracks to the reconstructed track points.

By repeating the calculation with different numbers of track iteration steps, we found that three iteration steps were sufficient. 
The bias of the displacement maps from the simulation does not decrease noticeably with additional iteration steps.


\subsection{Boundary Condition}
\label{subsec:Dboundary}

Independent of the E-field in the TPC, any ionization along a track at the anode position would be reconstructed at the same position.
Thus, in the displacement map, there is no spatial distortion at the anode.


Spatial displacements in the TPC are caused by differences between the actual E-field the electrons drift in, and the E-field used in reconstruction.
The reconstructed spatial position and the true spatial position are associated with each other by the position at the read-out planes with equivalent drifting time.
The displacement vectors connect them directly without explicit knowledge of the position at the read-out planes.
The spatial displacement vectors are related to the accumulated E-field distortion from the position of the ionization to the read-out planes.
Ionized electrons that are produced at the anode are immediately read out with no time to be affected by E-field distortions. 

As mentioned in section~\ref{subsec:LaserScan}, the coverage of the laser scans in MicroBooNE is restricted near the anode.
The gap in between the anode and the edge of the laser coverage is roughly \SI{50}{\centi\metre}. 
Considering the main sources of E-field distortion, space charge from cosmic rays and the detector deformation, we expect no rapid change in the E-field distribution.
Thus we anticipate only small changes of spatial displacements with respect to the spatial coordinates.

We impose the boundary condition that the spatial displacement is null at the anode.
With interpolation, the effective region of the displacement maps increases by about 20\%.
The validation of the displacement maps with the boundary condition included can be found in section~\ref{subsec:VerificationDMap}.


\subsection{Interpolation}
\label{subsec:Interploation}

In order to obtain the spatial displacement vectors on the regular grid, we interpolate them from the ones associated with the track points.
The regular grid points are located at the center of each bin in the spatial displacement maps.
Each bin has the same size of approximately \SI[product-units=repeat]{10x10x10}{\centi\metre}.
 
As mentioned in section~\ref{subsec:LaserScan}, laser beams do not reach the area blocked by the field-cage rings and the area near the anode.
Given the bin size and the area of missing coverage, we expect the E-field distortion only causes small variations in spatial displacement along the regular grid.
Interpolation could provide a fair estimation of the spatial displacement in the low coverage area. 

Interpolation is implemented in two steps.
We first create a mesh of all the track points by using Delaunay triangulation.
Then barycentric coordinates of the grid points in the corresponding Delaunay triangulation unit are computed, and they are further applied to achieve the spatial displacement vectors on the grid points.

The interpolation method is also used in the intermediate step of track iteration (section~\ref{subsubsec:iter}), and in the calculation of the drift velocity and E-field maps.

%
%
%
%

\subsubsection{Mesh with Delaunay Triangulation}
\label{subsubsec:mesh}

For the correction map, the mesh base is the collection of all the reconstructed track points.
For the distortion map, the mesh base is the collection of all the true track points which are the ends of the displacement vectors.

Three-dimensional Delaunay triangulation meshes a volume with tetrahedrons as unit elements.
This guarantees that the whole volume of interest can be filled without gaps. 
The vertices of a unit tetrahedron are four reconstructed track points (true track points) for interpolating correction vectors (distortion vectors).

\begin{figure}[htbp]
\centering 
\includegraphics[width=.7\textwidth]{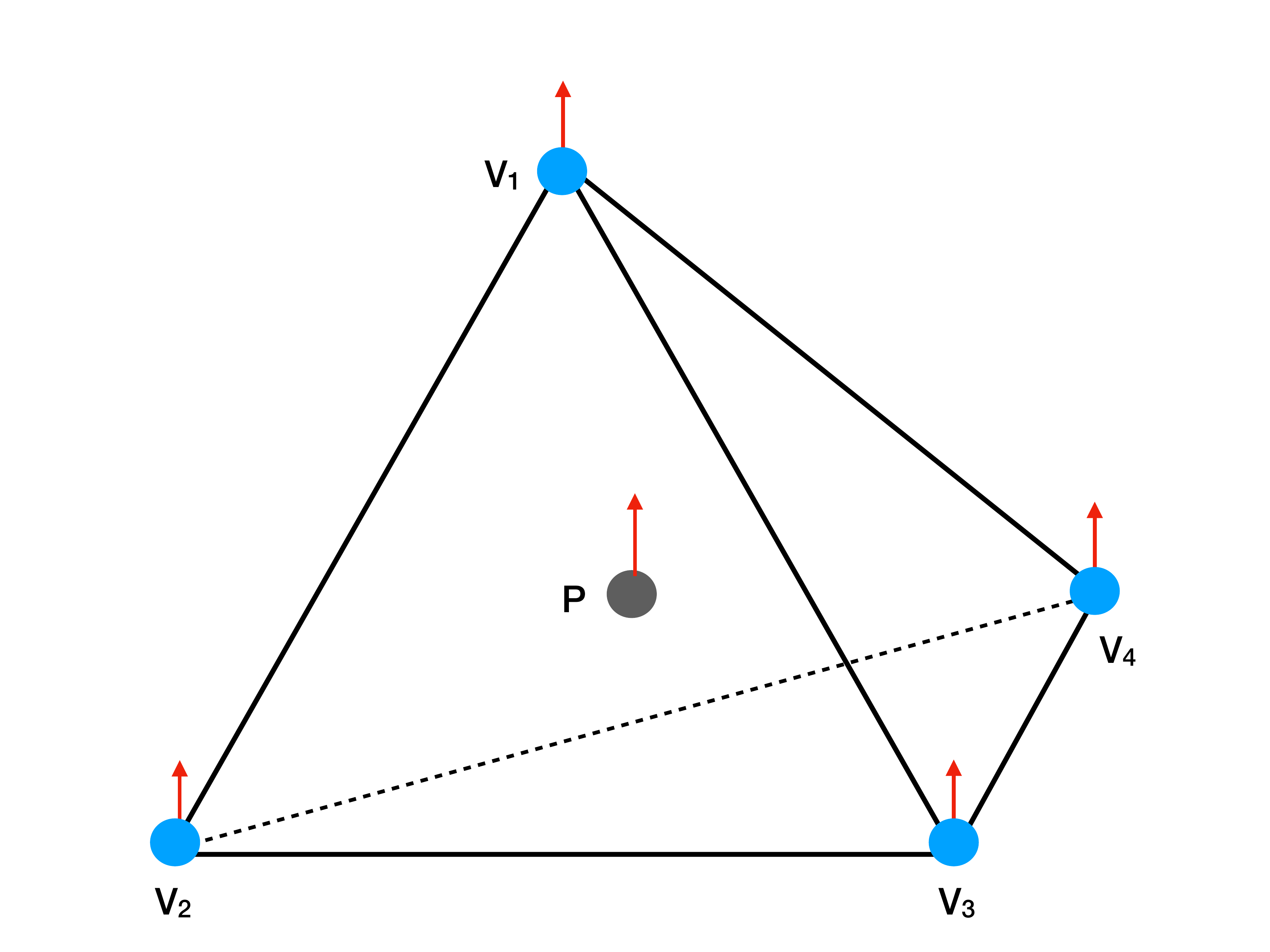}
\qquad
\caption{\label{fig:tetrahedron}
Diagram of a unit tetrahedron of the Delaunay Triangulation.
In our application, such tetrahedrons are used as mesh units for interpolating physical quantities represented by vectors.
The blue mesh points are the start of the mesh vectors and correspond to the points on the true tracks. 
The red arrows represent the distortion vectors.
We interpolate to the regular grid point \textbf{P} (gray).
}
\end{figure}

Delaunay triangulation has the property that the surrounding sphere of any unit tetrahedron must not contain any other mesh points. 
Thus, a unit tetrahedron usually has similar size as the surrounding units and its vertices are relatively close to each other with the given density of the mesh points.
Therefore, the interpolated values are eligible to represent the local characteristics such as the spatial displacement.
A diagram of a tetrahedron, a unit of the mesh in Delaunay triangulation, as shown in figure~\ref{fig:tetrahedron}.

We use \textit{CGAL 3D triangulation}~\cite{cgal:pt-tds3-18b}~\cite{cgal:pt-t3-18b} for meshing with Delaunay triangulation.

\subsubsection{Barycentric Coordinates}
\label{subsubsec:barycentric}

Interpolation at a certain position, $\mathbf{P}$, uses the reference of the tetrahedron which encloses $\mathbf{P}$.
The barycentric coordinates system is the reference system that we use.

With the position information of four tetrahedron vertices $\mathbf{V_1}(x_1, y_1, z_1)$, $\mathbf{V_2}(x_2, y_2, z_2)$, $\mathbf{V_3}(x_3, y_3, z_3)$, $\mathbf{V_4}(x_4, y_4, z_4)$, the coordinates can be calculated,
$$
\begin{pmatrix}
  \lambda_1 \\
  \lambda_2 \\
  \lambda_3
 \end{pmatrix}
 = \mathbf{T^{-1}}(\mathbf{P}-\mathbf{V_4}) ,
$$
where $\mathbf{T}$ is the matrix,
$$
\mathbf{T} = \begin{pmatrix}
  x_1 - x_4 & x_2 - x_4 & x_3 - x_4 \\
  y_1 - y_4 & y_2 - y_4 & y_3 - y_4 \\
  z_1 - z_4 & z_2 - z_4 & z_3 - z_4 
 \end{pmatrix},
$$
$\lambda_{1,2,3,4}$ are the barycentric coordinates of $\mathbf{P}$. The remaining $\lambda_4$ can be found by $\sum \lambda =1$. 

Effectively, $\lambda_{1,2,3,4}$ are the ratios of the tetrahedron volumes $\mathbf{PV_2V_3V_4}$, $\mathbf{PV_1V_3V_4}$, $\mathbf{PV_1V_2V_4}$, $\mathbf{PV_1V_2V_3}$, relative to the tetrahedron volume $\mathbf{V_1V_2V_3V_4}$.
This also applies when $\mathbf{P}$ is located on the surface or the edge or the vertex of the tetrahedron.
The volumes $\mathbf{PV_2V_3V_4}$, $\mathbf{PV_1V_3V_4}$, $\mathbf{PV_1V_2V_4}$ and $\mathbf{PV_1V_2V_3}$ may be null.

Finally, the value at the position of interest, $\Delta r$, can be interpolated by  
$$
\Delta r = \sum_i \lambda_i \Delta r_i ,
$$
where $\Delta r_i$ are the values at each vertex.

An example of interpolating the spatial displacement vector $\mathbf{d}=(dX, dY, dZ)$ is 
\begin{align*}
dX &= \lambda_1 dX_1 + \lambda_2 dX_2 +\lambda_3 dX_3 +\lambda_4 dX_4 ,
\\
dY &= \lambda_1 dY_1 + \lambda_2 dY_2 +\lambda_3 dY_3 +\lambda_4 dY_4 ,
\\
dZ &= \lambda_1 dZ_1 + \lambda_2 dZ_2 +\lambda_3 dZ_3 +\lambda_4 dZ_4 ,
\end{align*}
where $\mathbf{d_{1,2,3,4}}(dX_{1,2,3,4}, dY_{1,2,3,4}, dZ_{1,2,3,4})$ are the spatial displacement vectors at the four vertices.

\subsection{Sub-Map Merging and Statistical Uncertainty}
\label{subsec:SubMapMerging_StatUncertainty}

Originally, only the overall closest four track points determine a displacement map bin center.
To increase data statistics and reduce bias from individual track points, we divide the original track set of $n$ points into $m$ subsets.
 Then a given displacement map bin center is determined by four different track points for each of the $m$ subsets.

For each track, $m$ reconstructed track points are placed into the same subset.
For example, the $1$st reconstructed track point of track \textbf{A1} together with the $(m+1)$-th, the $(2m+1)$-th and so on, are grouped to compose a new track $\mathbf{A1_1}$ for subset $1$.
In subset $i$, a new track $\mathbf{A1_i}$ contains the $i$-th, $(m+i)$-th, ..., reconstructed points of the original track \textbf{A1}.
All the reconstructed points of \textbf{A1} can be found in a single subset.
Track $\mathbf{A1_n}$ ($n$ can be any integer from $1$ to $m$) shares the same true laser track as the one that corresponds to track \textbf{A1}.
The same procedure applies to every track in the original set.
Every subset has the same number of tracks as the original one.

The track set division is done before the calculation of displacement vectors.
Thus, in each subset, we still separate sample \textbf{a} from the upstream laser sub-system and sample \textbf{b} from the downstream laser sub-system.
Each subset can produce a distortion map and a correction map which have similar coverage as the ones from the original track set.
Therefore, with \textit{m} subsets, we have \textit{m} sub-distortion-maps and \textit{m} sub-correction-maps.

With the interpolation procedure explained in section~\ref{subsec:Interploation}, the displacement vector on a grid point only depends on the tetrahedron which encloses the grid point.
This means that only four track points and their displacement vectors determine the representative displacement vector in a \SI[product-units=repeat]{10x10x10}{\centi\metre} bin.
Without subset division, the typical spacing between track points is $\sim$\SI{3}{\milli\meter} for reconstructed tracks. 
Hence, in a large area, the track points density is high and most of them do not contribute to the displacement map.

Considering the density of the track points, we divide the original track set into 50 subsets for both the simulated track set and the laser data track set.
The number of subsets is chosen to be 50, so that the statistics in each bin are large enough, while maintaining the shape and characteristics of the typical unit tetrahedron.
We then average the corresponding 50 sub-maps to issue a final map.
The displacement vector on a grid point of the final map is the mean of 50 values calculated at the same point from sub-maps.
Therefore, the displacement vector on a grid point is calculated with the participation of up to 200 track points and their displacement vectors.
The averaging makes it unlikely that some individual track points with large bias could negatively affect the result on a grid point.

Due to small differences in the interpolation meshes of 50 subsets, the region with valid interpolated displacement may vary by a small amount at the edges. 
The standard deviation of the 50 $X$, $Y$, $Z$ components of the displacement vectors are taken as the statistical uncertainty of the final map.

\subsection{Bias Study and Systematic Uncertainty}
\label{subsec:VerificationDMap}


We use tracks generated through simulation to calculate displacement maps.
The bias is defined as the difference between the simulation truth and the calculated displacement map from the simulated tracks.

The fact that the biases of the displacement maps from toy simulation are small demonstrates the effectiveness of the methodology.
The bias of the displacement maps from complete laser simulation is taken as the systematic uncertainty for the displacement maps determined with data.

We use the distortion map as an example to explain the bias study.
As stated previously, the distortion map and the corresponding correction map describe similar spatial displacement which can be converted to each other.
For both the distortion map and the correction map, we use the same TPC boundary as described in section~\ref{subsec:LaserScan}, considering that the reconstruction is focused on a region of nominal active volume.
The bins with calculated displacement show the rough coverage of the laser, which might be slightly smaller than the actual coverage due to the bin size.
If the distortions of the E-field lead to the ionized electrons appearing to be out of the TPC active volume, the displacement vector in the edge bins of both the distortion map and the correction map may be difficult to represent.
In that case, the correction map shows a smaller area than the actual TPC active volume.
 
In the following, we define the $X$, $Y$ and $Z$ components of the calculated distortion vector from the simulation as $dX_\mathrm{calc}$, $dY_\mathrm{calc}$ and $dZ_\mathrm{calc}$, and magnitude of the calculated distortion vector as $d_\mathrm{calc}$.
The $X$, $Y$ and $Z$ components of the distortion vector from the simulation truth are defined as $dX_\mathrm{true}$, $dY_\mathrm{true}$ and $dZ_\mathrm{true}$, and magnitude of the distortion vector from the simulation truth is called $d_\mathrm{true}$.


\subsubsection{Bias Study Using Toy Simulation}
\label{subsubsec:ToySimResult}



\begin{figure}[htbp]
\centering 
\includegraphics[width=1.0\textwidth,trim=0 0 0 0,clip]{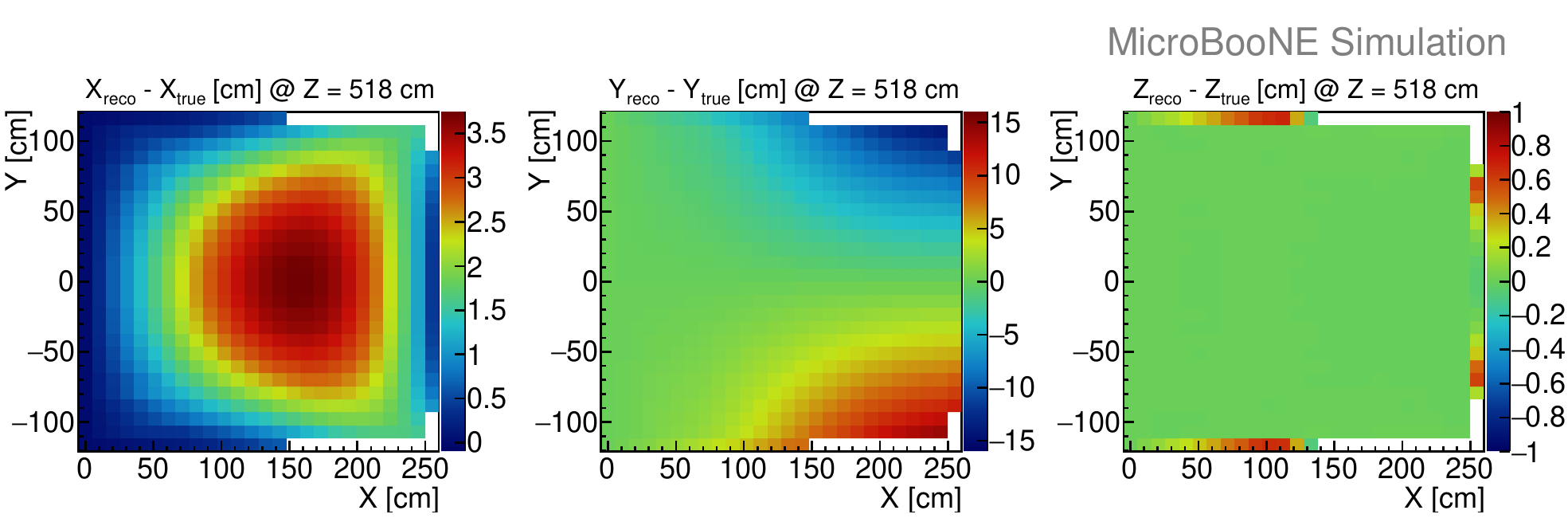}
\qquad
\caption{\label{fig:ToySim_Dmap_centralZ}
Calculated distortion map from a toy simulation showing a $X$-$Y$ slice at $Z=\SI{518}{\centi\metre}$ (central $Z$).
All three components are shown: $dX = X_\mathrm{reco} - X_\mathrm{true}$ (left), $dY = Y_\mathrm{reco} - Y_\mathrm{true}$ (middle), and $dZ = Z_\mathrm{reco} - Z_\mathrm{true}$ (right).}
\end{figure}

Figure~\ref{fig:ToySim_Dmap_centralZ} is the calculated distortion map from the toy simulation at $Z$ = \SI{518}{\centi\metre} (central $Z$), for the components $dX = X_\mathrm{reco} - X_\mathrm{true}$, $dY = Y_\mathrm{reco} - Y_\mathrm{true}$, $dZ = Z_\mathrm{reco} - Z_\mathrm{true}$, which are the result of the displacement calculation, applying the boundary condition and interpolation methods from section~\ref{subsec:correctionVec}, \ref{subsec:Dboundary} and \ref{subsec:Interploation}.
At central $Z$, $dX$ has a maximum distortion at $X \approx \SI{170}{\centi\metre}$ of about $\sim$\SI{4}{\centi\metre}.
The maximum distortion $dX$ at the central $Z$ is in the cathode direction because of space charge, which is the main contribution to the E-field distortion in MicroBooNE, accumulating along the drift direction.
At the region closest to the anode and the cathode, $dX$ is negligible.
The distortion $dY$ close to the cathode has a value up to $\sim$\SI{15}{\centi\metre} and points inward.
$dY$ is almost symmetric with respect to $Y=0$.
$dZ$ is generally small and uniform. 

\begin{figure}[htbp]
\centering 
\includegraphics[width=1.0\textwidth,trim=0 0 0 0,clip]{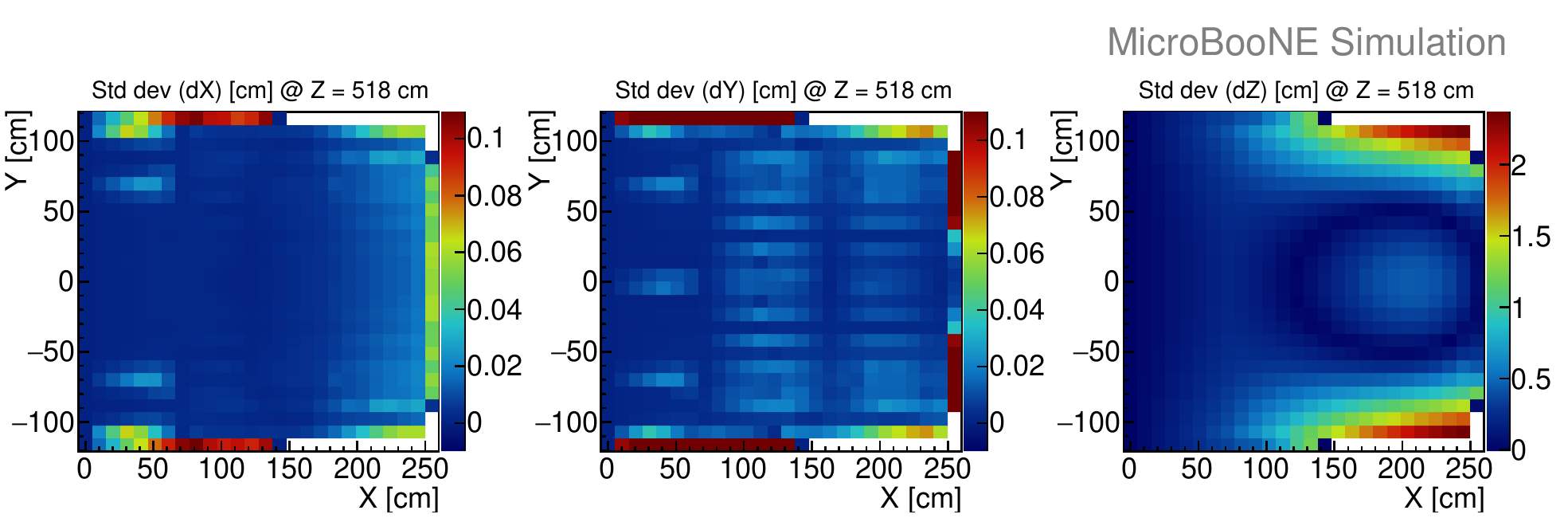}
\qquad
\caption{\label{fig:ToySim_Dmap_statErr_centralZ} 
Standard deviation of the calculated distortion for $dX$, $dY$ and $dZ$ at $Z=\SI{518}{\centi\metre}$ (central $Z$) in the toy simulation.}    
\end{figure}

Figure~\ref{fig:ToySim_Dmap_statErr_centralZ} shows the corresponding standard deviation of the calculated distortion in each bin at central $Z$ (section~\ref{subsec:SubMapMerging_StatUncertainty}).
In general, the standard deviations of $dX$, $dY$, $dZ$ are found to be small.
In comparison to other distortion components, the larger standard deviation in $dZ$ implies that the calculation of the displacement vectors is affected by the track angle.

\begin{figure}[htbp]
\centering 
\includegraphics[width=1.0\textwidth]{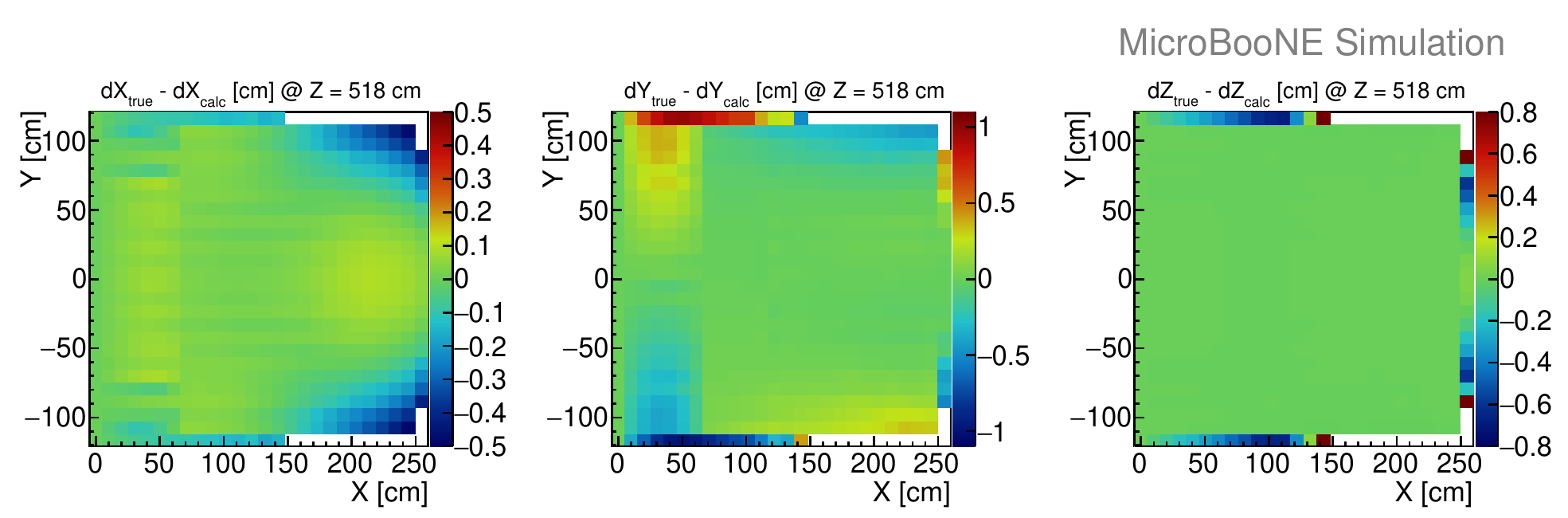}
\qquad
\caption{\label{fig:ToySim_Diff_centralZ} 
Bias in $\Delta dX = dX_\mathrm{true} - dX_\mathrm{calc}$, $\Delta dY = dY_\mathrm{true} - dY_\mathrm{calc}$, $\Delta dZ = dZ_\mathrm{true} - dZ_\mathrm{calc}$ from the toy simulation shown for a slice at a central slice at $Z=\SI{518}{\centi\metre}$.
In each bin the bias is relatively small compared to the distortion shown in figure~\ref{fig:ToySim_Dmap_centralZ}.}
\end{figure}

The difference between the simulation truth and the calculated distortion is taken as the bias of the algorithm, excluding the reconstruction process.
The bias $\Delta dX = dX_\mathrm{true} - dX_\mathrm{calc}$, $\Delta dY = dY_\mathrm{true} - dY_\mathrm{calc}$, $\Delta dZ = dZ_\mathrm{true} - dZ_\mathrm{calc}$ at central $Z$ is displayed in figure~\ref{fig:ToySim_Diff_centralZ}.
We observe good agreement, with only a small bias. 
The poor performance at the TPC edges is due to a combination of a lack of track coverage and effects of interpolation.


\subsubsection{Bias Study Using Complete Laser Simulation}
\label{subsubsec:LaserSimResult}

%

\begin{figure}[htbp]
\centering 
\includegraphics[width=1.0\textwidth,trim=0 0 0 0,clip]{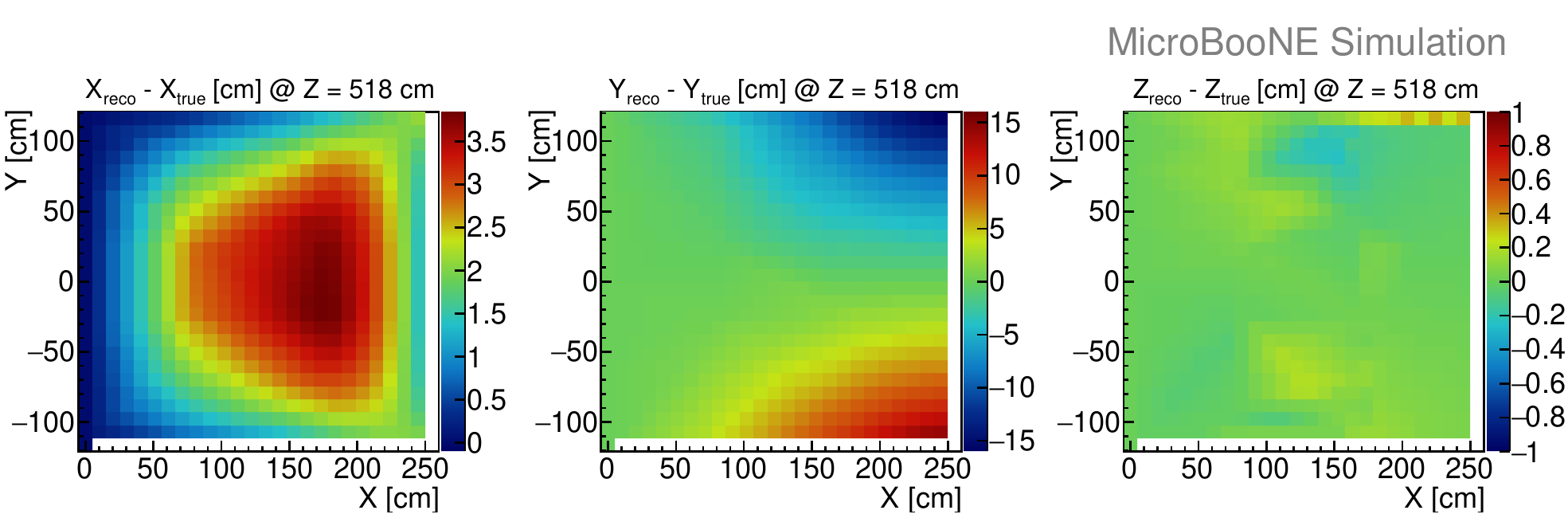}
\qquad
\caption{\label{fig:LaserSim_Dmap_centralZ} 
Calculated distortion map from the complete laser simulation at a central slice at $Z=\SI{518}{\centi\metre}$. 
Shown is $dX = X_\mathrm{reco} - X_\mathrm{true}$ (left), $dY = Y_\mathrm{reco} - Y_\mathrm{true}$ (middle), and $dZ = Z_\mathrm{reco} - Z_\mathrm{true}$ (right).}
\end{figure}

The calculated distortion map determined by the complete laser simulation at the central $Z$ slice is presented in $dX$, $dY$ and $dZ$ components in figure~\ref{fig:LaserSim_Dmap_centralZ}.
This distortion map is determined from the same E-field simulation, and thus input spatial distortion, as the one determined in the toy simulation study.
For this central $Z$ slice, the main features of the distortion are similar.
Compared to figure~\ref{fig:ToySim_Dmap_centralZ}, the maximum distortions in $dX$ and $dY$ have very similar magnitudes and similar positions, while the coverage near the cathode and at the bottom of the TPC are slightly different.
A possible reason for the difference is the behavior in reconstruction with more diffuse charge deposition.

\begin{figure}[htbp]
\centering 
\includegraphics[width=1.0\textwidth,trim=0 0 0 0,clip]{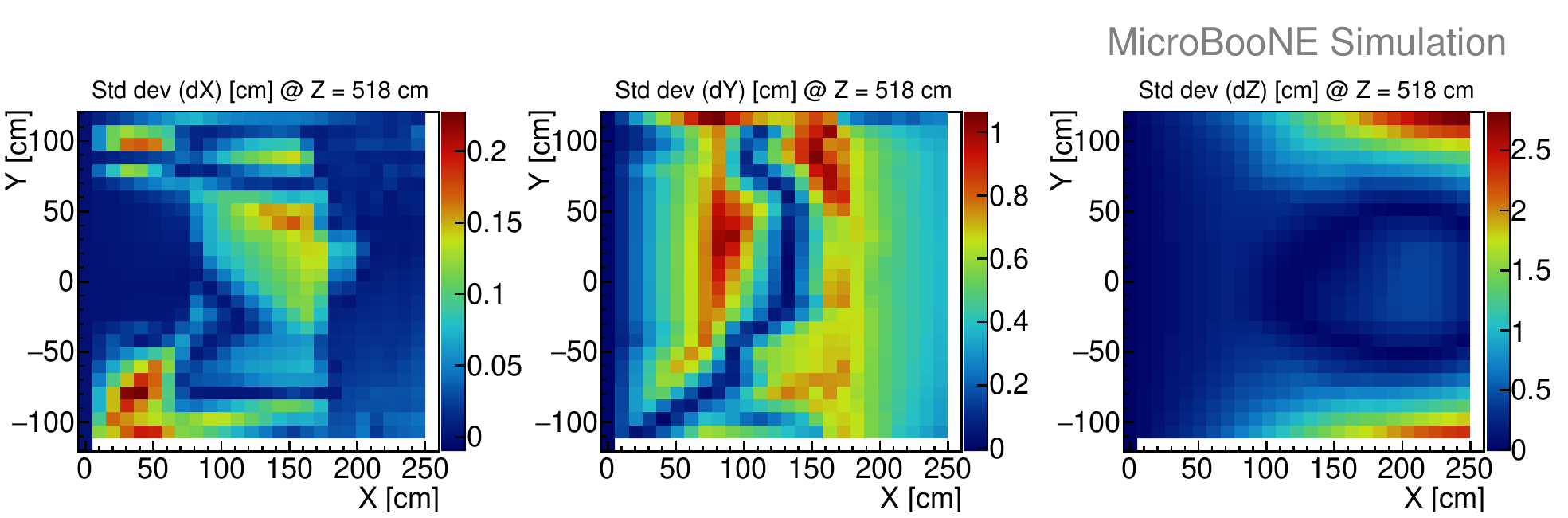}
\qquad
\caption{\label{fig:LaserSim_Dmap_statErr_centralZ} Standard deviation of the calculated distortion for $dX$, $dY$, and $dZ$ for a central slice at $Z=\SI{518}{\centi\metre}$ in the complete laser simulation.}    
\end{figure}

Figure~\ref{fig:LaserSim_Dmap_statErr_centralZ} illustrates the measured standard deviations in each bin at central $Z$.
Compared to the standard deviations of $dX$, $dY$, $dZ$ from the toy simulation (figure~\ref{fig:ToySim_Dmap_statErr_centralZ}), the complete laser simulation standard deviations fluctuate more, but in general remain reasonably small.
The larger standard deviation in $dZ$ (at central $Z$) at the top and bottom of the TPC close to the cathode is at the same scale as the toy simulation.
We attribute this increase in the statistical fluctuation of distortions to the track reconstruction.

\begin{figure}[htbp]
\centering 
\includegraphics[width=1.0\textwidth]{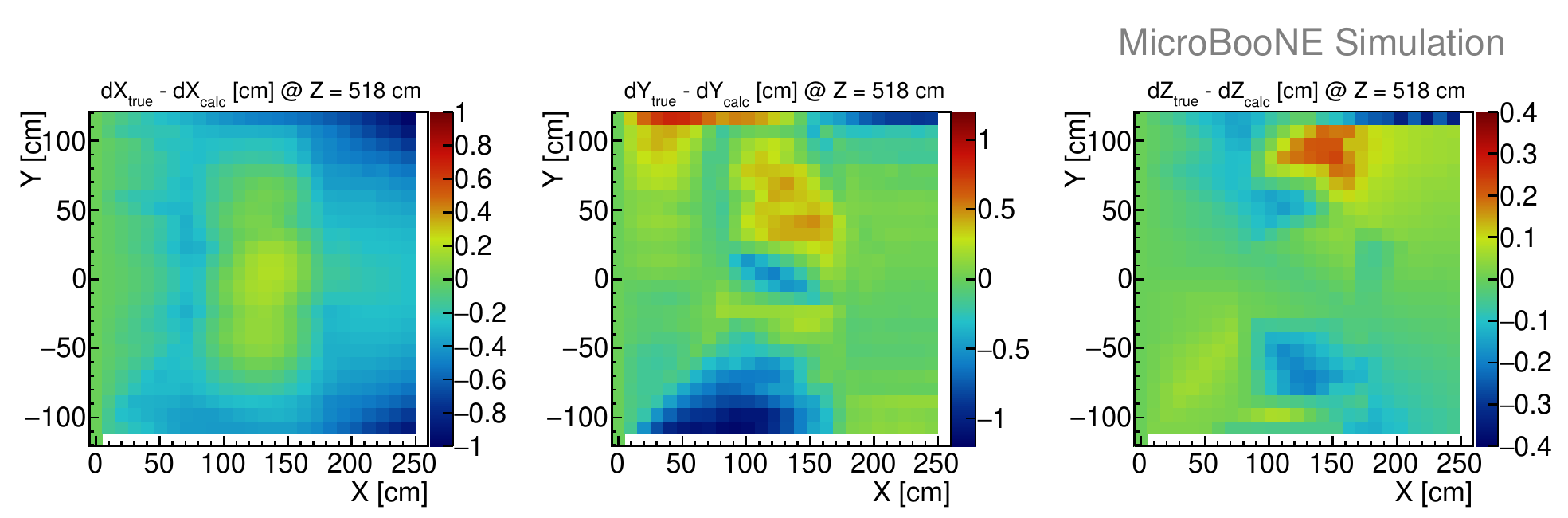}
\qquad
\caption{\label{fig:LaserSim_Diff_centralZ} 
Bias calculated as $\Delta dX = dX_\mathrm{true} - dX_\mathrm{calc}$, $\Delta dY = dY_\mathrm{true} - dY_\mathrm{calc}$, and $\Delta dZ = dZ_\mathrm{true} - dZ_\mathrm{calc}$ from the complete laser simulation at a central slice at $Z=\SI{518}{\centi\metre}$.
This bias shows the overall effect of the distortion calculation, the track reconstruction, and the selection.
In most of the bins, the bias is relatively small compared to the distortion.}
\end{figure}

The biases $\Delta dX = dX_\mathrm{true} - dX_\mathrm{calc}$, $\Delta dY = dY_\mathrm{true} - dY_\mathrm{calc}$ and $\Delta dZ = dZ_\mathrm{true} - dZ_\mathrm{calc}$ at central $Z$ is shown in figure~\ref{fig:LaserSim_Diff_centralZ}.
In comparison with the bias from toy simulation (figure~\ref{fig:ToySim_Diff_centralZ}), the magnitude of the bias slightly increases, but are still at the sub-centimetre level.
The slightly asymmetric distributions of $\Delta dY$ and $\Delta dZ$ in figure~\ref{fig:LaserSim_Diff_centralZ} also suggest an asymmetry introduced by reconstruction.
The bias here is also taken as the systematic uncertainty of the displacement calculation.


\subsubsection{Coverage of the TPC Volume with the UV-Laser Method}
\label{subsubsec:DmapValidRegion}

\begin{figure}[htbp]
\centering 
\includegraphics[width=1.0\textwidth,trim=0 0 0 0,clip]{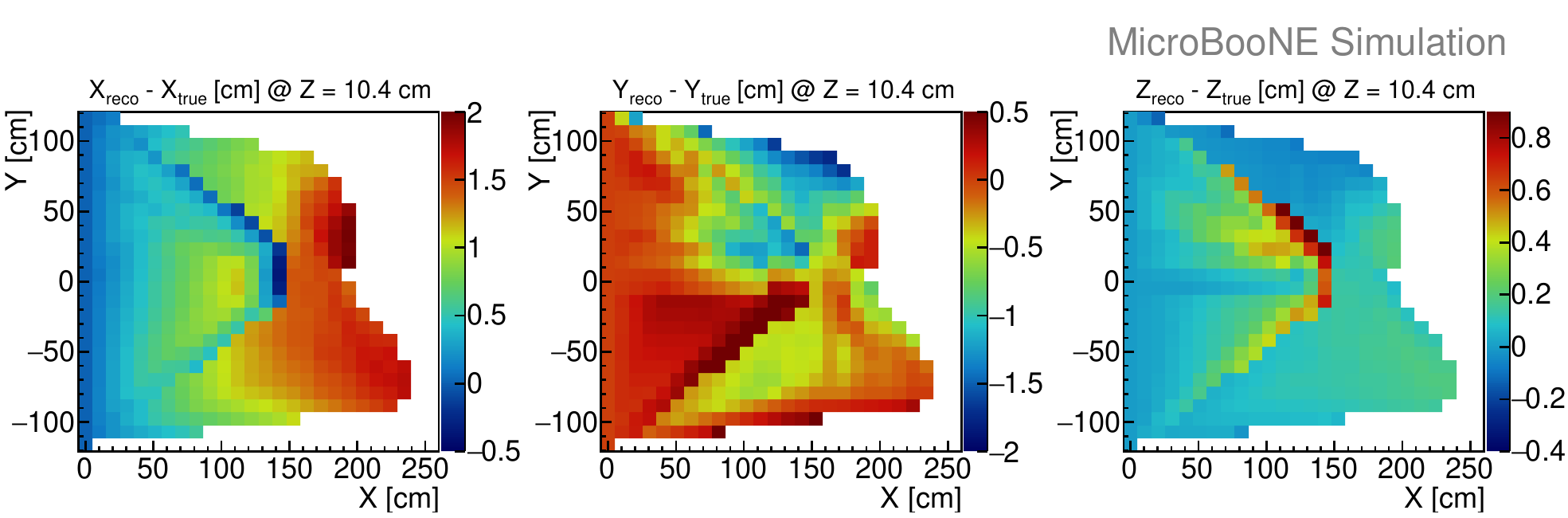}
\qquad
\caption{\label{fig:LaserSim_Dmap_edgeZ} 
Calculated distortion map from the complete laser simulation at an upstream slice at $Z=\SI{10.4}{\centi\metre}$.
Shown are $dX = X_\mathrm{reco} - X_\mathrm{true}$ (left), $dY = Y_\mathrm{reco} - Y_\mathrm{true}$ (middle), and $dZ = Z_\mathrm{reco} - Z_\mathrm{true}$ (right).
The blank parts are where the laser does not have coverage. 
A distinctive pattern from the center of this upstream slice in $Z$ to the top and bottom of the anode is noticeable in each component view.}
\end{figure}

\begin{figure}[htbp]
\centering 
\includegraphics[width=1.0\textwidth,trim=0 0 0 0,clip]{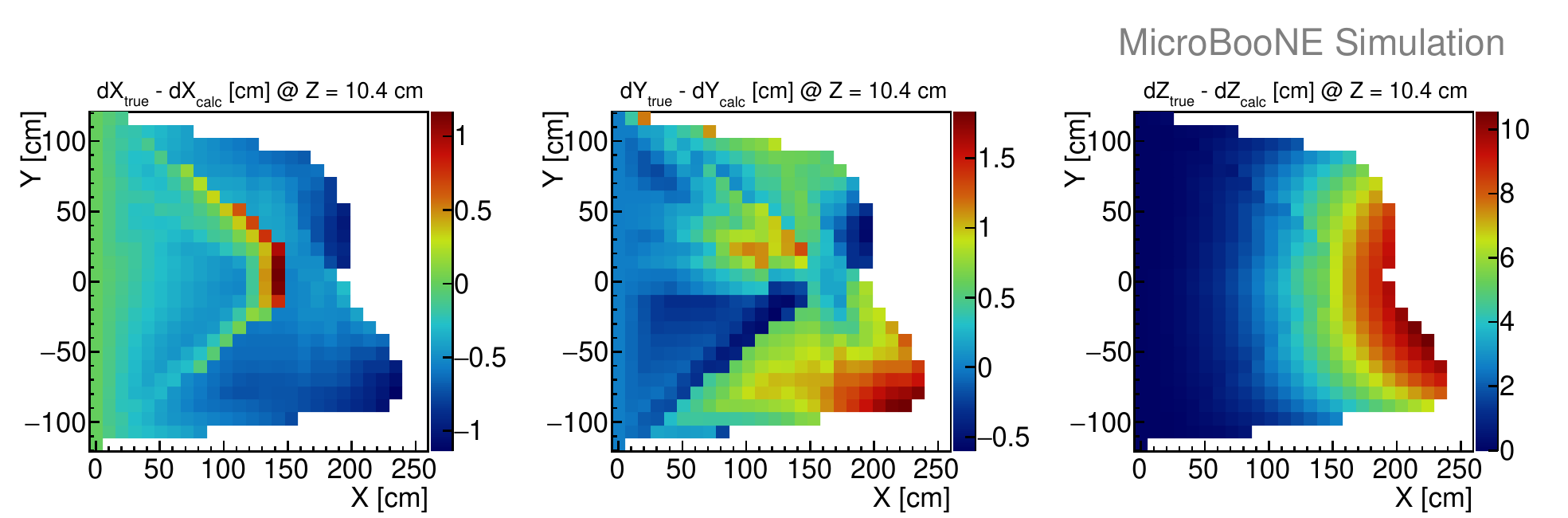}
\qquad
\caption{\label{fig:LaserSim_Dmap_Diff_edgeZ} 
Bias of the calculated distortion from the complete laser simulation at an upstream slice at $Z=\SI{10.4}{\centi\metre}$.
Shown are $\Delta dX = dX_\mathrm{true} - dX_\mathrm{reco}$, $\Delta dY = dY_\mathrm{true} - dY_\mathrm{reco}$ and $\Delta dZ = dZ_\mathrm{true} - dZ_\mathrm{reco}$.
}
\end{figure}

The distortion map calculated by the complete laser simulation is generally very close to the true distortion map.
However, at upstream and downstream edges of the TPC a sharp edge in the calculated distortion from the center of the $X$-$Y$ to the top and bottom ends of the anode is noticeable in all views, as shown in figure~\ref{fig:LaserSim_Dmap_edgeZ}.
Moreover, the bias of the same upstream $Z$ slice (figure~\ref{fig:LaserSim_Dmap_Diff_edgeZ}) shows a larger bias at the center of the $X$-$Y$ plane where the sharp edge is seen in figure~\ref{fig:LaserSim_Dmap_edgeZ}.
This is due to the specific placement of the laser mirrors in MicroBooNE.
The bias $\Delta dY$ are relatively large at lower $Y$ and $\Delta dZ$ reaches $\sim$\SI{10}{\centi\metre} at high $X$.

Since cosmic ray muons are also used in MicroBooNE to determine the distortion map through a separate technique, we are able to fill in the regions where the UV-laser method has larger uncertainties.
We define a region of validity for the UV-laser technique as the region with small uncertainty. 
As these regions are at the center of the TPC and the method using cosmic ray muons could compensate for gaps at the edges of the TPC; the combination is expected to improve the overall maps.

In simulation the distortion is continuous by design, the sharp edge in reconstructed distortion is an artificial effect caused by the algorithm.
The closest-point projection technique introduces an angular dependence in the initial determination of the displacement vectors.
The track iteration does reduce this bias, but is less effective at the upstream and downstream edges, which is due to the placement of the mirrors and thus the start points of the laser tracks.
Where these enter the TPC, they are close to collinear and point to the same spot on the mirror.
In figure~\ref{fig:coverage} we show that the starting points of the laser tracks are mostly located in an area not accessible by tracks from the other side.
Also, tracks which reach the other end in $Z$ usually have small angle differences compared to all tracks.
Imposing a boundary condition at the anode, together with interpolation, extends the local discontinuity from two points to a sharp triangle-like region in the distortion map.

\begin{figure}[htbp]
\centering 
\includegraphics[width=1.0\textwidth,trim=0 0 0 0,clip]{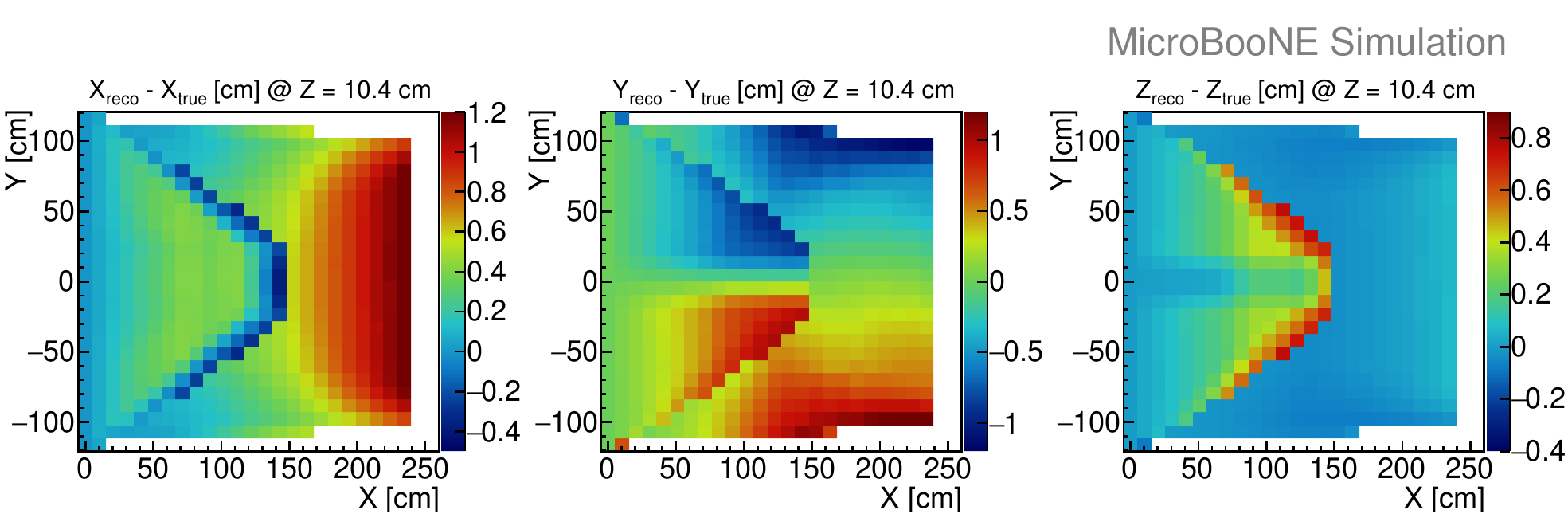}
\qquad
\caption{\label{fig:ToySim_Dmap_edgeZ} 
Calculated distortion map from the toy simulation at an upstream slice at $Z=\SI{10.4}{\centi\metre}$. 
The distinctive pattern in $YZ$ is more pronounced and in the same region as the one from the complete laser simulation.}
\end{figure}

\begin{figure}[htbp]
\centering 
\includegraphics[width=1.0\textwidth,trim=0 0 0 0,clip]{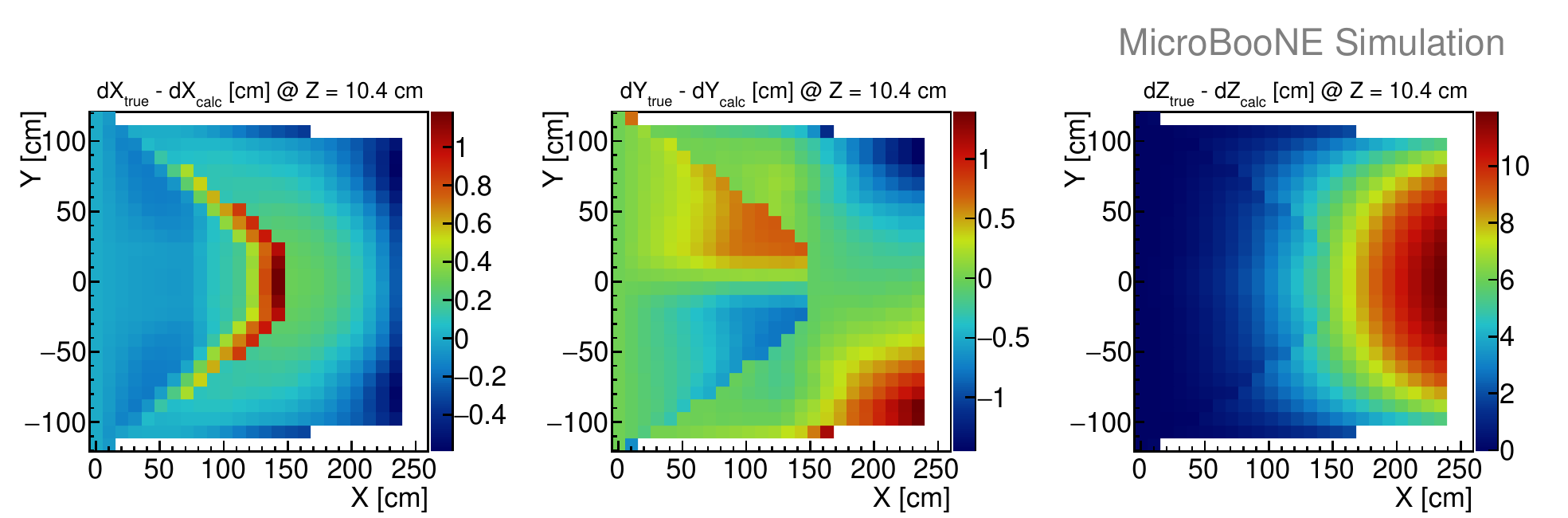}
\qquad
\caption{\label{fig:ToySim_Dmap_Diff_edgeZ} 
Bias of the calculated distortion from the toy simulation at an upstream slice at $Z=\SI{10.4}{\centi\metre}$, corresponding to the distortion maps shown in figure~\ref{fig:ToySim_Dmap_edgeZ}.
}
\end{figure}

Figures~\ref{fig:ToySim_Dmap_edgeZ} and \ref{fig:ToySim_Dmap_Diff_edgeZ} show the distortion and the bias determined by the toy simulation in the same $Z$ slice.
The sharp triangle-like region, the large bias $\Delta dY$ close to the top and bottom near the cathode, and the large bias $\Delta dZ$ near the cathode, are very pronounced.

\begin{figure}[htbp]
\centering 
\includegraphics[width=1.0\textwidth,trim=0 0 0 0,clip]{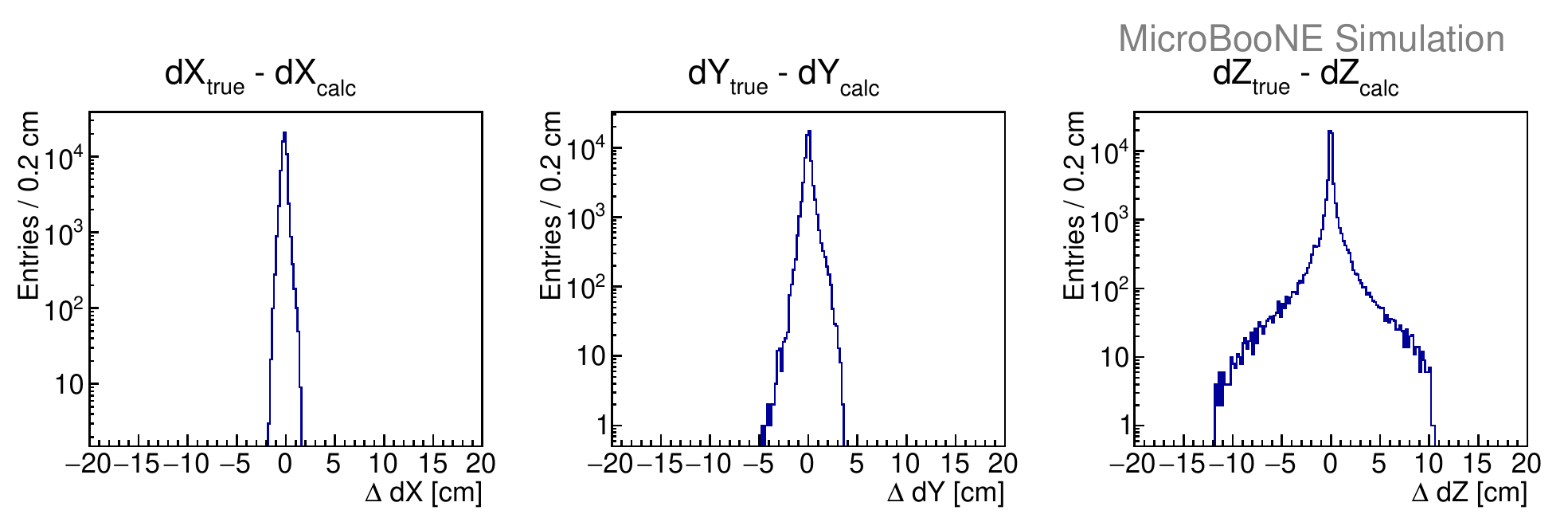}
\qquad
\caption{\label{fig:LaserSim_Dmap_Diff_wholeTPC} 
Bias of the calculated distortions in three components $\Delta dX$, $\Delta dY$ and $\Delta dZ$ from the complete laser simulation in the region with coverage. The bias distributions of $\Delta dY$ and $\Delta dZ$ are not symmetric with respect to the peak at 0.}    
\end{figure}

\begin{figure}[htbp]
\centering 
\includegraphics[width=1.0\textwidth,trim=0 0 0 0,clip]{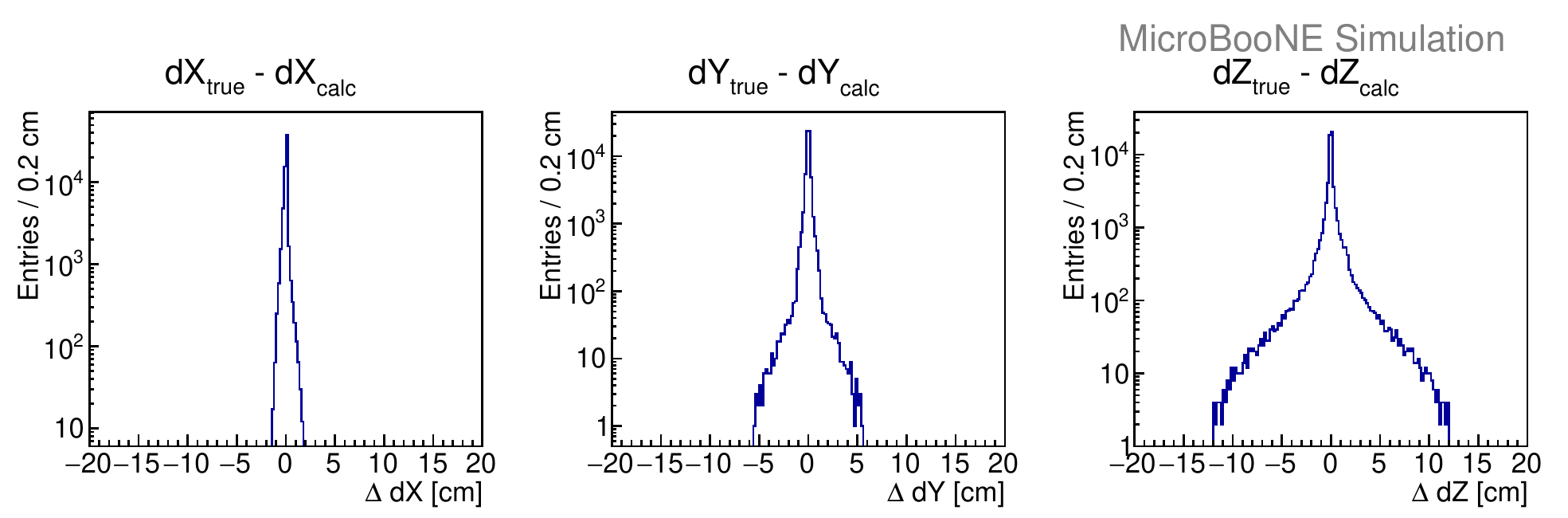}
\qquad
\caption{\label{fig:ToySim_Dmap_Diff_wholeTPC}
Bias of the calculated distortions in three components $\Delta dX$, $\Delta dY$ and $\Delta dZ$ from the toy simulation in the region with coverage.
The bias distributions are symmetric around the peak at 0.}    
\end{figure}

\begin{figure}[htbp]
\centering 
\includegraphics[width=0.6\textwidth]{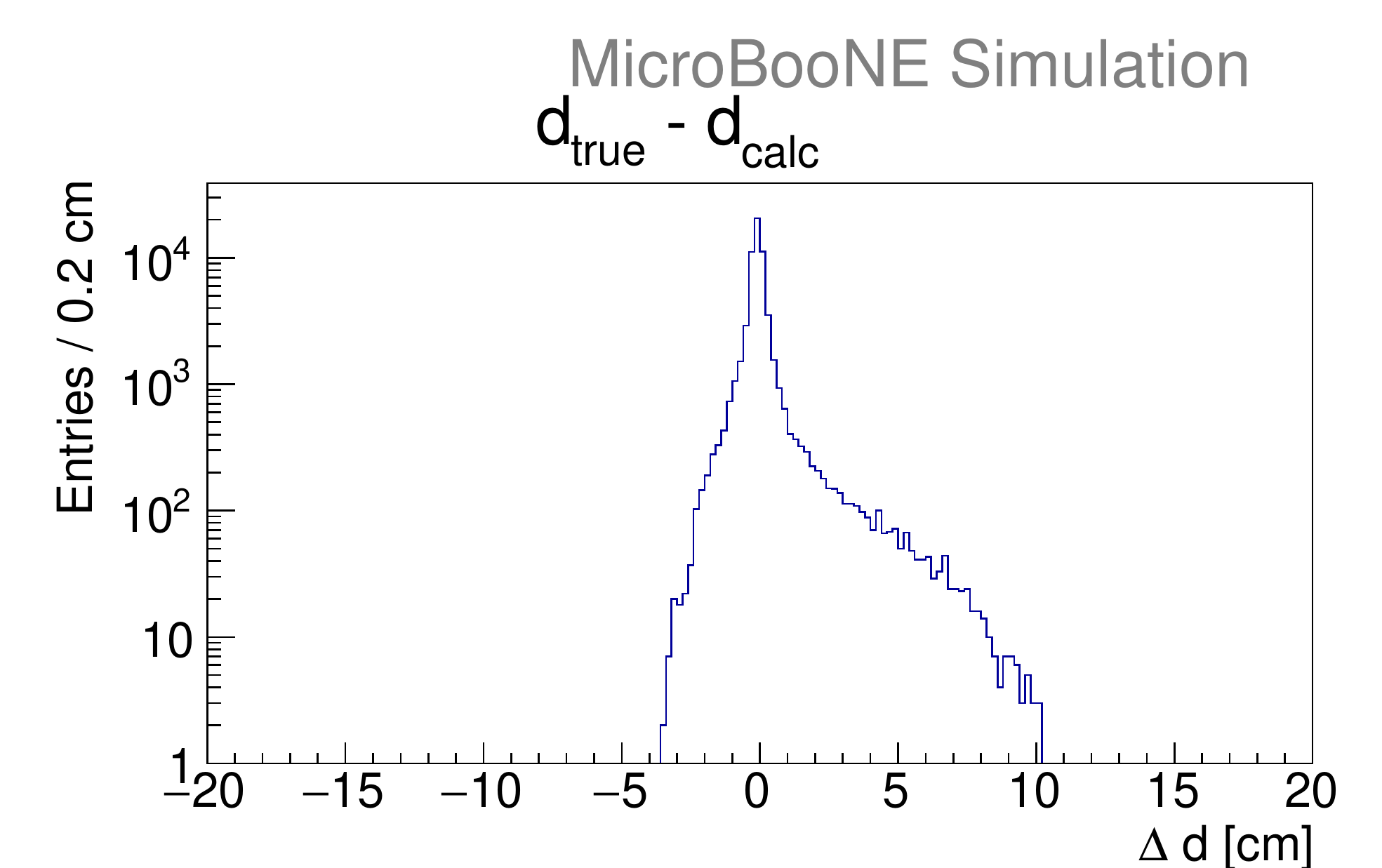}
\qquad
\caption{\label{fig:LaserSim_Diff_D_wholeTPC} 
The magnitude $d$ of the distortion vector, where $\Delta d = d_\mathrm{true} - d_\mathrm{calc}$ is the bias of the calculated distortion from the complete laser simulation of the map region with laser coverage. 
A excessive tail is populated at $\Delta d > 0$.
It is expected that overall the calculation distortion is smaller than the true distortion, because of the step of closest-point projection.}
\end{figure}

We proceed to define a more limited region of validity of the displacement map based on the bias and resolution maps.
Figure~\ref{fig:LaserSim_Dmap_Diff_wholeTPC} shows the bias of the calculated distortion compared to the complete laser simulation truth in three components $\Delta dX = dX_\mathrm{true} - dX_\mathrm{calc}$, $\Delta dY = dY_\mathrm{true} - dY_\mathrm{calc}$, and $\Delta dZ = dZ_\mathrm{true} - dZ_\mathrm{calc}$, for each bin.
The distribution of $\Delta dY$ has tails up to \SI{\pm 5}{\centi\metre}, and the distribution of $\Delta dZ$ has tails up to \SI{\pm 10}{\centi\metre}.
The bias $\Delta dX$, $\Delta dY$, $\Delta dZ$ derived by toy simulation are shown in figure~\ref{fig:ToySim_Dmap_Diff_wholeTPC}.
The bias distributions are similar in figures~\ref{fig:LaserSim_Dmap_Diff_wholeTPC} and \ref{fig:ToySim_Dmap_Diff_wholeTPC}.
The bias distributions of $\Delta dY$ and $\Delta dZ$ from the complete laser simulation (figures~\ref{fig:LaserSim_Dmap_Diff_wholeTPC}) are slightly asymmetric with respect to the peak at 0.
In comparison, the bias distributions from toy simulation (figures~\ref{fig:ToySim_Dmap_Diff_wholeTPC}) are more centered around zero.
The small difference is due to the additional bias introduced by the reconstruction as determined in the complete laser simulation, where $d = \sqrt{dX^2 + dY^2 + dZ ^2}$ is the length of the displacement vector.
The full distribution of biases $\Delta d = d_\mathrm{true} - d_\mathrm{calc}$ determined by the complete laser simulation over the entire TPC volume is shown in figure~\ref{fig:LaserSim_Diff_D_wholeTPC}.
A noticeable excess population of $\Delta d$ is positive because the calculation tends to underestimate $d_\mathrm{calc}$ by the closest point projection initiated method.

\begin{figure}[htbp]
\centering 
\includegraphics[width=1.0\textwidth,trim=0 0 0 0,clip]{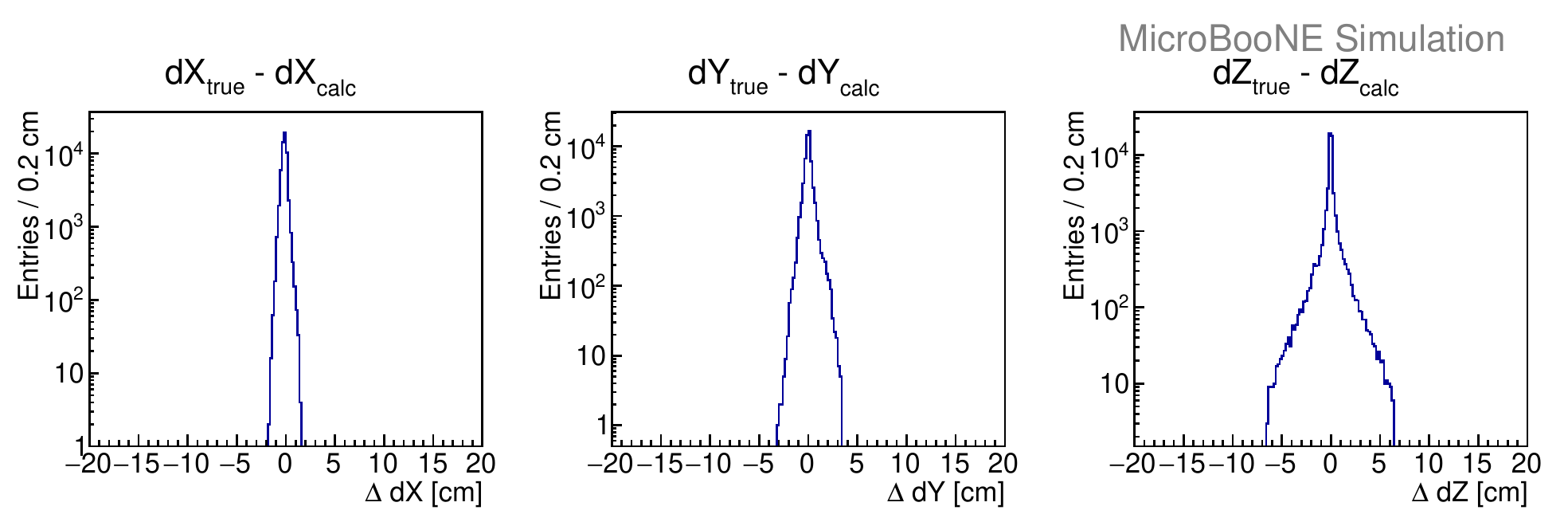}
\qquad
\caption{\label{fig:LaserSim_Diff_Dxyz_cutZ5} 
Bias of the calculated distortion in three components $\Delta dX$, $\Delta dY$ and $\Delta dZ$ from the complete laser simulation.
In the region of $ \SI{46.8}{\centi\metre} \leq Z \leq \SI{990}{\centi\metre}$, the bias distributions are centered at 0 and symmetric.
The spread of the bias distributions are also smaller compared to the ones including high $Z$ and low $Z$ ends.}    
\end{figure}

Figure~\ref{fig:LaserSim_Diff_Dxyz_cutZ5} shows the bias in $dX$, $dY$, $dZ$ excluding the TPC edge at low $Z$ ($ \SI{0}{\centi\metre} \leq Z \leq \SI{46.8}{\centi\metre}$) and high $Z$ ($ \SI{990}{\centi\metre} \leq Z \leq \SI{1036.8}{\centi\metre}$).
Biases in $dY$ and $dZ$ decreases in comparison to figure~\ref{fig:LaserSim_Dmap_Diff_wholeTPC}.
They are almost symmetric and are more consistently around zero.
As expected, a noticeable amount of underestimated $d$ are excluded when the valid region is limited to $ \SI{46.8}{\centi\metre} \leq Z \leq \SI{990}{\centi\metre}$.

\begin{figure}[htbp]
\centering 
\includegraphics[width=0.6\textwidth]{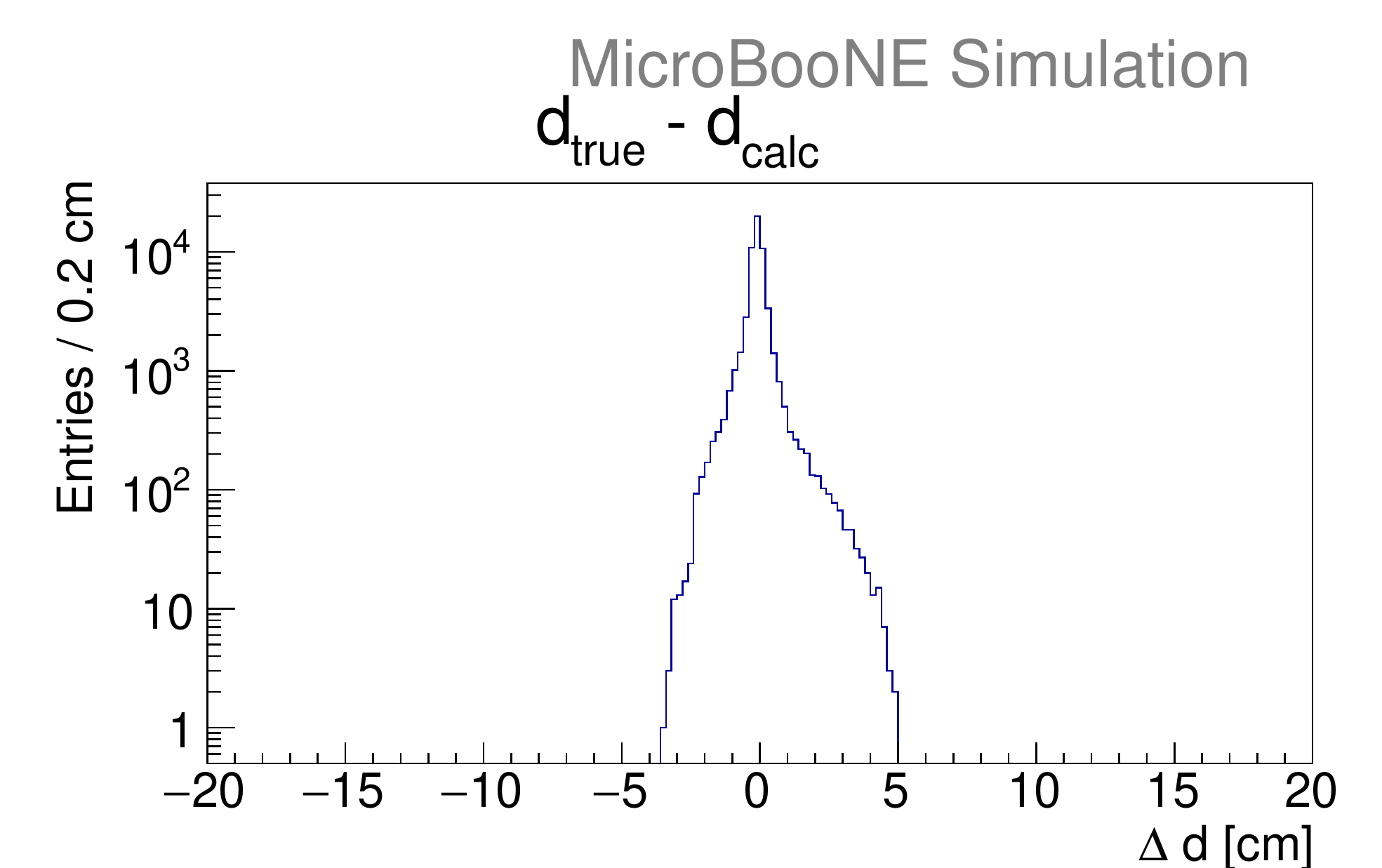}
\qquad
\caption{\label{fig:LaserSim_Diff_D_cutZ5}
Magnitude $\Delta d$ of the bias of the calculated distortion from the complete laser simulation.
The region of $ \SI{46.8}{\centi\metre} \leq Z \leq \SI{990}{\centi\metre}$ in the TPC is included, excluding high $Z$ and low $Z$ coverage. 
With this selection the tail of $\Delta d > 0$ is largely reduced.
}
\end{figure}

A further restriction on $Z$ to be between $\SI{99.8}{\centi\metre}$ and $\SI{937}{\centi\metre}$ results in a narrow bias of $\Delta dX$, $\Delta dY$ and $\Delta dZ$.
The standard deviation of $\Delta dY$ and $\Delta dZ$ are  \SI{0.40}{\centi\metre} and  \SI{0.47}{\centi\metre} respectively, which is compatible with standard deviation of $\Delta dX$ (\SI{0.25}{\centi\metre}).
The standard deviation of $\Delta d$ is also reduced to \SI{0.37}{\centi\metre}.
The tails of these bias distributions typically end around $\pm\SI{3}{\centi\metre}$.
A stricter restriction to the region of validity does not significantly narrow the bias, so the final valid region is $ \SI{99.8}{\centi\metre} \leq Z \leq \SI{937}{\centi\metre}$ and the full length in $X$ and $Y$.

\begin{figure}[htbp]
\centering 
\includegraphics[width=1.0\textwidth,trim=0 0 0 0,clip]{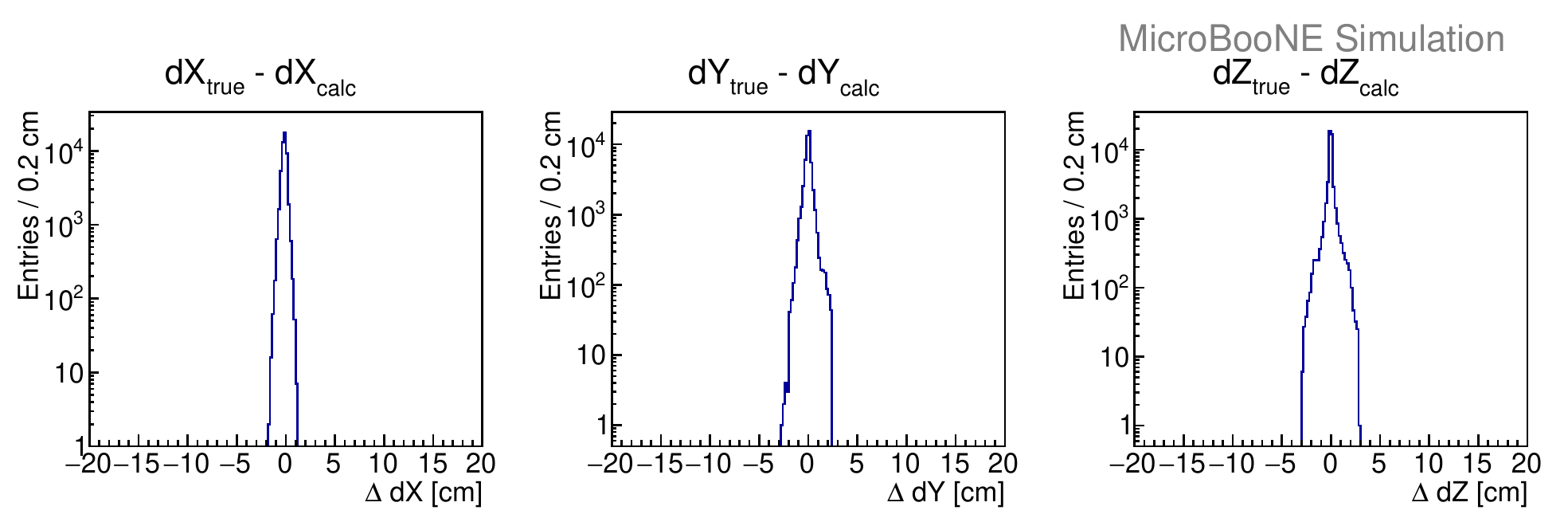}
\qquad
\caption{\label{fig:LaserSim_Diff_Dxyz_cutZ10} 
Bias of the calculated distortion in the three components $\Delta dX$, $\Delta dY$, and $\Delta dZ$ from the complete laser simulation.
Only the region of $ \SI{99.8}{\centi\metre} \leq Z \leq \SI{937}{\centi\metre}$ is included.
The bias distribution are now closer to symmetric around the peak at 0.
The standard deviations of each components are $\Delta dX \sim \SI{0.2}{\centi\metre}$, $\Delta dY \sim \SI{0.4}{\centi\metre}$, and $\Delta dZ \sim \SI{0.4}{\centi\metre}$.}    
\end{figure}

\begin{figure}[htbp]
\centering 
\includegraphics[width=0.6\textwidth]{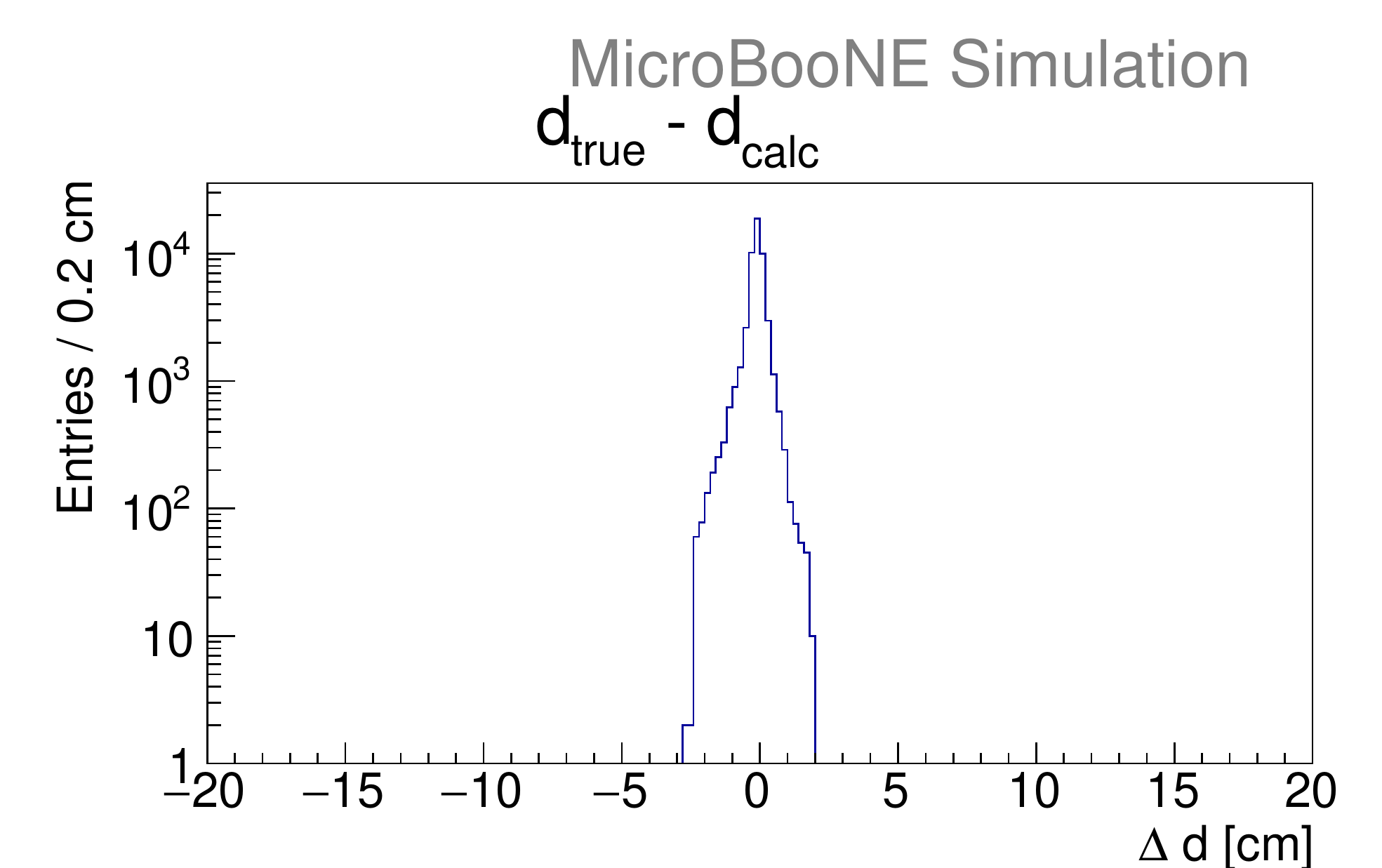}
\qquad
\caption{\label{fig:LaserSim_Diff_D_cutZ10}
Magnitude $\Delta d$ of the bias of the calculated distortion from the complete laser simulation, with $ \SI{99.8}{\centi\metre} \leq Z \leq \SI{937}{\centi\metre}$.
$\Delta d$ of each bin in the region of validity is relatively small.
The standard deviation is $\Delta d \sim \SI{0.4}{\centi\metre}$.}
\end{figure}

The methodology for calculating displacement maps, which has been validated by toy and complete laser simulations, can be regarded as a valid technique.
Room for improvement lays in the calculation of displacement vectors and laser track reconstruction.

\section{E-field and Drift Velocity Maps}
\label{sec: Emap}
Although the E-field for drifting ionized electrons is designed to be uniform in between the cathode and anode planes, this is difficult to achieve in practice.
Both the TPC geometry and space charge effect can distort the E-field.
To better identify charged particles and their locations in the TPC, a precise mapping of the local E-field at each point is necessary.

The E-field is calculated and then formed in the TPC active volume in true spatial coordinates on an adjustable regular grid.
We choose 26 bins along $X$, 26 bins along $Y$ and 101 bins along $Z$ with the TPC edge at the center of the edge-most bins.
With interpolation, the E-field can be determined at any position in the valid map area.

In the following, we describe how the E-field is extracted from distortions maps derived from UV-laser data. 
In principle, the technique can be also applied on any calibration data source.

\subsection{Field lines}
\label{subsec:FieldLine}
E-field lines extend from the anode to the cathode.
Ideally, ionized electrons generated anywhere along a field line drift to the same position at the anode.
Since the ionized electrons on a field line share the same read-out position, they would be reconstructed along a line which passes that read-out position and is perpendicular to the anode plane.

\begin{figure}[htbp]
\centering 
\includegraphics[width=0.8\textwidth]{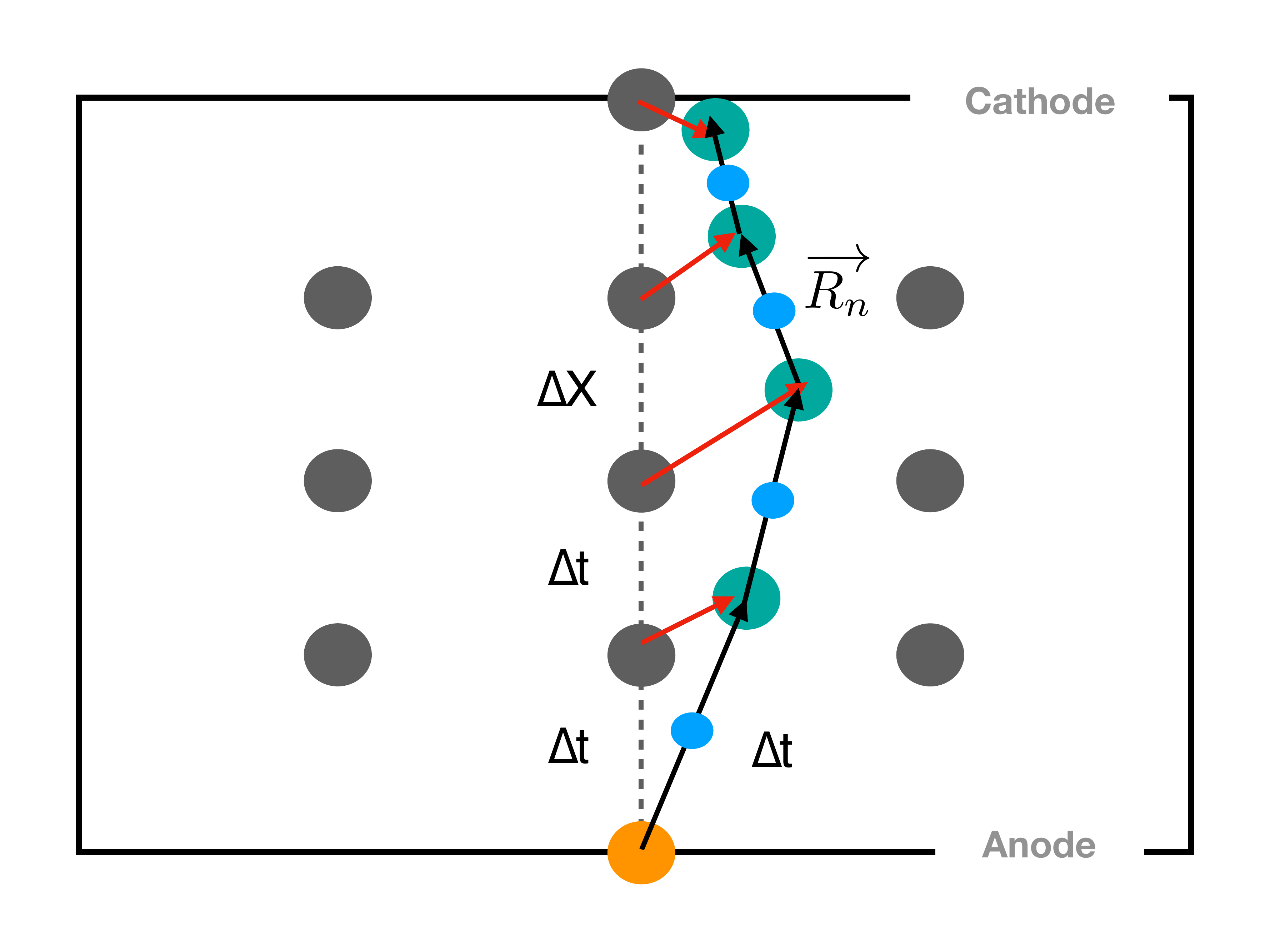}
\qquad
\caption{\label{fig:E_schematic} 
 Diagram showing how the E-field is calculated. 
	The gray dots are the regular grid points in reconstructed spatial coordinates.
        They are spaced by $\Delta x = v_0 \Delta t$, corresponding to $\Delta t$ with the nominal drift velocity $v_0$. 
	The turquoise points are the positions of the actual energy deposition in true spatial coordinates.
        As the time sampling of the TPC is fixed, these are also spaced by $\Delta t$.
	The red arrows represent the correction vector pointing from the reconstructed to the corresponding true positions. 
	Ionized electrons drifting from any turquoise dots along a field line are read out at the same orange point at the anode. 
	Thus the line connecting these points, with $\vec{R _ \mathrm n}$ the spatial vector at each step, follows the E-field.
	The black arrows indicate the direction of the derived E-field.}
\end{figure}

We take the correction map, which has a regular grid in reconstructed spatial coordinates, as the input to the E-field calculation.
In figure~\ref{fig:E_schematic}, the gray dots are regular grid points in reconstructed spatial coordinates.
The dots along the dashed line all correspond to the same read-out position (orange).
The red arrows are the correction vectors on those gray grid points, and the turquoise dots are the corresponding true positions.
We organize the turquoise dots by the distance of their corresponding reconstructed grid points to their read-out position.
The connection from that read-out position to the first green dot, represents the field line in this step.
Similarly, the continuous connection from a turquoise dot to the next turquoise dot forms the field line.


\begin{figure}[htbp]
\centering 
\includegraphics[width=1.\textwidth]{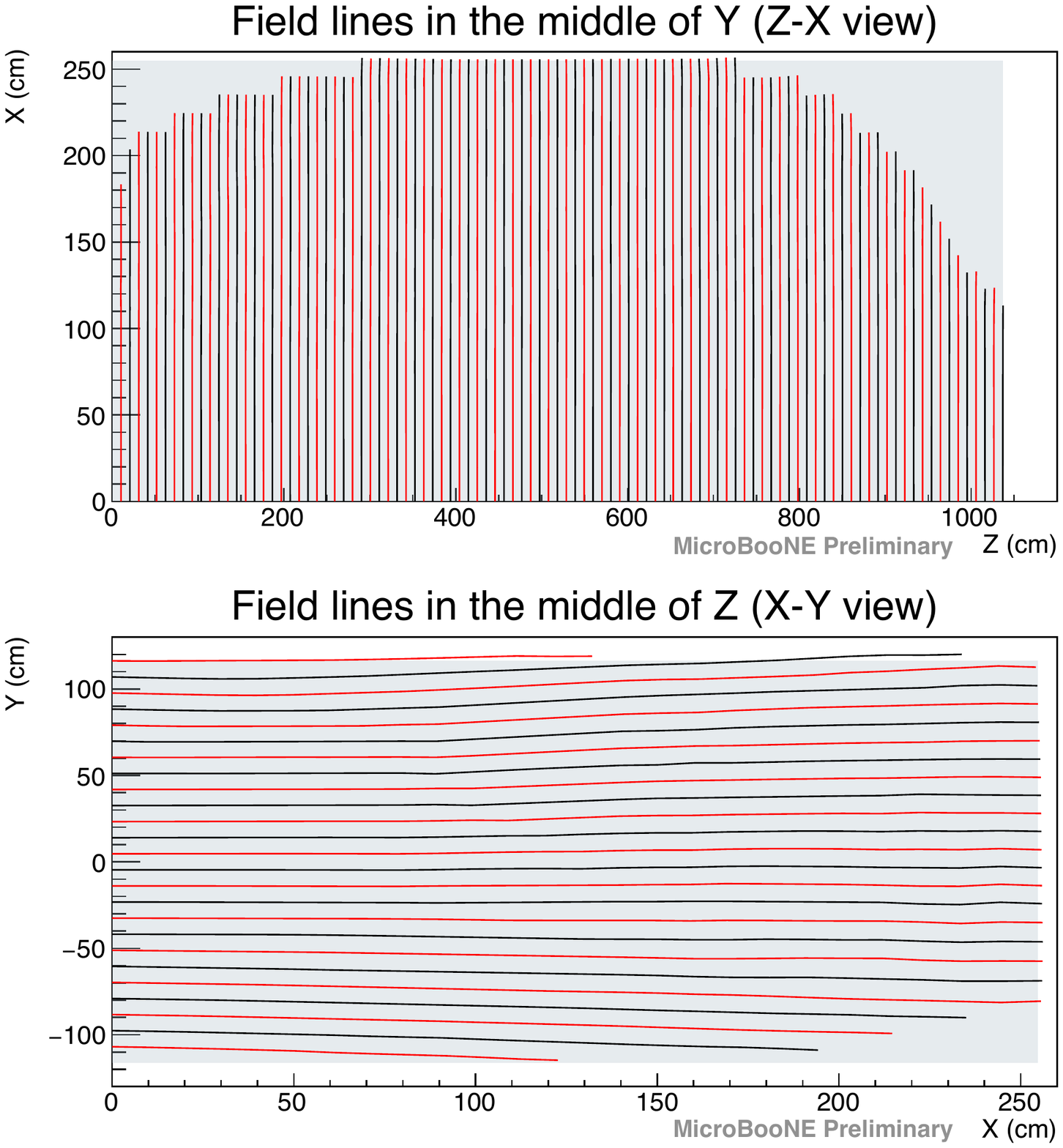}
\qquad
\caption{\label{fig:FieldLines} 
The upper plot shows the projection of the field lines at a central slice in $Y$ ($Y = 0$), and the bottom plot shows the projection of field lines at central $Z$ ($Z= \SI{518}{\centi\metre}$). 
The field lines are derived from laser data. 
Different line colors are used to distinguish neighboring field lines. 
The gray box shows the projection of the TPC active volume in true spatial coordinates.}
\end{figure}

Figure~\ref{fig:FieldLines} shows the field lines derived from laser data at $Y=0$ (central $Y$) and at $Z = \SI{518}{\centi\metre}$ (central $Z$).
The field lines are smooth. 
No kinks and no crossing lines are observed.
The distortion in the direction of the E-field is pronounced at central $Z$.
Close to the cathode, the E-field bends outwards.
Thus, in reconstructed spatial coordinates, the corner near the top and bottom edges of the cathode is not seen.
The gray boxes in figure~\ref{fig:FieldLines} indicate the projection of the TPC active volume.
In principle, the field lines should be contained in the TPC active volume.
The top two field lines at central $Z$ (bottom plot) are outside the gray box by $\sim$\SI{1}{\centi\metre}.
The E-field exists outside the TPC active volume around the boundaries because the field-cage rings are a few centimetres away from the active (instrumented by wires) volume boundaries.
The E-field near the field-cage rings is typically stronger.
This effect may be physical, but the position uncertainties in this region are large, on the order of \SI{1}{\centi\metre}.
Also, the field lines extend slightly beyond the gray box in the $X$ direction.
We obtain $X$ coordinates via the nominal drift velocity along $X$ in the TPC.
The nominal drift velocity ($v_0$) is \SI{1.098}{\milli\metre \per \micro\second}.
We take the cathode position as \SI{254.8}{\centi\metre} from an estimate of the TPC drift length in MicroBooNE at a LAr temperature of (\SI{89}{\kelvin}).
This value has an uncertainty of about \SI{1}{\centi\metre} along $X$.

Some field lines in figure~\ref{fig:FieldLines} do not extend across the entire drift length due to a lack of laser coverage.
The asymmetry in $Z$ (upstream and downstream) may be due to the difference in the two laser positions and pulse energy settings.
While the asymmetry in $Y$ is likely due to the fact that the dichroic mirrors have decreasing reflectance efficiency for larger incident angle.

\subsection{Extraction of the Local Drift Velocity}
\label{subsec:Driftv}

We illustrate the calculation of the local drift velocity using figure~\ref{fig:E_schematic}.
We define a step as going from the read-out position to the first true track point position, or from one track point to the next in true coordinates.
Ionized electrons from the first true track point drift to the anode plane and thus the readout position in a time $\Delta t$.
The reconstruction procedure restores the position to the first point, assuming a uniform E-field perpendicular to the readout plane at a distance $X_1=|v_0| \cdot \Delta t$.
All the reconstructed track points are spaced $\Delta X =|v_0| \cdot \Delta t$ from each other, as they are on a regular grid in reconstructed spatial coordinates.
The nominal drift velocity, $v_0$, is taken as constant along $X$, so the time difference from any gray point to the next gray point is also $\Delta t$.
The \textit{n}-th track point in reconstructed coordinates is associated to the \textit{n}-th true point, where \textit{n} is an integer between 1 and 24 for the grid used here.
The time which is used in the reconstruction of the \textit{n}-th point is $\mathit{n} \Delta t$, so the drift time from the \textit{n}-th point to the readout point is also $\mathit{n} \Delta t$.
Similarly, the time associated to the $(n+1)$-th reconstructed or true points is $(n+1) \Delta t$.
Therefore, the time of any step \textit{n} is $\Delta t$, which is the same as the time in between two neighboring gray points along $X$.

We assume the local drift velocity is constant during each step.
The step vector $\vec{R_\mathit{n}}$ points from the \textit{n}-th to the $(n+1)$-th true point.
Eq.~\ref{eq:deltaX} shows the relation of $|\vec{R_\mathit{n}}|$ and $\Delta X$,
\begin{equation}
\label{eq:deltaX}
\frac{|\vec{R_\mathit{n}}|}{\Delta X} = \frac{|\vec{v_\mathit{n}}|\Delta t}{|\vec{v_0}|\Delta t} .
\end{equation}
Thus, the local drift velocity can be calculated as:
\begin{equation}
\label{eq:deltaV}
 |\vec{v_\mathit{n}}| = \frac{|\vec{R_\mathit{n}}|}{\Delta X} |\vec{v_0}| .
\end{equation}
The local drift velocity $\vec{v_\mathit{n}}$ has the opposite direction to $\vec{R_\mathit{n}}$.
We assign $\vec{v_\mathit{n}}$ to the center of step \textit{n}.

If the correction vector for the \textit{n}th reconstructed point contains large uncertainty, it only affects $\vec{R_\mathit{n}}$ and $\vec{R_{n+1}}$.
Thus, only the local drift velocity $\vec{v_n}$ and $\vec{v_{n+1}}$ are influenced.

If there is no spatial displacement, any local drift velocity, $\vec{v_\mathit{n}}$, would be the same as the nominal drift velocity, $\vec{v_0}$.
If the correction vectors on the reconstructed points are ideal, the magnitude of the assumed nominal drift velocity, $|\vec{v_0}|$, is irrelevant for the measurement of the velocity map.
Any deviation would be absorbed by the correction map and the step vector $\vec{R_\mathit{n}}$ would remain unchanged.
The local drift velocity, $|\vec{v_\mathit{n}}| = \frac{|\vec{R_\mathit{n}}|}{\Delta t} = \frac{|\vec{R_\mathit{n}}|}{\Delta X'}  |\vec{v'_0}| = \frac{|\vec{R_\mathit{n}}|}{\Delta X}  |\vec{v_0}|$, also would not change.
However, the displacement vectors calculated as described in section~\ref{subsec:correctionVec} have an angular dependence on the  laser tracks.
Thus, the magnitude of the nominal drift velocity $|\vec{v_0}|$ can be relevant to the local drift velocity calculation.

The local drift velocity vectors on the regular grid are then interpolated from the drift velocities of the local steps as described in section~\ref{subsec:Interploation}.

\subsection{Extraction of the Local E-field}
\label{subsec:Emap}
\begin{figure}[htbp]
\centering 
\includegraphics[width=1.\textwidth]{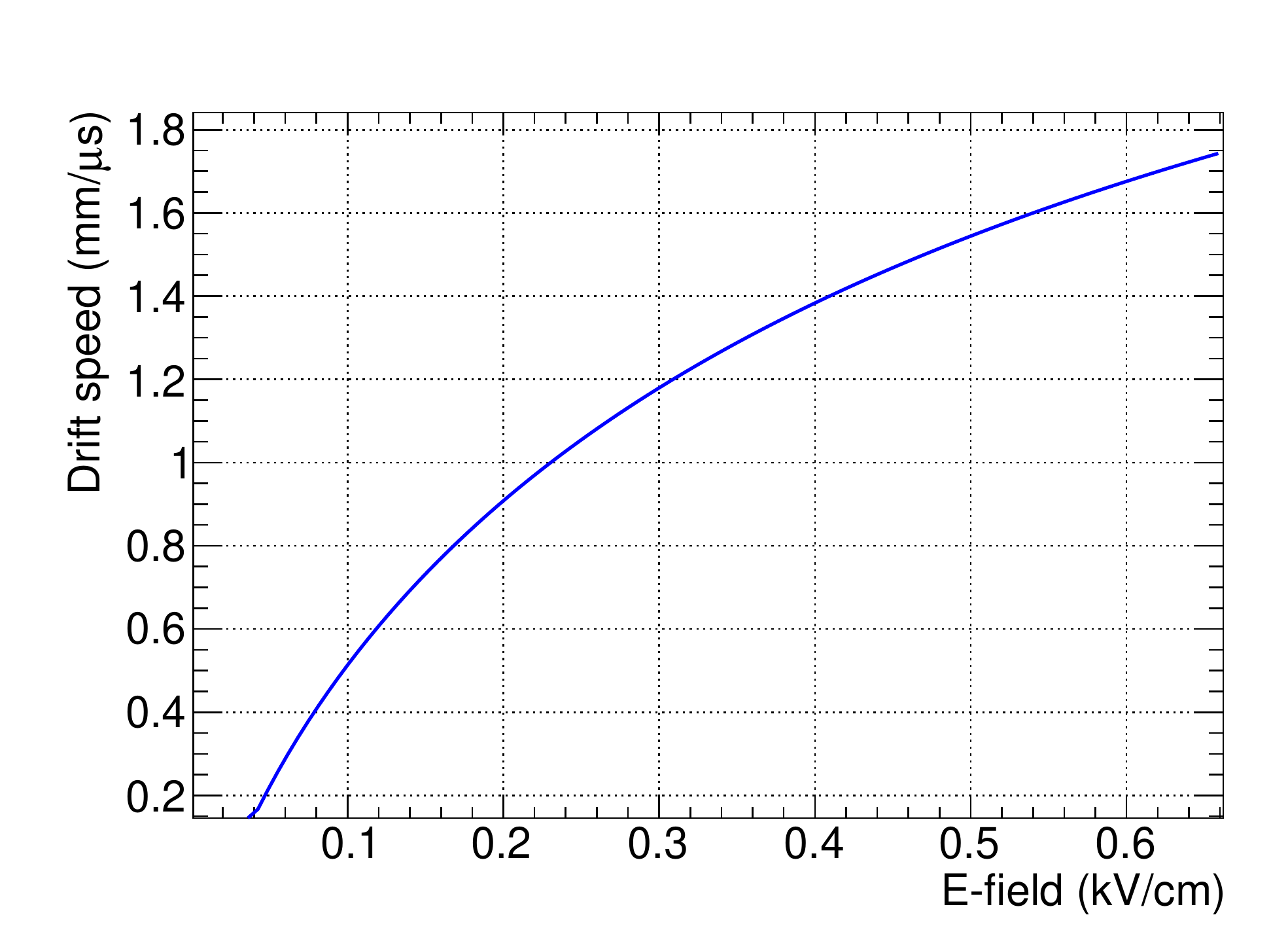}
\qquad
\caption{\label{fig:driftV} 
Drift speed as a function of E-field ~\cite{Amoruso:2004ti, Walkowiak:2000wf}, in the range of \SIrange {0.04}{0.66}{\kilo\volt\per\centi\metre}.
}
\end{figure}

The local E-field is derived from the local drift velocity map.
The direction of the E-field is the same as the step vector $\vec{R_\mathit{n}}$.
The magnitude of the local E-field is determined from the corresponding local drift velocity by using the relation shown in figure~\ref{fig:driftV}.
The drift speed increases from $\sim$\SI{0.20}{\milli\metre\per\micro\second} to $\sim$\SI{1.75}{\milli\metre\per\micro\second} when the corresponding E-field increases from $\sim$\SI{0.02}{\kilo\volt\per\centi\metre} to $\sim$\SI{0.66}{\kilo\volt\per\centi\metre}.
The drift speed as function of E-field is taken from the LArSoft package and from ref.~\cite{Amoruso:2004ti, Walkowiak:2000wf}. 
The drift speed as a parametric function of the E-field and LAr temperature is given in ref.~\cite{Walkowiak:2000wf}.
In MicroBooNE, the LAr temperature is stably maintained at \SI{89}{\kelvin}, which we use in the drift speed function.

The drift speed has a non-linear, monotonic dependence on the E-field.
Thus one can determine the local E-field from the measured local drift speed magnitude and direction.
We then use bisection to numerically determine the magnitude of the local E-field.
When the distortion of the E-field is not dramatic, the change of drift speed is nearly linear around the nominal field $|\vec{E_0}| = \SI{0.274}{\kilo\volt\per\centi\metre}$.

We assign the local E-field vectors to the center of the steps (blue points) shown in figure~\ref{fig:E_schematic}.
They share the same positions as the local drift velocities of the steps.
We also set the average E-field of two neighboring steps to the midpoint of those two steps.
The start of the first step and the end of the last step use the local E-field of the first and last step.
The local E-field on the regular grid is formed by interpolation as described in section~\ref{subsec:Interploation}. 


\subsection{Validation of the E-field determination}
\label{subsec:Emap_verification}
The space charge simulation in LArSoft, described in section~\ref{sec:TrackSim}, contains an input E-field map, a correction map and a distortion map.
We take the correction map from the space charge simulation to calculate a E-field map.
The calculated E-field map is then compared to the simulation truth to evaluate the bias introduced by this method.
The bias is defined as the bin-to-bin difference of calculated E-field and E-field from the simulation truth.
The bias also folds uncertainties of the E-field map and the correction map from the space charge simulation.
The E-field map from simulation truth has 21 bins in $X$, 21 bins in $Y$ and 81 bins in $Z$.
The first and last bin centers of each direction are located at the TPC limits.
To avoid introducing different bias, we use the same binning for the calculated E-field map in validation.

\begin{figure}[htbp]
\centering 
\includegraphics[width=1.\textwidth]{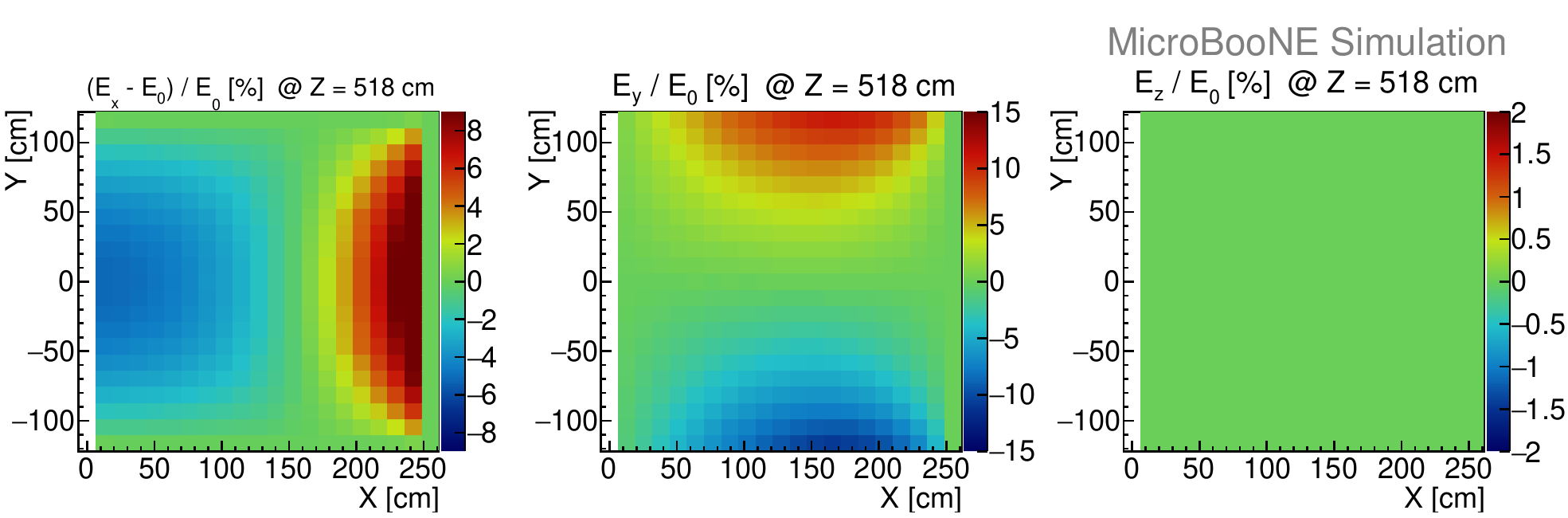}
\qquad
\includegraphics[width=1.\textwidth]{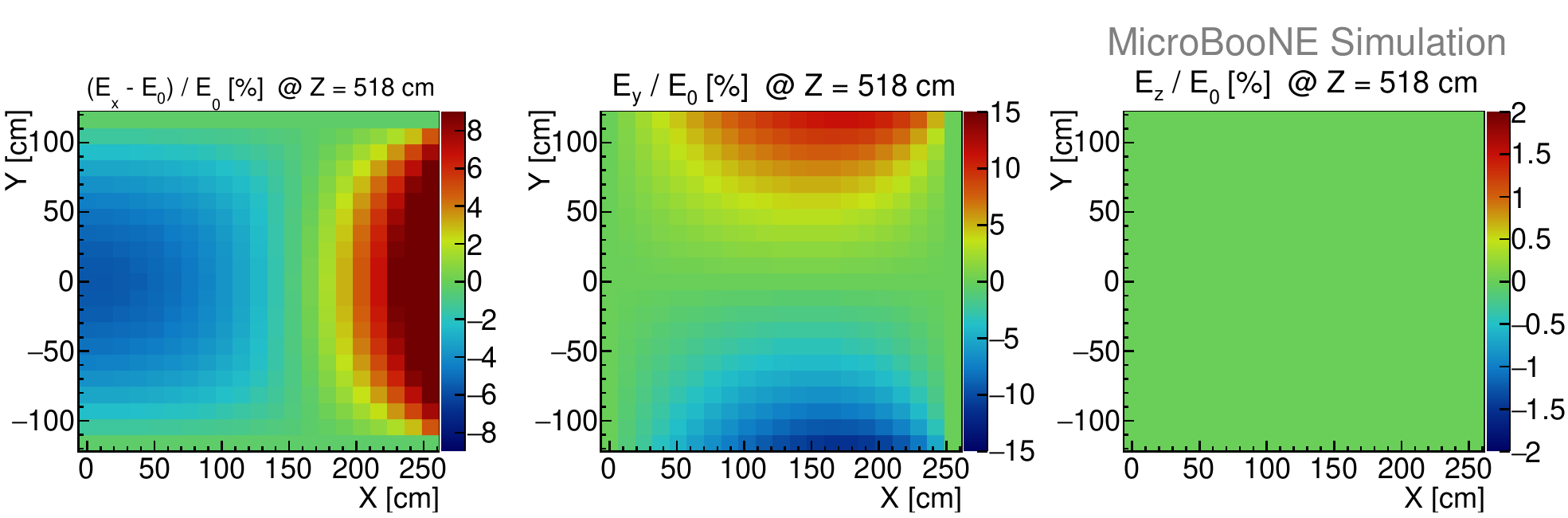}
\caption{\label{fig:ESCE_central} 
Relative E-field distortions (in percent) with respect to the nominal E-field in components $(E_X - E_0) / E_0 [\%]$ (left), $E_Y / E_0 [\%]$ (middle) and $E_Z / E_0 [\%]$ (right).
The upper row shows the result of the calculated E-field from the correction map..
The lower row shows the E-field from the space charge simulation, which is used as simulation truth here.
Both of them show the E-field distortion at a central slice at $Z=\SI{518}{\centi\metre}$, in the TPC true spatial coordinates.
}
\end{figure}

\begin{figure}[htbp]
\centering 
\includegraphics[width=1.\textwidth]{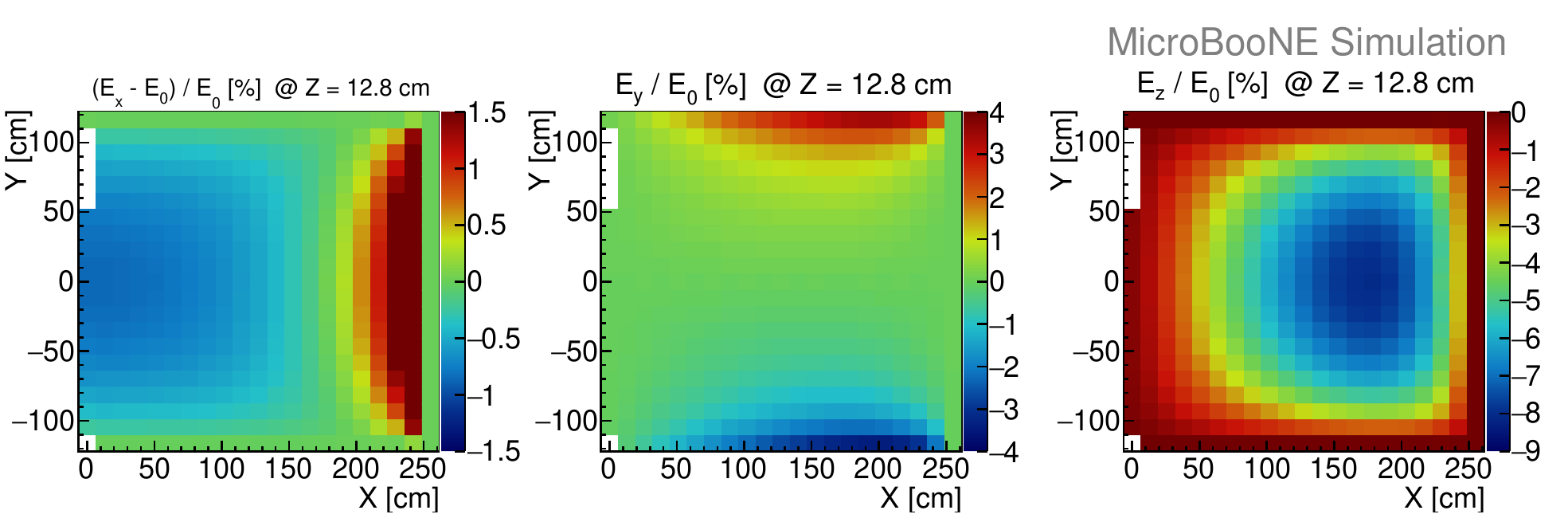}
\qquad
\includegraphics[width=1.\textwidth]{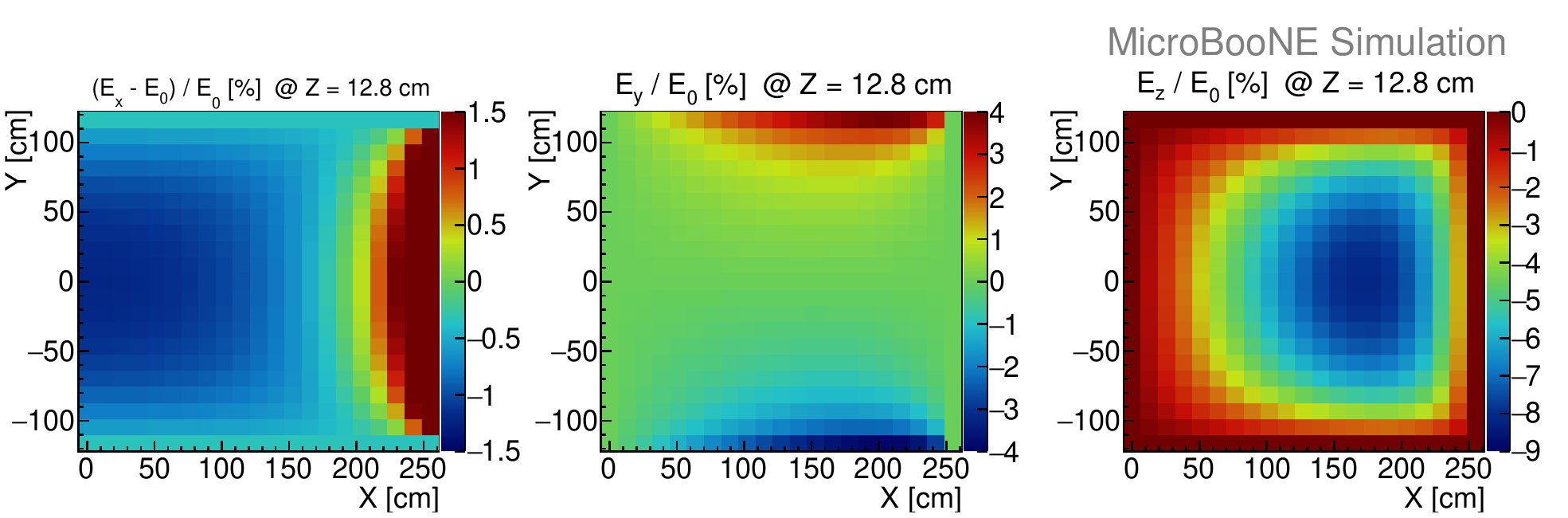}
\caption{\label{fig:ESCE_edge} 
Relative E-field distortion (in percent) with respect to the nominal E-field in components $(E_X - E_0) / E_0 [\%]$ (left), $E_Y / E_0 [\%]$ (middle), $E_Z / E_0 [\%]$ (right).
The upper row is the result of the calculated E-field from the correction map.
The lower row is the E-field from the space charge simulation, which is used as simulation truth here.
Both of them show the E-field distortion at an upstream slice at $Z=\SI{12.8}{\centi\metre}$ in the TPC true spatial coordinates.
}
\end{figure}

Figures~\ref{fig:ESCE_central} and \ref{fig:ESCE_edge} show the calculated E-field (top) and the E-field from the simulation truth (bottom) at the central $Z$ and an upstream $Z$ slice, respectively.
We present 2D slices of the distortion in the E-field map as a percentage with respect to the nominal E-field $E_0 = 274$ V/cm in components, $(E_X - E_0) / E_0 [\%]$ (left), $E_Y / E_0 [\%]$ (middle), $E_Z / E_0 [\%]$ (right).
At central $Z$, near the cathode, $E_X$ is larger than $E_0$ by about 9\%, and near the anode $E_X$ is smaller than $E_0$ by about 6\%.
At central $Z$, the maximum value of $E_Y$ is about 15\% of $E_0$ and points outwards at the top and the bottom of the TPC.
At central $Z$, $E_Z$ is relatively small.
The E-field distortion in the $Z$ direction is less than 1\% of $E_0$.
At a upstream $Z$ slice, near the cathode, $E_X$ exceeds $E_0$ by about 1.5\% and near the anode $E_X$ is below $E_0$ by about 1\%. The vertical component $E_Y$ is about 4\% of $E_0$ and points outwards at the top and the bottom of the TPC.
The component along the beam, $E_Z$, points to the outside of the TPC.
The maximum distortion in $E_Z$ is about 10\% of $E_0$ and the position of the maximum is at $X \sim \SI{160}{\centi\metre}$ and $Y \sim \SI{0}{\centi\metre}$.
The general shape of the calculated E-field distortion agrees well with simulation truth (compare figures~\ref{fig:ESCE_central} and \ref{fig:ESCE_edge}).
The small coverage problem in the bins at the anode and the cathode occurs because the correction map of simulation truth does not correct spatial coordinates to those region in true spatial coordinates.


\begin{figure}[htbp]
\centering 
\includegraphics[width=1.0\textwidth,trim=0 0 0 0,clip]{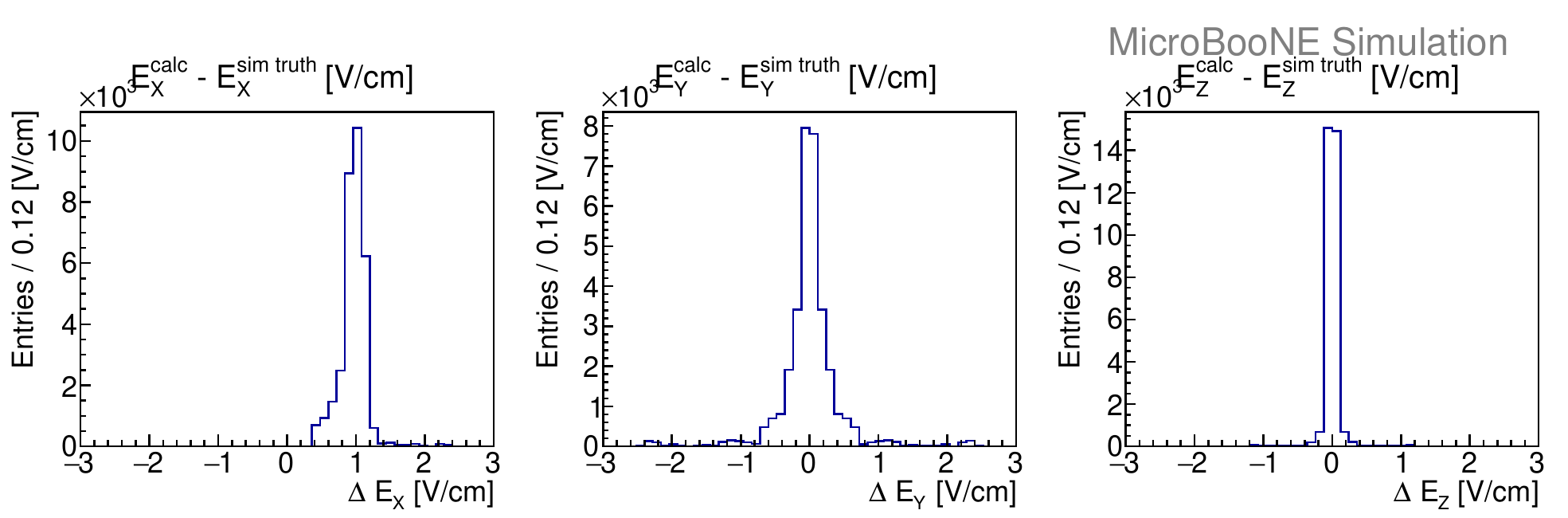}
\qquad
\caption{\label{fig:EDiff_xyz_wholeTPC} 
Distributions of bias components $E_X^\mathrm{calc} - E_X^\mathrm{sim\ truth}$, $E_Y^\mathrm{calc} - E_Y^\mathrm{sim\ truth}$ and $E_Z^\mathrm{calc} - E_Z^\mathrm{sim\ truth}$ in \SI{}{\volt\per\centi\metre}.
Each entry of the distribution corresponds to the  bias in one bin.
The distributions of $\Delta E_Y$ and $\Delta E_Z$ are narrowly peaked around 0.
Mean of $\Delta E_X$ is \SI{0.96}{\volt\per\centi\metre}.
The slight bias in $E_X$ could be due to the nominal E-field and the drift distance used in the input space charge simulation.
}    
\end{figure}

\begin{figure}[htbp]
\centering 
\includegraphics[width=0.6\textwidth]{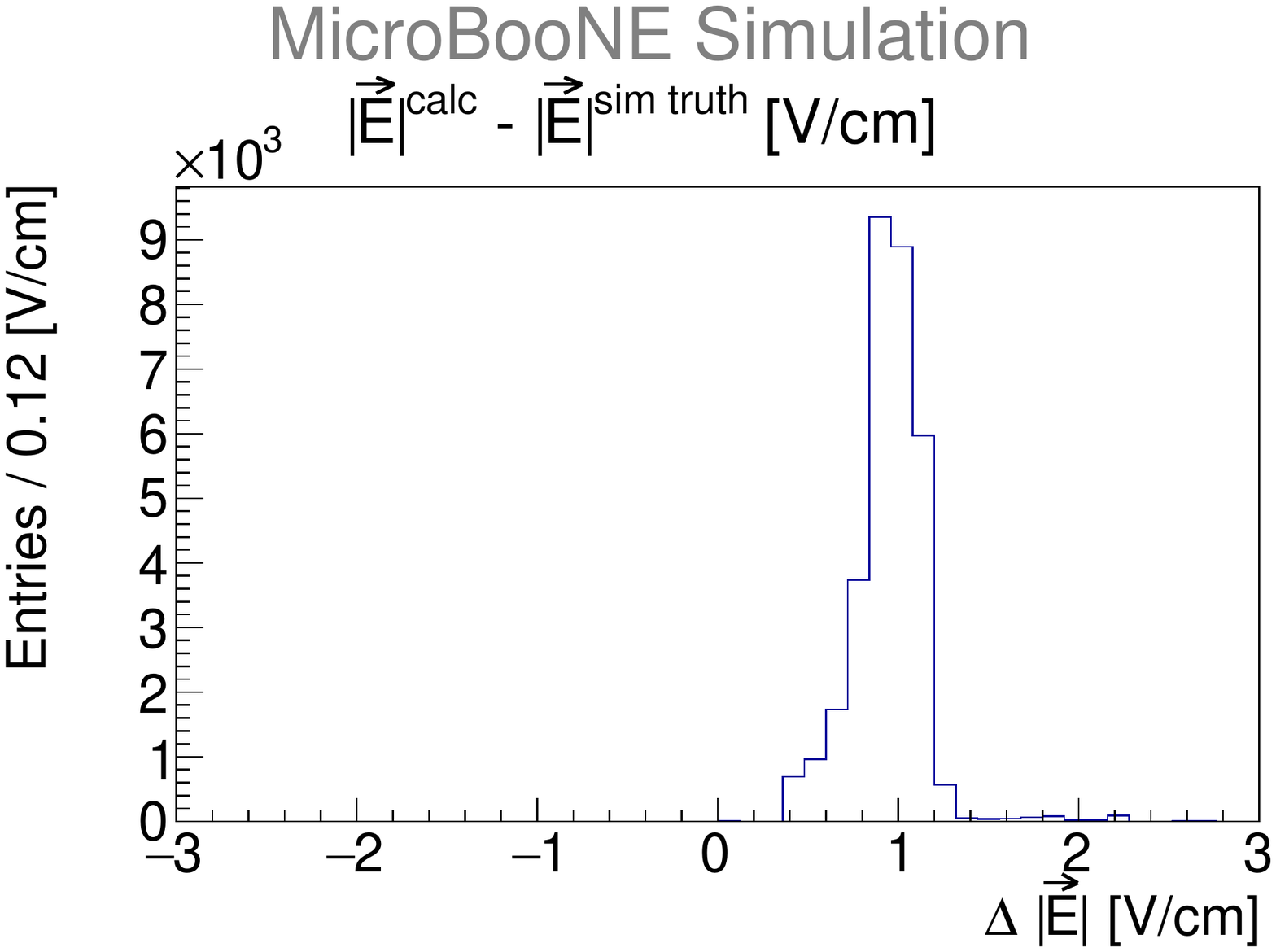}
\qquad
\caption{\label{fig:EDiff_mag_wholeTPC} 
Bias distributions of the local E-field magnitude $|\vec{E}|^\mathrm{calc} - |\vec{E}|^\mathrm{sim\ truth}$ in \SI{}{\volt\per\centi\metre}.
Each entry of the distribution is the bias in one bin.
The calculated E-field is very close to the simulated truth within \SI{0.95}{\volt\per\centi\metre}.
The small bias in the local E-field magnitude is dominated by the bias in $E_X$.
}
\end{figure}

Figure~\ref{fig:EDiff_xyz_wholeTPC} shows the absolute bias of the calculated E-field compared to the input E-field from the simulation truth in \SI{}{\volt\per\centi\metre}.
The distribution of $X$, $Y$ and $Z$ components of the bias are displayed.
The $Y$ and $Z$ distributions are all narrowly peaked around zero.
The bias of $E_X$ has a mean value of \SI{0.96}{\volt\per\centi\metre} and the standard deviation of the bias in the $X$, $Y$ and $Z$ components are \SI{0.201}{\volt\per\centi\metre}, \SI{0.451}{\volt\per\centi\metre} and \SI{0.142}{\volt\per\centi\metre}, respectively.
Figure~\ref{fig:EDiff_mag_wholeTPC} shows the absolute bias of the magnitude of the E-field in \SI{}{\volt\per\centi\metre}.
The mean and the standard deviation of the bias distribution are \SI{0.95}{\volt\per\centi\metre} and \SI{0.074}{\volt\per\centi\metre} respectively.
The bias in E-field magnitude is dominated by the bias in $E_X$, which is sensitive to the nominal E-field and the drift distance used in the input space charge simulation. 
The small bias is also visible in figure~\ref{fig:ESCE_central} and \ref{fig:ESCE_edge}.
This may be caused by our limited knowledge of the cold cathode position.
Overall, the bias between the reconstructed and simulated E-fields is negligible compared to the expected E-field distortions, which validates the method of calculating E-field and drift velocity.


\section{Measurements in MicroBooNE}
\label{sec:measurement_results}

Using laser data collected with the MicroBooNE experiment, and the track reconstruction and selection described in section~\ref{sec:Reco_select}, we extract the spatial distortion and correction maps, and the corresponding drift velocity and E-field maps.

\subsection{Distortion Maps}
\label{subsec:LaserDataResult}

With the method described in section \ref{sec:Dmap}, we derive the distortion and correction maps determined from UV-laser data.
We present 2D slices of the 3D maps and the corresponding uncertainties. 

\begin{figure}[htbp]
\centering 
\includegraphics[width=1.0\textwidth]{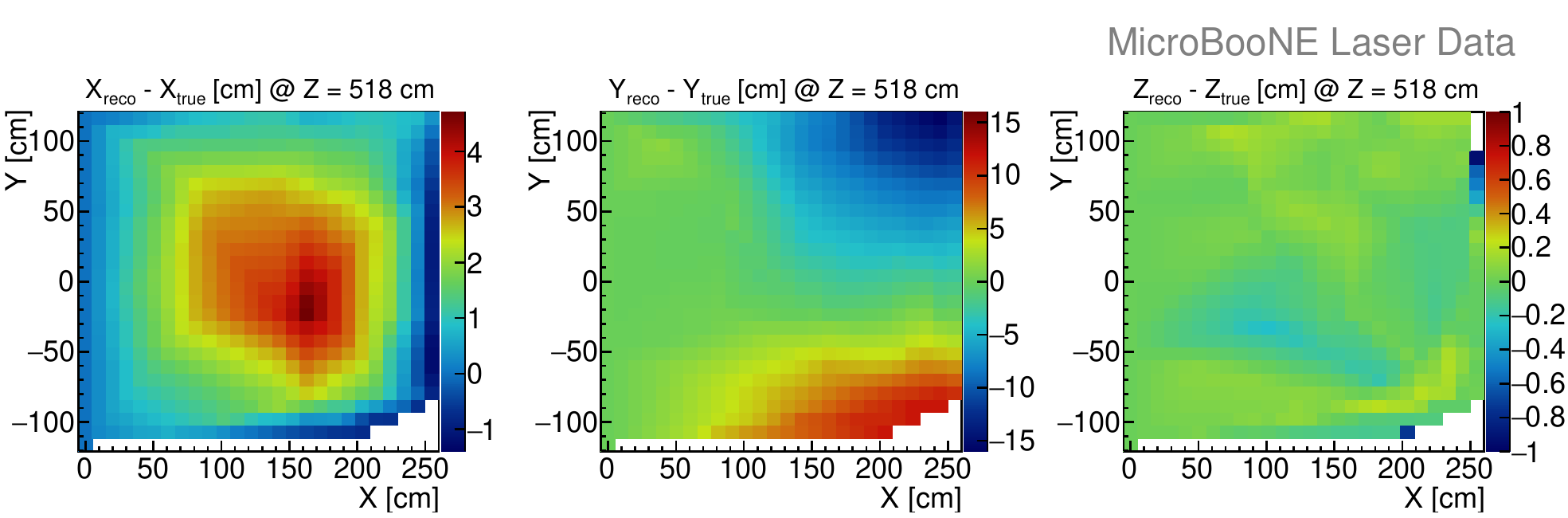}
\qquad
\caption{\label{fig:Dmap_LaserData}
Calculated distortion map from laser data at a central slice in $Z$.
Shown are the three components $dX = X_\mathrm{reco} - X_\mathrm{true}$ (left), $dY = Y_\mathrm{reco} - Y_\mathrm{true}$ (middle), and $dZ = Z_\mathrm{reco} - Z_\mathrm{true}$ (right).
In this $Z$ slice the maximum $dX$ is around \SI{4}{\centi\metre} and the maximum $dY$ is about $\pm$ \SI{15}{\centi\metre}.
The corresponding figures from simulation are figure~\ref{fig:ToySim_Dmap_centralZ} and figure~\ref{fig:LaserSim_Dmap_centralZ}.}
\end{figure}

Figure~\ref{fig:Dmap_LaserData} shows at central $Z$ the spatial distortion of $dX$, $dY$ and $dZ$ calculated from laser data. 
The corresponding figures from simulation are figure~\ref{fig:ToySim_Dmap_centralZ} and figure~\ref{fig:LaserSim_Dmap_centralZ}.
The maximal $dX$ observed in data is about \SI{4}{\centi\metre}, located around $X \approx \SI{160}{\centi\metre}$ and $Y \approx \SI{0}{\centi\metre}$.
$dX$ is larger than the simulation and beyond the calculated uncertainty, which indicates that space charge in the simulation may be underestimated. 
Close to the cathode, the upper and lower edges of $Y$ are distorted by about \SI{15}{\centi\metre} inwards in the TPC.
$dZ$ is relatively small and uniform.
The overall shape of the spatial distortion is similar to the simulated shape.

\begin{figure}[htbp]
\centering 
\includegraphics[width=1.0\textwidth]{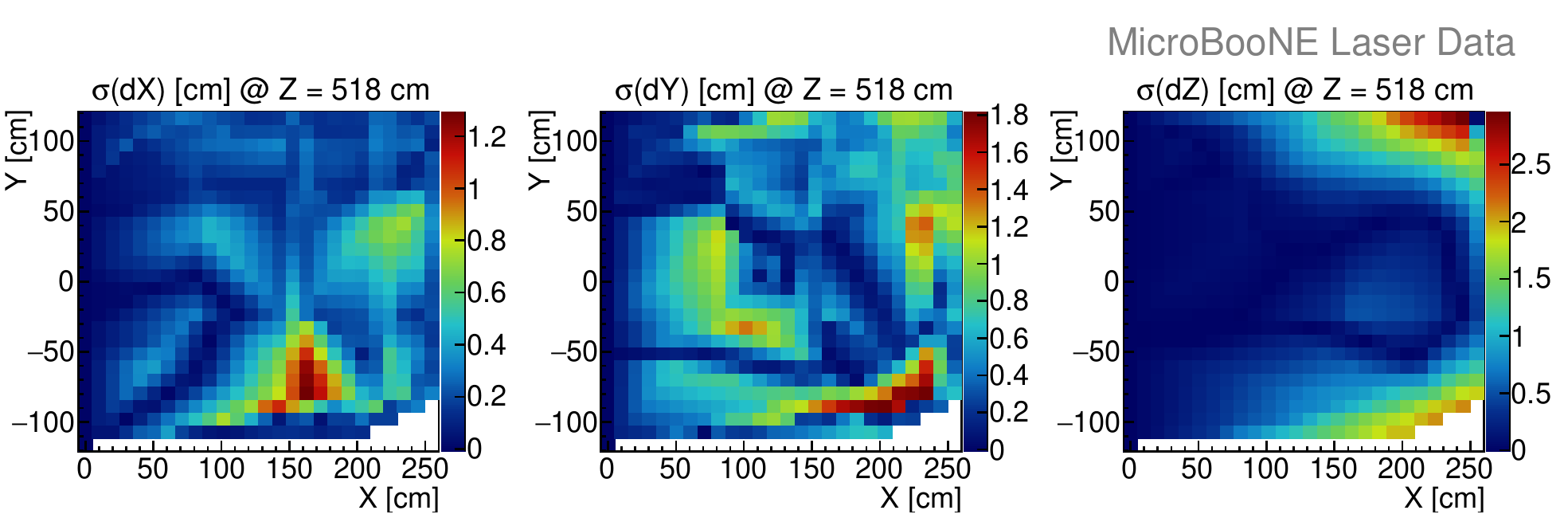}
\qquad
\caption{\label{fig:Dmap_LaserData_statErr}
Standard deviations ($\sigma$) of the spatial distortion $dX$, $dY$, $dZ$ corresponding to figure~\ref{fig:Dmap_LaserData}.}
\end{figure}

The standard deviation in each bin is calculated from 50 sub-maps.
The standard deviation for the central $Z$ slice is shown in figure~\ref{fig:Dmap_LaserData_statErr}.
It is of a similar size as the standard deviation calculated from the complete laser simulation (figure~\ref{fig:LaserSim_Dmap_statErr_centralZ}).
This is taken as the statistical uncertainty for spatial distortion in each bin.
The systematic uncertainty is estimated from the bias of the true and calculated displacement in the complete laser simulation (figure~\ref{fig:LaserSim_Diff_centralZ}).


\begin{table}
\caption{
  	Spatial distortion measured from laser data for selected representative points in the TPC.
	}
\begin{small}
\begin{center}
    \begin{tabular}{ | c || c | c | c |}
    \hline
    Position ($X$, $Y$, $Z$)  [cm] & $dX\pm\mathrm{stat.}\pm\mathrm{syst.}$ [cm] & $dY\pm\mathrm{stat.}\pm\mathrm{syst.}$ [cm] & $dZ\pm\mathrm{stat.}\pm\mathrm{syst.}$ [cm] \\ 
    \hhline{|=||=|=|=|}
    (174.11, -4.65, 518.40) & $4.22\pm 0.34 \pm 0.10$ &$ -2.00 \pm 0.26 \pm 0.60$ & $-0.01 \pm 0.44 \pm 0.11$ \\ \hline
    (30.72, -4.65, 518.40) & $1.15 \pm 0.02 \pm 0.05$ &$ -0.02 \pm 0.52 \pm 0.04$ & $-0.01\pm 0.06 \pm 0.01$ \\ \hline
    (225.32, -60.45, 518.40) & $1.11 \pm 0.58 \pm 0.37$ &$ 6.51 \pm 1.04 \pm 0.05$ & $0.09 \pm 0.68 \pm 0.14$ \\ \hline
    (225.32, 60.45, 518.40) & $1.46 \pm 0.38 \pm 0.38$ &$ -8.20 \pm 0.82 \pm 0.06$ & $-0.01 \pm 0.48 \pm 0.05$ \\ 
    \hline   
    \end{tabular} 
\end{center}
\end{small}
	\label{tab:spatial_distortion}
\end{table}






Selected spatial distortion values, with their uncertainties, are listed in table~\ref{tab:spatial_distortion}. 
Given the scale of the spatial distortions, it is crucial to correct for them in
order to achieve more accurate tracking and calorimetric information from the TPC.

\subsection{Drift Velocity Map from Laser Data}
\label{subsec:vmap}
We calculate the local drift velocity from the correction map, which is based on laser calibration data, as described in section \ref{subsec:Driftv}.

\begin{figure}[htbp]
\centering 
\includegraphics[width=1.0\textwidth]{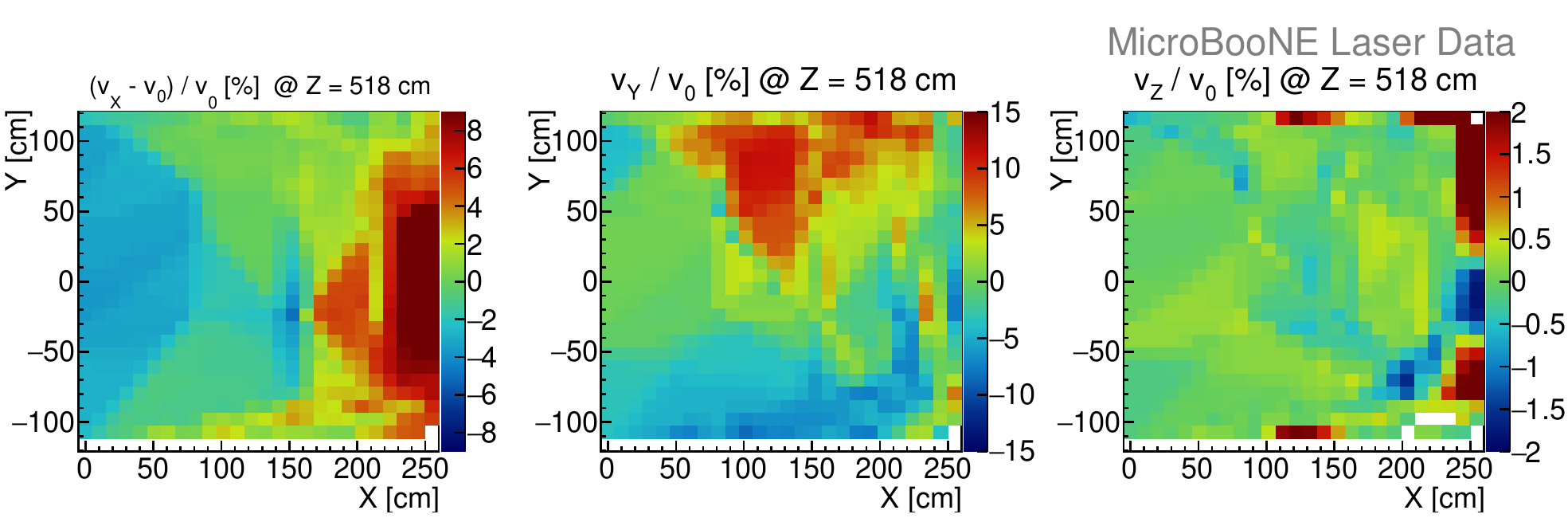}
\qquad
\caption{\label{fig:vmap_LaserData}
Relative drift velocity components (in percent)for the three coordinates $(v_X - v_0)/v_0 [\%]$ (left), $v_Y / v_0 [\%]$ (middle), and $v_Z / v_0 [\%]$ (right) at a central slice in $Z$.
The drift velocity has a maximal distortion from the nominal drift velocity $v_0$ of about 10$\%$.
}
\end{figure}

Figure~\ref{fig:vmap_LaserData} shows the drift velocity components with respect to the nominal drift velocity $v_0$.
In MicroBooNE, $v_0$ is \SI{1.098}{\milli\metre\per\micro\second}.
The nominal drift velocity is estimated by looking at the end of cathode-piercing tracks and measuring the drift time for the ionized electrons to travel from the cathode plane to the anode plane.
The uncertainty on $v_0$ comes from the determination of the cold cathode position and the residual space charge effect at the cathode.

At central $Z$, the drift velocity component along $X$ is about 10\% larger than $v_0$ near the cathode and $v_X$ is about 6\% smaller than $v_0$ near the anode.
The area with larger $v_X$ is slightly smaller than the area with smaller $v_X$.
In the upper and lower region, $v_Y$ is about \SI{0.1}{\milli\metre\per\micro\second} pointing outwards.
The vertical drift velocity, $v_Z$, is relatively small and uniform at central $Z$.


The statistical uncertainty of the drift velocity is derived from the correction map.
For each bin of the correction map, 500 correction vectors are generated according to a normal distribution with $\mu $ the central value of the correction vector and $\sigma $ the standard deviation, to produce 500 correction maps.
We then propagate these 500 correction maps into velocity maps and in each bin we obtain 500 velocity components.
We take the standard deviation from a fit to the distributions in each bin as the statistical uncertainty of the velocity measurement.
As we are eventually interested in the E-field, more details on the uncertainties are given in section~\ref{subsec:Emap}.

The systematic uncertainty of the local drift velocity is again determined with a bias study using the complete laser simulation.
The bias is the difference between the calculated drift velocity and the drift velocity from the input simulation.
The calculated drift velocity is taken from the correction map estimated using the complete laser simulation.
The drift velocity in simulation is the drift velocity taken from the relationship shown in figure~\ref{fig:driftV}, corresponding to the E-field from the space charge simulation.
A set of representative resulting drift velocities and uncertainties measured with laser data is listed in table~\ref{tab:drift_velocity}.

\begin{table}
\caption{
  Representative local drift velocities measured using MicroBooNE laser data. 
  In all cases, the uncertainty is dominated by statistical uncertainty, which is at least an order of magnitude larger than the systematic uncertainty.
}
\begin{small}
\begin{center}
    \begin{tabular}{ | c || c | c | c |}
    \hline
    Position ($X$, $Y$, $Z$) [cm] & $v_X\pm\mathrm{stat.}$ [\SI{}{\milli\metre\per\micro\second}] & $v_Y\pm\mathrm{stat.}$ [\SI{}{\milli\metre\per\micro\second}] & $v_Z\pm\mathrm{stat.}$ [\SI{}{\milli\metre\per\micro\second}] \\ \hhline{|=||=|=|=|}
    (30.58, -4.65, 518.40) & $1.060\pm 0.001 $ &$ 0.001 \pm 0.015 $ & $0.001 \pm 0.001$ \\ \hline
    (224.22, -4.65, 518.40) & $1.171 \pm 0.011 $ &$ -0.021 \pm 0.022 $ & $-0.001\pm 0.015$ \\ \hline
    (122.30, -79.05, 518.40) & $1.086 \pm 0.012 $ &$ -0.046 \pm 0.003$ & $0.001 \pm 0.006$ \\ \hline
    (122.30, 79.05, 518.40) & $1.101 \pm 0.002$ &$ 0.123 \pm 0.008$ & $0.002 \pm 0.011$ \\ 
    \hline
    \end{tabular}
\end{center}
\end{small}
\label{tab:drift_velocity}
\end{table}


\subsection{E-field Map measured from Laser Data}
\label{subsec:Emap-data}
We calculate the absolute local E-field from the correction map derived from laser data as described in section~\ref{subsec:Emap}.

\begin{figure}[htbp]
\centering 
\includegraphics[width=1.\textwidth,trim=0 0 0 0,clip]{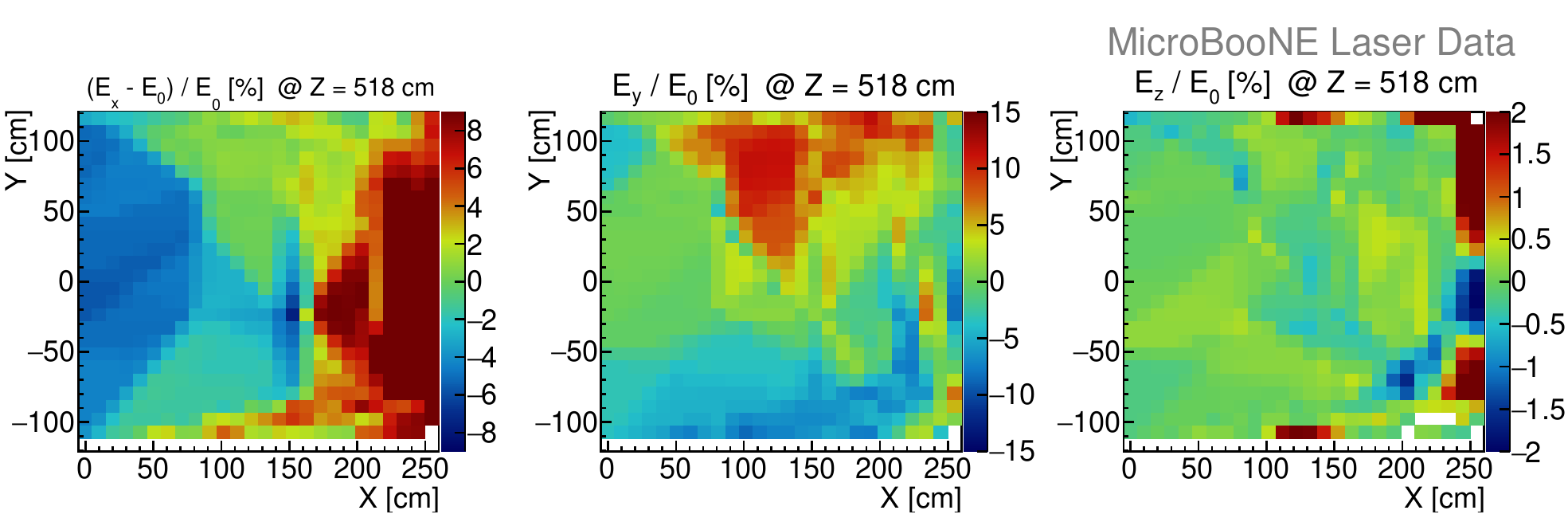}
\qquad
\caption{\label{fig:Emap-LaserData} 
Relative E-field components (in percent) for the three coordinates  $(E_X - E_0)/E_0 [\%]$ (left), $E_Y / E_0 [\%]$ (middle) and $E_Z / E_0 [\%]$ (right) at a central slice in $Z$.
The maximal relative E-field distortion is about 10\%.
}
\end{figure}

Figure~\ref{fig:Emap-LaserData} shows the components of the E-field map relative to the nominal E-field at central $Z$: $(E_X - E_0)/E_0 [\%]$, $E_Y / E_0 [\%]$ and $E_Z / E_0 [\%]$.
The nominal E-field in MicroBooNE is $E_0=\SI{273.9}{\volt\per\centi\metre}$ along the drift direction $X$, which is derived from the high voltage applied between the cathode and the anode and the estimated drift length in liquid argon.

At central $Z$, the E-field component along the drift direction $X$ is about 10\% larger than $E_0$ near the cathode and is about 6\% smaller than $E_0$ near the anode.
The area with weaker $E_X$ is slightly larger than the area with stronger $E_X$.
In the upper region, $E_Y$ is about \SI{30}{\volt\per\centi\metre} pointing upwards.
In the lower region, $E_Y$ is about \SI{15}{\volt\per\centi\metre} pointing downwards.
$E_Z$ is relatively small and flat at central $Z$, mostly within -2--\SI{2}{\volt\per\centi\metre}.
The lowest row in $Y$ is empty due to lack of laser coverage.

The dominant contribution to the E-field distortion in MicroBooNE is space charge arising from cosmic rays.
Muons from cosmic rays enter the TPC at a rate of $\sim$\SI{10}{\kilo\hertz}.
Positive charge, which drifts $\sim10^5$ times slower than the electrons in the same E-field, builds up in the form of an ion cloud near the cathode.
If ionized electrons travel through the ion cloud from the cathode side, they will be boosted by the E-field from the attraction of the ion cloud.
Along $X$ direction, the E-field near the cathode plane is typically larger than the nominal E-field.
If ionized electrons travel in the ion cloud, they are attracted from the ion cloud from both the $+X$ and $-X$ directions.
The position where the attraction reaches equilibrium in the drift direction, $E_X$ is close to $E_0$.
If the ionized electrons travel further near the anode, the attraction from the ion cloud pulls the electrons backwards.
With the static E-field from the ion cloud in the opposite direction of $E_0$, the total $E_X$ is smaller due to distance.
Similarly, ionized electrons are attracted to the ion cloud in all directions, and indeed we observe $E_Y$ pointing outwards in the upper and lower regions of the TPC.
The asymmetry of $E_Y$ in the upper and lower regions may be due to the cosmic ray muons entering the TPC at the top, with fewer ions created at the bottom.
Additionally the poor laser coverage near the bottom of the TPC may contributes to this.
Furthermore, convection of the liquid argon inside the TPC can move ion charges.
The E-field along the beam direction $E_Z$ is very small and uniform, because the space charge distribution is almost symmetric with respect to the central $Z$.

%
%

\begin{figure}[htbp]
\centering 
\includegraphics[width=0.8\textwidth,trim=0 0 0 0,clip]{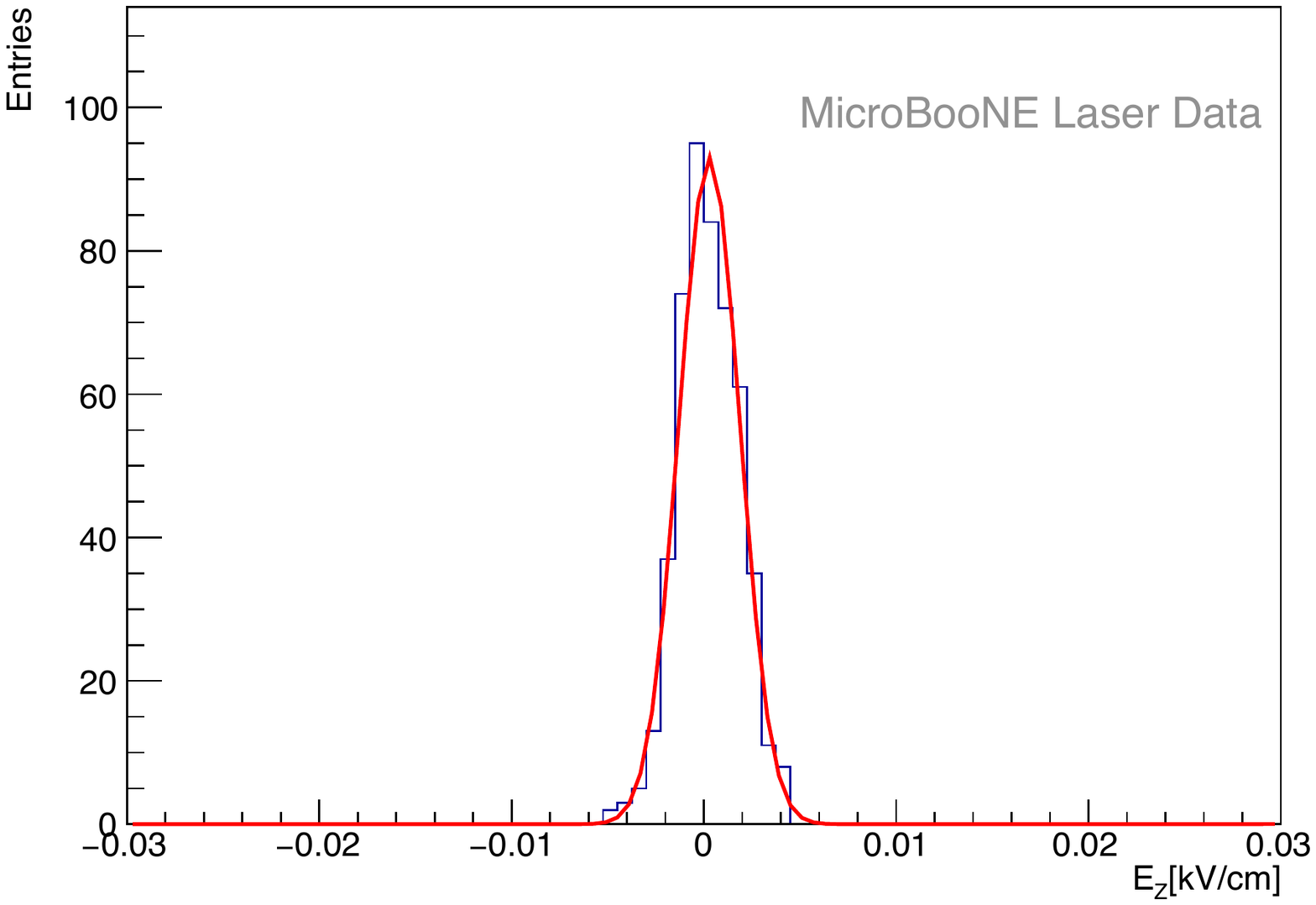}
\qquad
\caption{\label{fig:Error_propagation} 
Distribution of the vertical E-field components $E_Z$ for 500 initial correction maps used in the  error propagation.
The red line shows the Gaussian fit.
}
\end{figure}

The statistical uncertainty of the measured E-field obtained using laser data is calculated by propagating the statistical uncertainty from the correction map in the same way as was done for the velocity map (Sec~\ref{subsec:vmap}).
An example of the distribution of E-field component in a bin can be seen in figure~\ref{fig:Error_propagation}.
We take the standard deviation from a fit to the distribution in every bin as the statistical uncertainty of the E-field measurement.
In figure~\ref{fig:driftV}, around the nominal E-field in MicroBooNE $E_0$, and with E-field distortion up to \SI{50}{\volt\per\centi\metre}, the drift speed varies almost linearly with respect to the magnitude of the E-field.
As the error propagation to the E-field shows Gaussian distributions in figure~\ref{fig:Error_propagation} we use the values from the fit.

\begin{figure}[htbp]
\centering 
\includegraphics[width=1.\textwidth,trim=0 0 0 0,clip]{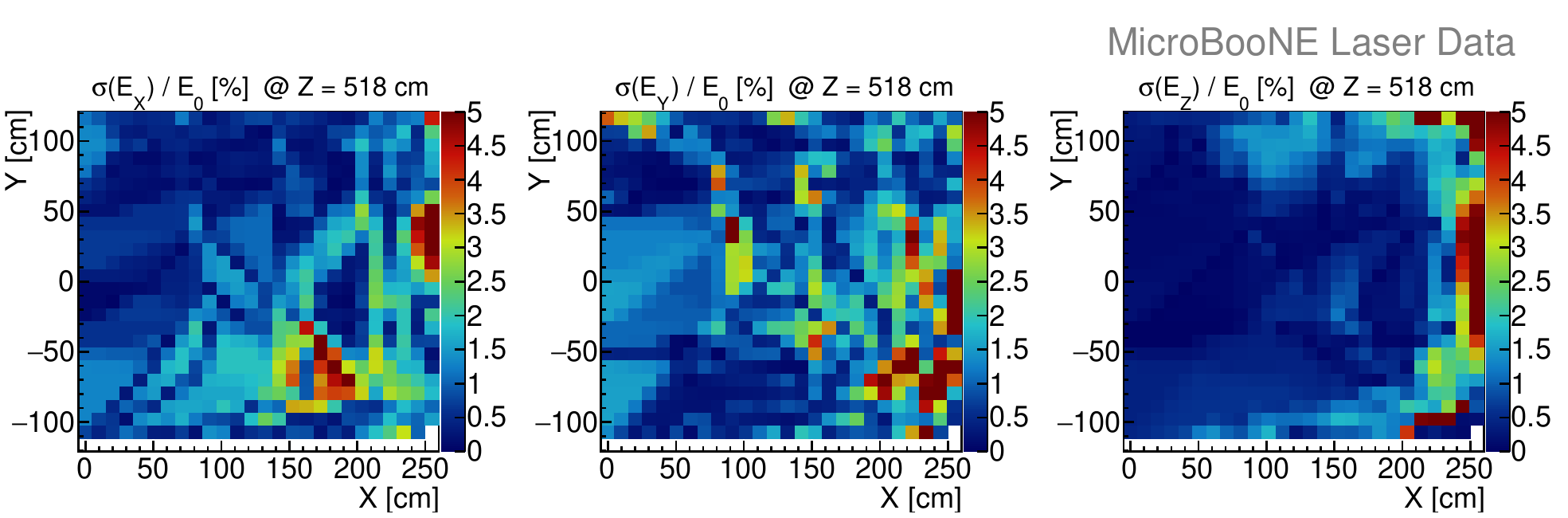}
\qquad
\caption{\label{fig:Emap-LaserData-StatErr} 
Standard deviation of the E-field components in each bin with respect to the nominal E-field in a central slice in $Z$.
From left to right the three components are shown: $\sigma(E_X)/E_0 [\%]$,  $\sigma(E_Y)/E_0 [\%]$ and  $\sigma(E_Z)/E_0 [\%]$.
The deviations are mostly within 2\% of the nominal E-field $E_0$, with the exception of some bins.
Here the standard deviation of the E-field in each bin is the statistical uncertainty of the E-field measurement with laser data.
}
\end{figure}

The statistical uncertainty shown as a percentage of $E_0$ at central $Z$ is shown in figure~\ref{fig:Emap-LaserData-StatErr}.
Typical statistical uncertainties of the E-field component in a bin are less than 2$\%$ of $E_0$.

\begin{figure}[htbp]
\centering 
\includegraphics[width=1.\textwidth,trim=0 0 0 0,clip]{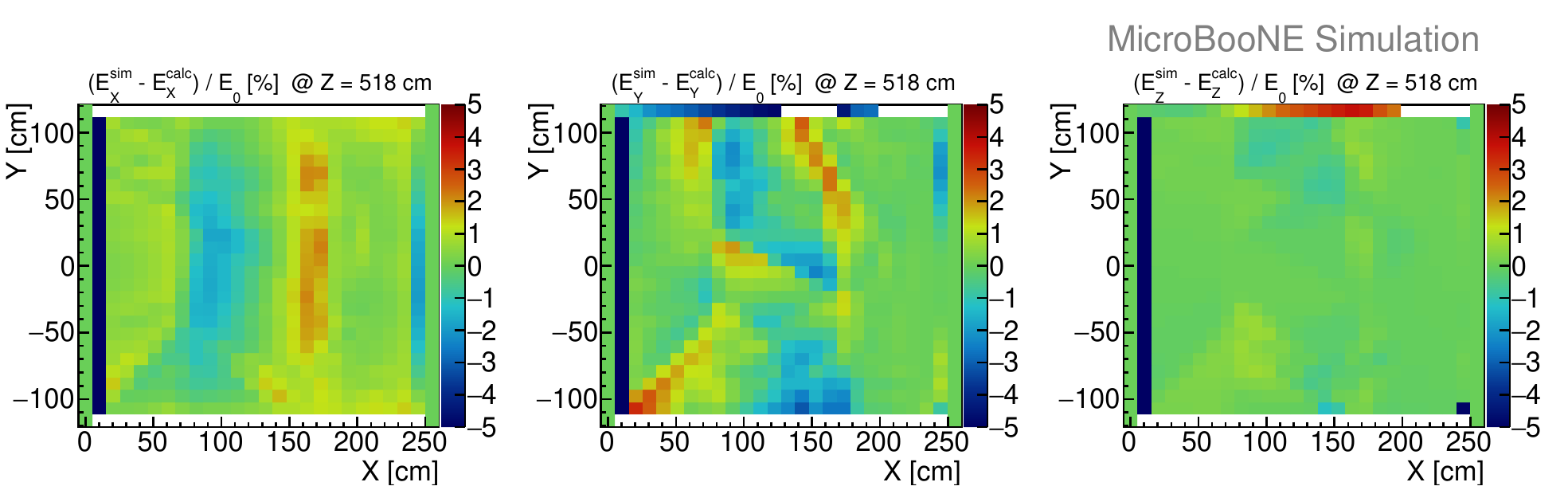}
\qquad
\caption{\label{fig:Emap-LaserData-SysErr} 
Bias of E-field components as determined from the complete laser simulation in percent of the nominal E-field at a central slice in $Z$.
From left to right the three coordinates are shown: $(E_X^\mathrm{sim}-E_X^\mathrm{calc})/E_0 [\%]$,  $(E_Y^\mathrm{sim}-E_Y^\mathrm{calc})/E_0 [\%]$ and  $(E_Z^\mathrm{sim}-E_Z^\mathrm{calc})/E_0 [\%]$.
The bias is mostly within 2\% of $E_0$.
The bias of the E-field determined by the complete laser simulation in each bin is used as systematic uncertainty for the E-field measurement with laser data.
}
\end{figure}

The systematic uncertainty of the E-field measurement performed with laser data is also determined by the bias study using the complete laser simulation.
The bias is defined as the difference between the simulated E-field (section~\ref{subsec:Emap_verification}) and the calculated E-field using the complete laser simulation described in section~\ref{subsubsec:LaserSimResult}.
The E-field in the simulation is derived from the correction map, based on the space charge simulation, and is binned in (21, 21, 81) bins in ($X$, $Y$, $Z$).
We interpolate the simulated E-field map in the same binning as the calculated one.
The deviations that we observe at the edge of the map are due to the tri-linear interpolation lacking surrounding cubes.
This systematic uncertainty, shown as a percentage of the nominal E-field $E_0$ at central $Z$, is shown in figure~\ref{fig:Emap-LaserData-SysErr}.
Typical systematic uncertainties in a bin are -2\% to 2\% $E_0$.

A set of representative E-field vectors measured by laser data is listed in table~\ref{tab:efield}.
The values are presented as relative E-field distortions in components of $\Delta E_X = E_X - E_0$,  $\Delta E_Y = E_Y$,  $\Delta E_Z = E_Z$.
The E-field distortion reaches $\sim$\SI{30}{\volt\per\centi\metre}, with typical uncertainties of a few \SI{}{\volt\per\centi\metre}.

\begin{table}
\caption{
  Representative E-field distortions measured by laser data.
}
\begin{small}
\begin{center}
    \begin{tabular}{ | c || c | c | c |}
    \hline
    Position ($X$, $Y$, $Z$) [cm] & $\Delta E_X\pm\mathrm{stat.}\pm\mathrm{syst.}$ [\SI{}{\volt\per\centi\metre}] & $\Delta E_Y\pm\mathrm{stat.}\pm\mathrm{syst.}$ [\SI{}{\volt\per\centi\metre}] & $\Delta E_Z\pm\mathrm{stat.}\pm\mathrm{syst.}$ [\SI{}{\volt\per\centi\metre}] \\ \hhline{|=||=|=|=|}
    (30.58, -4.65, 518.40) & $-14.76\pm 0.30 \pm 1.06$ &$ 0.13 \pm 3.75 \pm 0.06$ & $0.13 \pm 0.35 \pm 0.01$ \\ \hline
    (224.22, -4.65, 518.40) & $30.70 \pm 4.90 \pm 1.09$ &$ -5.51 \pm 5.69 \pm 0.31$ & $-0.20\pm 3.81 \pm 0.01$ \\ \hline
    (122.30, -79.05, 518.40) & $-4.51 \pm 4.92 \pm 1.29$ &$ -11.45 \pm 0.84 \pm 1.33$ & $0.19 \pm 1.53 \pm 0.03$ \\ \hline
    (122.30, 79.05, 518.40) & $2.23 \pm 0.71 \pm 0.29$ &$ 30.85 \pm 1.95 \pm 0.17$ & $0.55 \pm 2.85 \pm 0.72$ \\ 
    \hline
    \end{tabular}
\end{center}
\end{small}
\label{tab:efield}
\end{table}

\section{Time Stability Studies}
\label{sec:efieldtime_studies}
The time stability of the E-field in the TPC is important in order to estimate the required frequency of the E-field measurements.
It also provides an assessment of additional uncertainties of the E-field measurement for a single measurement which are not included in section~\ref{subsec:Emap}.

A time stability study of the E-field over longer time scales of months or years will be part of an upcoming publication.
In this section, we report the results of a time stability study performed over a few hours.
One of the major contributions to the E-field distortion in MicroBooNE is charge build-up from cosmic induced ions.
The positive charges induced by a single cosmic ray muon, which enters the MicroBooNE detector, typically stay in the TPC for $\mathcal{O}(\SI{}{\minute})$.
Furthermore, the LAr is constantly circulated through filters to keep the concentration of electronegative impurities low, and additionally, there is convection through heat input through the cryostat.
These convections could possibly move the charges in the LAr.

%

To investigate the stability of the E-field, we pulse the laser with the mirror set to point at a constant direction across the TPC in $Z$ over a few hours.
The frequency of the laser beam pulse is kept low, at $\sim$\SI{}{\milli\hertz}, so multi-photon ionization from the laser would not contribute to space charge.
To avoid the possible uncertainties from laser track reconstruction, explained in section~\ref{subsec:Reconstruction}, we conduct the time stability study by looking at the raw waveform signals from the wire readout.
If the E-field varies over time, the position (time in the waveform) of the charge deposition of the laser in the TPC also changes due to variations in the drift velocity and path.
Thus, the position of the laser induced waveforms in each wire over time is a good representation of the time stability.

We observe the amplitude and the sampling time of the laser signals (peak) in the waveform, where the amplitude is measured in ADC counts and the time in sample ticks.
From each raw waveform, an ADC baseline, calculated for that waveform, is subtracted.
The ADC unit is then converted to \SI{}{\milli\volt} by a known calibration factor.
The sample ticks are transformed to \SI{}{\micro\second} using the fact that every ADC sampling time takes \SI{500}{\nano\second}.
The sampling time of the laser signals is taken with respect to the laser-trigger time, given by a diode in the laser boxes and recording the laser pulse emission.
For this time stability study, only waveform signals from the collection plane are used.

\begin{figure}[htbp!]
\centering 
\includegraphics[width=0.7\textwidth]{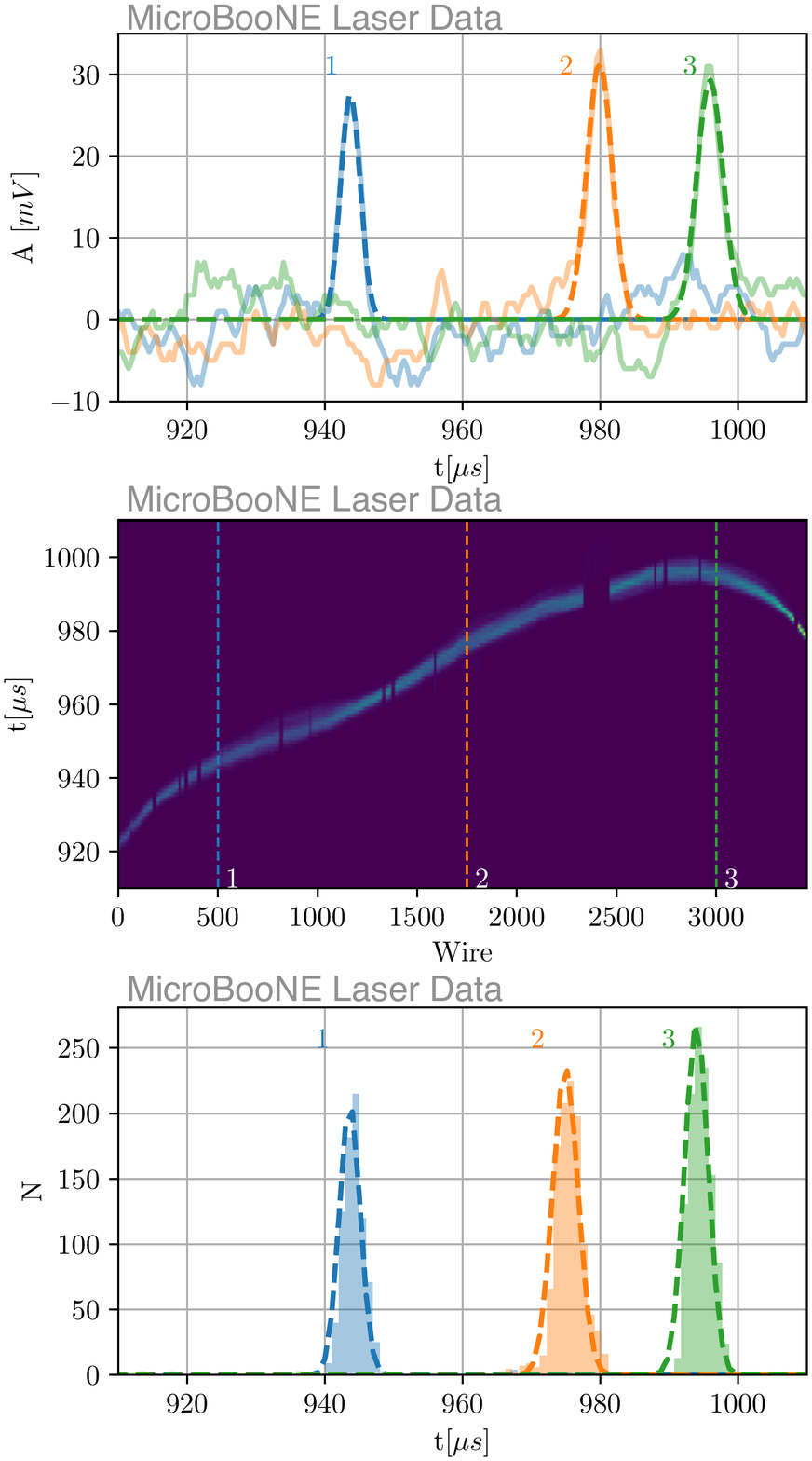}
\qquad
\caption{\label{fig:time_histo}
\textit{Top:} Example of raw waveforms of selected wires in an event: wire 500 (blue, labeled as case \textit{1}), wire 1750 (orange, labelled as case \textit{2}) and wire 3000 (green, labelled as case \textit{3}).
Each waveform is fitted by a Gaussian and the central value of the Gaussian defines the peak sampling time. 
\textit{Middle:} Overlap of 1500 laser pulses.
The tracks are nearly straight with a vertical displacement of about 6.6 cm over a track lengths of 10.4 m.
The horizontal axis in the figure is the wire number and the vertical axis is the drift time.
The color illustrated is the sum of amplitudes of the 1500 waveform.
Gaps in the color stream (laser beam) correspond to unresponsive wires in the MicroBooNE read-out.
The positions of the three wires \textit{1}, \textit{2} and \textit{3} are indicated in the plot.
\textit{Bottom:} Histogram of peak sampling times on wire \textit{1}, \textit{2} and \textit{3}.
Each histogram has 1500 entries and each entry is the peak sampling time of a waveform on that wire from a laser pulse.
Gaussian fits to each histogram result in means and standard deviations of $\mu_1= \SI{943.63}{\micro\second}$, $\sigma_1 = \SI{1.52}{\micro\second}$, $\mu_2 = \SI{974.92}{\micro\second}$, $\sigma_2 = \SI{1.83}{\micro\second}$ and $\mu_3 = \SI{993.95}{\micro\second}$, $\sigma_3 = \SI{1.70}{\micro\second}$.
}
\end{figure}

First we show the variation of the peak sampling time from three typical wires.
The middle plot in figure~\ref{fig:time_histo} shows the overlap of 1500 laser pulses from downstream to upstream (right to left).
The three wires correspond to different regions in the TPC.
The top plot of figure~\ref{fig:time_histo} shows one representative waveform out of the 1500 laser events from each chosen wire.
The baselines of these three waveforms have already been subtracted.
As can be seen in the figure, the peak amplitudes of the laser signal in these three wires are similar.
The slightly smaller peak amplitude of the first wire is due to the fact that it represents a signal from the laser beam after it traveled in LAr for more than \SI{10}{\metre}, \SI{3.75}{\metre} more than than the next selected wire.
The waveforms are fit by Gaussian functions and the mean of the Gaussian is defined as the peak sampling time.
The bottom plot of figure~\ref{fig:time_histo} shows the histograms of peak sampling times in each of the selected wires.
Each histogram only contains peak sampling times of the waveforms from a single wire.
The peak sampling time in the top plot contributes to one entry in the corresponding histogram.
We then perform a further Gaussian fit to each histogram and obtain $\mu_1= \SI{943.63}{\micro\second}$, $\sigma_1 = \SI{1.52}{\micro\second}$, $\mu_2 = \SI{974.92}{\micro\second}$, $\sigma_2 = \SI{1.83}{\micro\second}$, $\mu_3 = \SI{993.95}{\micro\second}$, and $\sigma_3 = \SI{1.70}{\micro\second}$ for the three wires, respectively.
The width, $\sigma$, of the Gaussian distribution indicates the spread in laser position in the TPC, which reflects the stability of the E-field.
In these studies the stability of the laser track itself can be neglected due to mechanical constraints and considerations.

Among all the wires of the collection plane, the largest variation from the Gaussian fit of the waveforms is found to be $\sigma_\mathrm{max} = \SI{1.93}{\micro\second}$.
On that wire, the best fit gives $\mu = \SI{956.23}{\micro\second}$, corresponding to a spread of the laser position on a single wire of $\hat{\sigma_\mathrm{max}} = \SI{2.2}{\milli\metre}$.
This implies a 0.2\% change in drift velocity along the $X$ direction ($v_X$) and a time-related variation of the E-field less than 0.3\%.

\begin{figure}[htbp!]
\centering 
\includegraphics[width=0.8\textwidth]{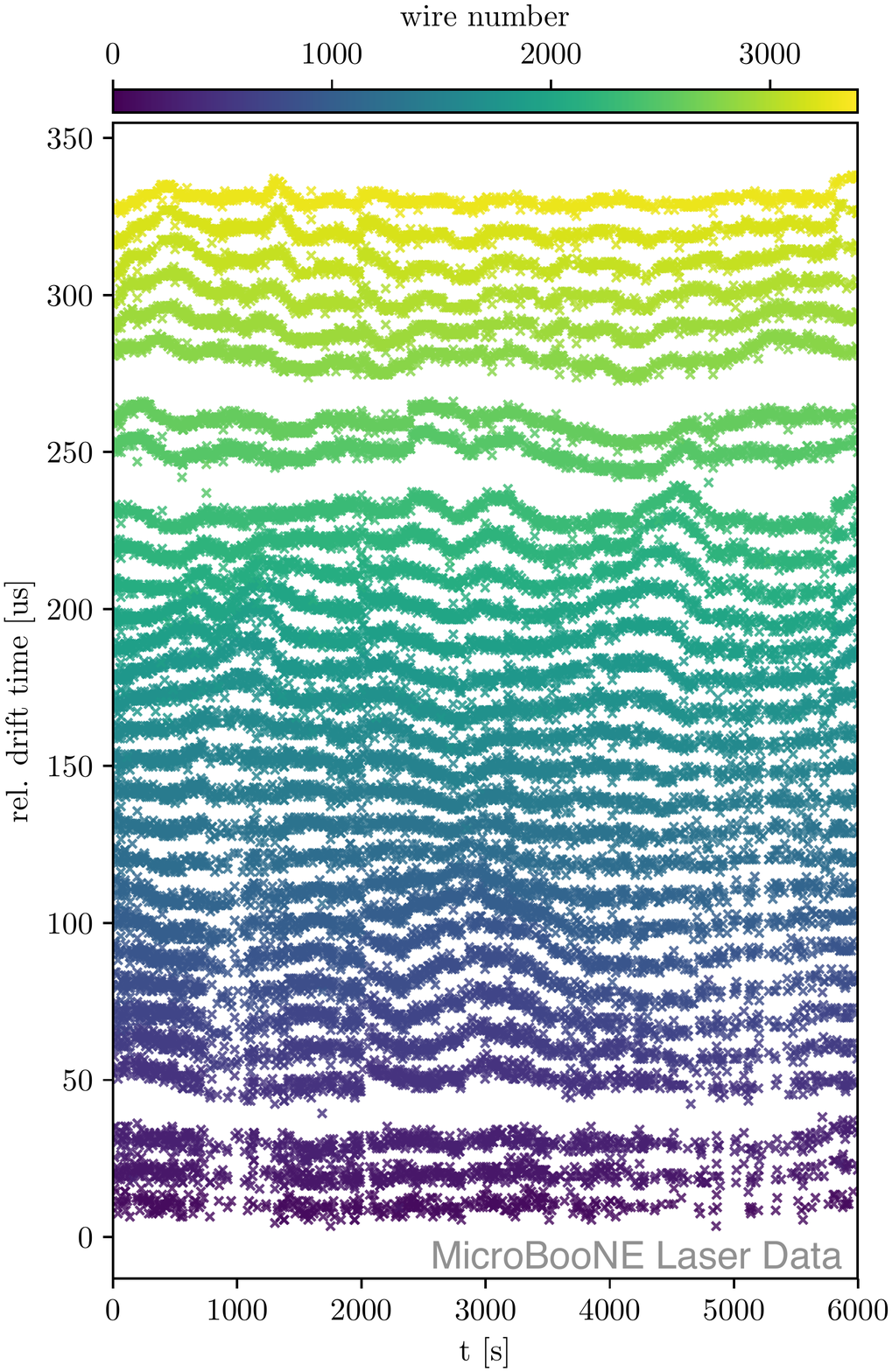}
\qquad
\caption{\label{fig:time_evolution}
Time evolution of the peak sampling time on different wires.
The horizontal axis is the time with respect to the first laser pulse.
The vertical axis shows the relative peak sampling time of different wires.
The same color indicates that peak sampling time is from the same wire.
We skip 100 wires in between every 2 wires that are shown.
For display reasons, the baseline of the peak sampling time on each displayed wire are increased by \SI{10}{\micro\second} recursively.
}
\end{figure}

Figure~\ref{fig:time_evolution} shows the peak sampling time of the laser signal for more wires over time.
The horizontal axis is the laser triggering time with respect to the first laser pulse.
The vertical axis shows the relative peak sampling time for the different wires.
For simplicity, we skip 100 wires ($\sim$\SI{30}{\centi\metre}) in between every wire that is shown.
The baseline of the peak sampling time is incremented by \SI{10}{\micro\second} from the previous wire that is shown for visibility.
The pattern of the peak sampling time over the presented 2 hour range shows correlations with respect to the neighboring wires in time.
This implies that the time-related variations are position dependent.
Thus, the variations in peak sampling time are likely to be indeed related to the E-field.

In the area with the largest distortion we find variations along $X$ of about \SI{2}{\milli\metre}, \SI{0.004}{\milli\metre\per\micro\second} and \SI{1.8}{\volt\per\centi\metre} in spatial displacement, drift velocity and E-field.
This is not considered to be significant and thus the E-field in the TPC is found to be stable over a few-hours timescale.

\FloatBarrier

\section{Conclusion}
\label{sec:conclusion}

It is vital for large LArTPCs, such as MicroBooNE and the future SBND and DUNE experiments, to have precise E-field measurements in order to minimize uncertainties in neutrino event reconstruction, since the E-field of the TPC directly impacts particle tracking, calorimetry, and light yield.  

For the first time in a large LArTPC, MicroBooNE has successfully implemented a fully steerable laser system, which can be operated remotely, and has developed a methodology to extract the local E-field, as well as the local drift velocity and spatial displacement, in the TPC active volume.
The charge induced by laser system is an excellent source to examine the TPC readout system.
We have observed laser tracks longer than \SI{10}{\metre} long spanning the length of the TPC.

By introducing laser beams along known paths, the spatial displacement of their reconstructed positions affected by E-field distortions can be measured.
We determined the E-field map of the MicroBooNE TPC by retracing the drift path of ionized electrons in the TPC E-field and comparing the derived space points to those reconstructed using a nominal E-field.
The measured E-field distortion reaches $\sim$\SI{30}{\volt\per\centi\metre} with respect to the nominal field of \SI{273.9}{\volt\per\centi\metre}, and the uncertainties are less than a few \SI{}{\volt\per\centi\metre}.
The measured maximal spatial displacement is $\sim$\SI{15}{\centi\metre}, and the uncertainties are typically less than \SI{1}{\centi\metre}.
In general, the shapes of distortions match the predicted distortions from space charge induced by cosmic ray muons, which is expected to be the largest contribution.
Additionally, a time stability study showed that the E-field distortion is stable over a time scale of a few hours.

With the laser setup in MicroBooNE, we measure the E-field, drift velocity and spatial displacement accurately in the bulk of the TPC, including throughout the fiducial volume used for most analyses.
The methods established in this work could be adopted by future experiments.
The results of these measurements are used to calibrate the position and scale of charge depositions in the MicroBooNE TPC.
Using the E-field measurement, we are able to correct for the effect of charge recombination and disentangle it from the correction needed for the electron life time. 
A good understanding of the drift velocity and E-field, as provided by this work, are essential to understand the impact of charge diffusion in a LArTPC.
The low level improvements to the charge calibration provided by this work will help to improve the reconstruction of track length, track direction, measurements of momentum by multiple Coulomb scattering and the vertex of neutrino interactions.
These improvements will all improve the performance of particle identification algorithms in MicroBooNE.
Future MicroBooNE analyses will use the E-field and spatial calibration based on this work.

\acknowledgments

This document was prepared by the MicroBooNE collaboration using the resources of the Fermi National Accelerator Laboratory (Fermilab), a U.S. Department of Energy, Office of Science, HEP User Facility. Fermilab is managed by Fermi Research Alliance, LLC (FRA), acting under Contract No. DE-AC02-07CH11359. MicroBooNE is supported by the following: the U.S. Department of Energy, Office of Science, Offices of High Energy Physics and Nuclear Physics; the U.S. National Science Foundation; the Swiss National Science Foundation; the Science and Technology Facilities Council of the United Kingdom; and The Royal Society (United Kingdom). Additional support for the laser calibration system and cosmic ray tagger was provided by the Albert Einstein Center for Fundamental Physics (Bern, Switzerland).

\bibliographystyle{JHEP}
\bibliography{bibliography.bib}








\end{document}

%% file: microboone-author-list-july2019-JINST.tex
\collaboration{MicroBooNE Collaboration}

\author[k]{C.~Adams}
\author[m]{M.~Alrashed}
\author[l]{R.~An}
\author[d]{J.~Anthony}
\author[dd]{J.~Asaadi}
\author[q]{A.~Ashkenazi}
\author[ii]{S.~Balasubramanian}
\author[j]{B.~Baller}
\author[r]{C.~Barnes}
\author[u]{G.~Barr}
% ...

% ...
\affiliation[d]{University of Cambridge, Cambridge CB3 0HE, United Kingdom}
% ...
\affiliation[j]{Fermi National Accelerator Laboratory (FNAL), Batavia, IL 60510, USA}
\affiliation[k]{Harvard University, Cambridge, MA 02138, USA}
\affiliation[l]{Illinois Institute of Technology (IIT), Chicago, IL 60616, USA}
\affiliation[m]{Kansas State University (KSU), Manhattan, KS, 66506, USA}
% ...
\affiliation[q]{Massachusetts Institute of Technology (MIT), Cambridge, MA, 02139, USA}
\affiliation[r]{University of Michigan, Ann Arbor, MI, 48109, USA}
% ...
\affiliation[u]{University of Oxford, Oxford OX1 3RH, United Kingdom}
% ...
\affiliation[dd]{University of Texas, Arlington, TX, 76019, USA}
% ...
\affiliation[ii]{Wright Laboratory, Department of Physics, Yale University, New Haven, CT, 06520, USA}

  \emailAdd{microboone\_info@fnal.gov}
\date{}
